\documentclass[aps,prb,twocolumn,amssymb,amsmath,floatfix,eqsecnum]{revtex4}
\usepackage{graphicx}
\usepackage{bm}
\usepackage{verbatim}
\usepackage{amsmath}
\usepackage{amssymb}

\usepackage[T1]{fontenc}
\usepackage{ae,aecompl}

\newcommand{\curH}{{\cal H}}
\newcommand{\curL}{{\cal L}}
\newcommand{\oh}{\frac{1}{2}}
\newcommand{\bk}{{\bf k}}

\newcommand{\ibr}{{\textbf{\em r}}}
\newcommand{\ibk}{{\textbf{\em k}}}

\newcommand{\ibQ}{{\textbf{\em Q}}}

\newcommand{\grad}{{\bm{\nabla}}}
\newcommand{\zh}{\hat{\bm z}}
\newcommand{\xh}{\hat{\bm x}}
\newcommand{\yh}{\hat{\bm y}}
\newcommand{\nh}{\hat{\bm n}}
\newcommand{\mh}{\hat{\bm m}}
\newcommand{\lh}{\hat{\bm \ell}}

\newcommand{\br}{{\bf r}}
\newcommand{\bu}{{\bf u}}
\newcommand{\bv}{{\bf v}}
\newcommand{\be}{\begin{equation}}
\newcommand{\ee}{\end{equation}}
\newcommand{\bea}{\begin{eqnarray}}
\newcommand{\eea}{\end{eqnarray}}
\newcommand{\bse}{\begin{subequations}}
\newcommand{\ese}{\end{subequations}}

\newcommand{\sumfrac}[2]{\genfrac{}{}{0cm}{3}{#1}{#2}}

\newcommand{\ve}{\vert}

\newcommand{\eps}{\varepsilon}

\def\rf#1{(\ref{#1})}
\def\rfs#1{Eq.~\rf{#1}}

\begin{document}
\title{Finite-momentum superfluidity and phase transitions in a
  $p$-wave resonant Bose gas}
\author{Sungsoo Choi}
\author{Leo Radzihovsky}
\affiliation{
Department of Physics, 
University of Colorado, 
Boulder, CO, 80309}
\date{\today}
\begin{abstract}
  We study a degenerate two-species gas of bosonic atoms interacting
  through a $p$-wave Feshbach resonance as for example realized in a
  $^{85}$Rb-$^{87}$Rb mixture.  We show that in addition to a
  conventional atomic and a $p$-wave molecular spinor-1 superfluidity
  at large positive and negative detunings, respectively, the system
  generically exhibits a finite momentum atomic-molecular
  superfluidity at intermediate detuning around the unitary point. We
  analyze the detailed nature of the corresponding phases and the
  associated quantum and thermal phase transitions.
\end{abstract}

\maketitle
\section{Introduction}
\subsection{Background and Motivation}
Since the experimental realization of Bose-Einstein condensation (BEC)
in trapped alkali-metal-atom gases~\cite{cornell.95,ketterle.95}, the
resulting burgeoning field of degenerate atomic gases has seen an
ever-expanding research activity. It has been fueled by the steady
advances in new experimental techniques to control and interrogate the
continually growing class of degenerate atomic systems.  A Feshbach
resonance (FR) has been one of these exceptionally fruitful
experimental ``knobs'' that lends exquisite tunability (via magnetic
field) of interactions in the ultracold atomic gases.  For fermionic
trapped gases, it enabled a realization of a fermionic atom-paired
$s$-wave superfluidity and exploration of its BEC-BCS crossover and
resonant universality~\cite{regal.04,
  zwierlein.04,kinast.04,chin.grimm.04,hulet.05,zwierlein.ketterle.05,
  timmermans.99,dieckmann.ketterle.02,ohara.thomas.02,salomon.03,hulet.03,duine.stoof.04,
  nozieres.85,sademelo.93,holland.01,ohashi.griffin.02,radzihovsky.04,levinsen.gurarie.06,radzihovsky.07}.

Motivated by the demonstration of $p$-wave FR in $^{40}$K and $^6$Li,
$p$-wave paired fermionic superfluidity has also been extensively
explored
theoretically~\cite{ho.diener.05,radzihovsky.05,radzihovsky.07,yip.05},
predicting to exhibit an even richer phenomenology.  A recent
laboratory production of $p$-wave Feshbach
molecules~\cite{gaebler.07,zhang.salomon.04} shows considerable
promise toward a realization of $p$-wave fermion-paired superfluidity
and the associated rich phenomenology~\cite{radzihovsky.07}, though
substantial challenges of stability
remain~\cite{gaebler.07,levinsen.cooper.gurarie.08}.

The bosonic counterparts have also been extensively explored and in
fact in the $s$-wave FR case of $^{85}$Rb~\cite{wieman.00} predate
recent fermionic developments.  As was recently
emphasized~\cite{radzihovsky.04.boson,radzihovsky.08.boson,romans.04},
in contrast to their fermionic analogs, which undergo a smooth BEC-BCS
crossover, resonant {\em bosonic} gases are predicted to exhibit
magnetic-field- and/or temperature-driven sharp phase transitions
between distinct molecular and atomic superfluid phases.  One serious
impediment to a laboratory realization of this rich physics is the
predicted~\cite{efimov.71,mueller.08} and observed~\cite{wieman.01}
instabilities of a resonantly attractive Bose gas sufficiently close
to a FR.  Nevertheless, a number of features of the
phase diagram are expected to be exhibited away from the resonance
and/or reflected in the nonequilibrium phenomenology (before the onset
of the instability) of a resonant Bose gas. Furthermore, recent
extension to an $s$-wave resonant Bose gas in an optical
lattice~\cite{zoller1.10,zoller2.10} demonstrated the stabilization
through a quantum Zeno mechanism proposed by Rempe~\cite{rempe.08},
which dates back to Bethe's~\cite{bethe} analysis of the triplet
linewidth in hydrogen.
The predictions~\cite{radzihovsky.04.boson, radzihovsky.08.boson, romans.04, zoller1.10, zoller2.10} 
have been supported by recent 
density matrix renormalization group~\cite{Ejima.Simons.11},
exact diagonalization~\cite{Bhaseen.Simons.09}, 
and quantum Monte Carlo~\cite{Bonnes.Wessel.11} studies,
as well extensions to two species~\cite{Bhaseen.Simons.09}.

Along with the ubiquitous $s$-wave resonances, recent experiments on a
$^{85}$Rb-$^{87}$Rb mixture have demonstrated an interspecies $p$-wave
FR at $B=257.8$ G~\cite{papp.thesis,papp.08}.
Although the consequences of this two-body $p$-wave resonance on the
degenerate many-body state of such a gas mixture has not been explored
experimentally, it provided the main motivation for our
recent~\cite{RCprl.09} and present studies.
We note that closely related studies of
BEC in $p$ (and higher) bands in optical lattices
have been carried out in Refs.~\onlinecite{kuklov.06, liu.06}.

The rest of the paper is organized as follows.  We conclude the
Introduction with a summary of our main results and their experimental
implications.  In Sec. \ref{sec:2channelmodel} we introduce a
microscopic two-channel $p$-wave FR model for a description of a
two-component Bose gas, as for example realized by a
$^{85}$Rb-$^{87}$Rb mixture.  Having related the parameters of the
model to two-body scattering experiments on a dilute gas, in Sec.
\ref{sec:symmetries} we present a general symmetry-based discussion of
phases and associated phase transitions expected in such an atomic gas
at finite density.  In Sec. \ref{sec:mft}, by minimizing the
corresponding imaginary-time coherent state action, we map out a
generic mean-field phase diagram for this system.  In Sec.
\ref{sec:fluctuations}, we supplement this Landau analysis with a
derivation of the corresponding Goldstone-mode Lagrangians and extract
from them the low-energy elementary excitations and dispersions
characteristic of each phase.  The true (beyond-mean-field) nature of
the quantum and thermal phase transitions is discussed in Sec.
\ref{sec:PhaseTransition}.  In Sec. \ref{sec:topol_defects} we
study the topological defects, vortices and domain walls, in each of
the phases. We make a more direct contact with cold-atom experiments
in Sec. \ref{sec:LDA} by using a local density approximation (LDA)
to include the effects of the trapping potential. We close with a
brief summary in Sec. \ref{sec:summary}.

\subsection{Summary of results}
\label{sec:results}
\begin{figure}[thb]
\includegraphics[width=8.5cm]{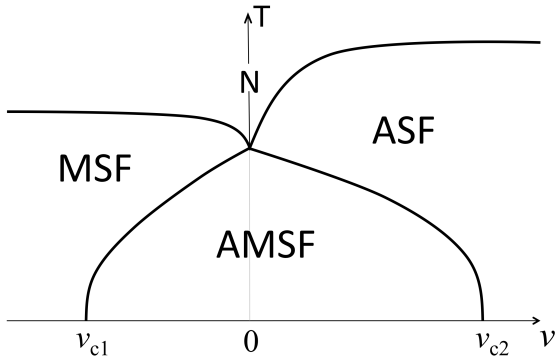} 
\caption{Schematic temperature-detuning phase diagram for a balanced
  two-species $p$-wave resonant Bose gas. As illustrated, it exhibits
  atomic (ASF), molecular (MSF), and atomic-molecular (AMSF)
  superfluid phases. The AMSF state is characterized by a
  $p$-wave, molecular, and finite-momentum $Q$ (see Fig.~\ref{fig:Q})
  atomic superfluidity.}
\label{fig:phasediagram}
\end{figure}

Before turning to the analysis of the system, we present the main
predictions of our work, a small subset of which was previously
reported in a Letter~\cite{RCprl.09}.  Our key results are
summarized by a FR temperature-detuning phase
diagram, illustrated in Fig.~\ref{fig:phasediagram}, and by the
properties of the corresponding phases and transitions.
We find that in addition to the normal (i.e., non-superfluid) 
high-temperature phase, the $p$-wave Feshbach-resonant two-component {\em
  balanced} Bose gas (e.g., equal mixture of $^{85}$Rb and $^{87}$Rb
atoms) generically exhibits three classes of superfluid phases: atomic
(ASF), molecular (MSF), and atomic-molecular (AMSF) condensates, where
atoms, $p$-wave molecules, and both are Bose-Einstein-condensed,
respectively. Our most interesting finding is that the AMSF phase,
sandwiched between (large positive detuning) ASF and (large negative
detuning) MSF phases, is necessarily a {\em finite-momentum} $\ibQ$ spinor
superfluid, akin to (but distinct from) a
supersolid~\cite{Andreev.69,Chester.70,Leggett.70,Kim.Chan.04}. It is
characterized by a momentum $\hbar \ibQ$, with its magnitude
\begin{equation}
Q=\alpha m \sqrt{n_m} \sim \sqrt{\gamma_p\ell n_m}\lesssim\sqrt{\gamma_p}/\ell,
\label{Q}
\end{equation}
tunable with a magnetic field [via FR detuning, $\nu$ that primarily
enters through the molecular condensate density $n_m(\nu)$], with
$\alpha$, $m$, $\ell$, and $\gamma_p$, respectively, the FR coupling,
atomic mass, average atom spacing, and a dimensionless measure of FR
width~\cite{radzihovsky.07}.
\begin{figure}[thb]
\includegraphics[width=6.2cm]{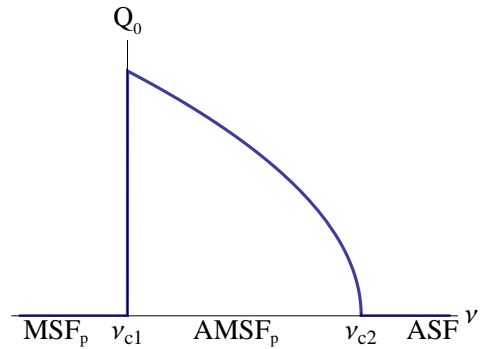} 
\caption{(Color online) Schematic momentum $Q(\nu)$ characteristic of the AMSF (polar) state,
  ranging between zero and the $p$-wave FR width-dependent value.}
\label{fig:Q}
\end{figure} 

As illustrated in the phase diagram (Fig.~\ref{fig:phasediagram}), the
ASF appears at a large positive detuning (weak FR attraction) and low
temperature, where one of the three combinations (ASF$_1$, ASF$_2$,
ASF$_{12}$) of the $^{85}$Rb and $^{87}$Rb atoms is Bose-Einstein-condensed
into a conventional, uniform superfluid and the $p$-wave
$^{85}$Rb-$^{87}$Rb molecules are energetically costly and therefore
appear only as gapped excitations.

\begin{figure}[b]
\includegraphics[width=7cm]{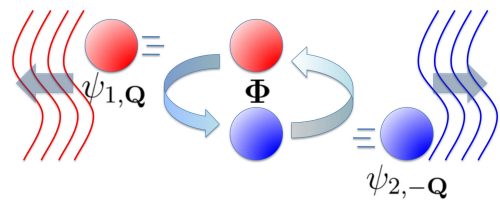} 
\caption{(Color online) A cartoon of a $p$-wave molecule decaying into two oppositely
  moving species of atoms, illustrating a resonant mechanism for a
  finite-momentum $\ibQ$ atomic superfluidity (indicated by wavy
  lines) in the AMSF phase.}
\label{cartoon}
\end{figure}

In the complementary regime of a large negative detuning, the
attraction between two flavors of atoms is sufficiently strong so as
to bind them into tight $p$-wave heteromolecules (e.g.,
$^{85}$Rb-$^{87}$Rb molecule), which at low temperature condense into
a $p$-wave superfluid, with atoms in the species-balanced case
existing only as gapped excitations.  In this tight-binding molecular
regime the gas reduces to a well-explored system of a spinor-$1$
condensate~\cite{stenger.ketterle.98,gorlitz.ketterle.03,chang.chapman.04,chang.chapman.05,kurn.08,
  ho.spinor.98,ohmi.machida.98,stamperkurn.98,matthews.cornell.98,
  bigelow.98,ho.yip.00,zhou.spinor.01,demler.zhou.02,barnett.demler.06,mukerjee.moore.06,
  mur-petit.06,lamacraft.07,mukerjee.moore.07}, with the spinor
corresponding to the relative orbital angular momentum $\ell=1$ of the
two constituent atoms of the $p$-wave molecule. Thus, for negative
detuning we predict the existence of $\ell_z=0$ ``polar'' (MSF$_{\rm
  p}$) and $\ell_z=\pm 1$ ``ferromagnetic'' (MSF$_{\rm fm}$) molecular
$p$-wave superfluid phases, with their relative stability determined
by the ratio $a_0/a_2$ of molecular spin-0 ($a_0$) to molecular spin-2
($a_2$) scattering lengths.  We find that this ratio and therefore the
first-order MSF$_{\rm p}$-MSF$_{\rm fm}$ transition are, in turn,
controlled with the $p$-wave FR detuning $\nu$, or
equivalently, the atomic $p$-wave scattering volume $v\sim 1/\nu$,
tunable with a magnetic field.

We emphasize that (in contrast to the $s$-wave
case~\cite{radzihovsky.04.boson,radzihovsky.08.boson,romans.04})
because a $p$-wave resonance does not couple a {\em uniform} atomic
condensate to the molecular one, a $p$-wave molecular condensate is
{\em not} automatically induced inside the ASF state.

The most distinctive signatures of these superfluids should be
directly detectable via time-of-flight shadow images, with the ASF
exhibiting an atomic condensate peak and the MSF displaying a $p$-wave
molecular one.  At higher densities in a trap, the bulk phase diagram
as a function of the chemical potential (see Fig.~\ref{fig:AMSFshells}
and ~\ref{fig:AMSFshellscut}) translates into shell structure of
distinct phases, which we estimated within the 
LDA~\cite{bloch.06,campbell.ketterle.06,vishveshwara.07}.

In addition to these fairly conventional {\em uniform} atomic and
molecular BECs, for intermediate detuning around a unitary point we
predict the existence of AMSF$_{\rm p}$ and AMSF$_{\rm fm}$
phases, characterized by a {\em finite} momentum $Q$ atomic
condensate~\cite{kuklov.06, RCprl.09}, 
that is a superposition of the two atomic species. Such a
generically supersolid
state~\cite{Andreev.69,Chester.70,Leggett.70,Kim.Chan.04} is always
accompanied by a $p$-wave molecular condensate, concomitantly induced
through the $p$-wave FR interaction. In addition to exhibiting an
off-diagonal long-range order (ODLRO) of an ordinary superfluid the
two AMSF$_{\rm p,fm}$ states (distinguished by the polar versus
ferromagnetic nature of their $p$-wave molecular condensates) spontaneously
partially break orientational and translational symmetries, akin to
polar and smectic liquid crystals~\cite{degennes.prost.95} and the
putative Fulde-Ferrell-Larkin-Ovchinnikov states of imbalanced paired
fermions~\cite{ff.64,lo.65, hulet.10,Sarrao.03, mizushima.05,SRprl.06,
  SRaop.07, RSreview.10}.

As illustrated in Fig.~\ref{fig:AMSFp}, in the AMSF$_{\rm p}$ state,
$\ibQ$ aligns along the quantization axis along which the molecular
condensate has a zero projection of its internal $\ell=1$ angular
momentum. For the case of the AMSF$_{\rm fm}$ state,
$\ibQ$ lies in the otherwise isotropic plane, transverse to the
$p$-wave molecular condensate axis, as illustrated in Fig.~\ref{fig:AMSFfm}.

In the narrow FR approximation we find that the AMSF$_{\rm p, fm}$
states are collinear, characterized by a single $\ibQ$ of a
Fulde-Ferrell-like form~\cite{ff.64}, as opposed to a $+\ibQ$ and
$-\ibQ$ Larkin-Ovchinnikov-like~\cite{lo.65} or other more complicated
crystalline forms, found in imbalanced paired fermionic
systems~\cite{SRprl.06, SRaop.07, RSreview.10}. However, because the
detailed spatial structure of the AMSF$_{\rm fm}$ (but not the
AMSF$_{\rm p}$ state) sensitively depends on the interactions (since it
spontaneously breaks symmetry transverse to the $\ell_z = 1$ axis), we
do not exclude a more general lattice structure in a more generic
beyond-mean-field model, which is best analyzed numerically.

The phase boundaries between this rich variety of phases can be
calculated for a narrow FR and in a dilute Bose-gas
limit, but are notoriously difficult to estimate in a strongly
interacting system, where they can only be qualitatively estimated
within a mean-field analysis.  In the former case the zero-temperature
phase boundaries are given by critical detunings:
\bse
\bea
\nu_c^{\rm MSF_p-AMSF_p} &=&
-\big(g_1+g_2-2g_{am}+\frac{m\alpha^2}{\hbar^2}\big)n_m, \,\,\,
\ \ \ \ \ \ \ \\
\nu_c^{\rm AMSF_p-ASF} &=& 
\big(2\lambda-g_{am}+\frac{m\alpha^2}{2\hbar^2}\big)n_a, \ \ \text{for} \ g_2<0, \nonumber \\
\eea
\ese
with similar expressions for transitions out of the ferromagnetic phases, 
which can be found in Eq.~\rf{fmCriticalnu}.
Here $g_i$'s and $\lambda_i$'s are molecular and atomic two-body
interaction pseudopotentials, respectively related to the background
molecular ($a_0^{bg}$ and $a_2^{bg}$) and atomic scattering lengths
($a^{bg}_{11}$, $a^{bg}_{22}$, and $a^{bg}_{12}$).

\begin{figure}[thb]
\includegraphics[width=7cm]{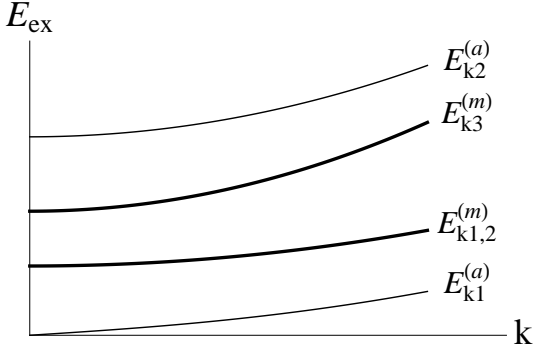}
\caption{Schematic ASF single BEC (ASF$_\sigma$) excitation spectrum.
The lowest curve is the atomic Bogoliubov mode and the upper curves 
are gapped atomic (thin) and molecular (thick) modes.}
\label{asfSingle}
\end{figure}

\begin{figure}[thb]
\includegraphics[width=7cm]{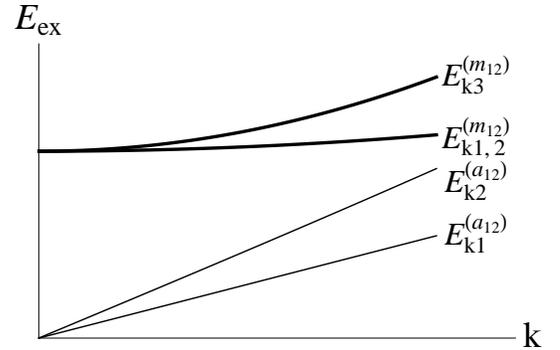} 
\caption{Schematic ASF double BEC (ASF$_{12}$) excitation spectrum.  There are
  two gapless atomic Bogoliubov modes (thin) as well as three gapped
  molecular modes (thick).}
\label{asfDouble}
\end{figure}

\begin{figure}[thb]
\includegraphics[width=7cm]{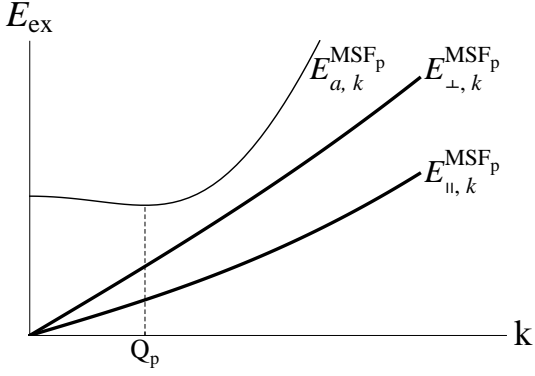} 
\caption{Schematic excitation spectrum for the
  MSF$_{\rm p}$. The doubly degenerate atomic spectrum (upper thin curve)
  exhibits a minimum gap at nonzero $k$, a precursor of finite-momentum 
  atomic condensation inside the AMSF$_{\rm p}$. The molecular spectra (thick curves),
  one longitudinal (lowest) and two degenerate transverse (middle)
  modes, are of Bogoliubov type.}
\label{msfexcPol}
\end{figure}

\begin{figure}[thb]
\includegraphics[width=7cm]{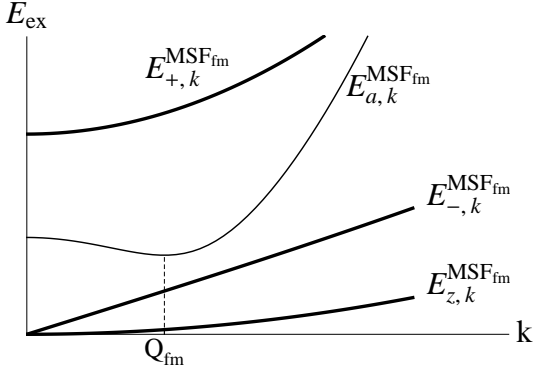} 
\caption{Schematic excitation spectrum for the
  MSF$_{\rm fm}$. The doubly degenerate atomic spectrum
  (thin curves) exhibits a minimum gap at nonzero $k$, a precursor of finite
  momentum atomic condensation. The molecular spectrum (thick curves)
  consists of a longitudinal gapless quadratic ferromagnetic spin-wave
  mode (lowest), a Bogoliubov sound mode, and a quadratic gapped mode.}
\label{msfexcFer}
\end{figure}

As any neutral superfluid, ASF, MSF, and AMSF are each characterized
by Bogoliubov modes, illustrated in Figs.~\ref{asfSingle},~\ref{asfDouble},~\ref{msfexcPol}, and~\ref{msfexcFer},
with long-wavelength acoustic ``sound'' dispersions,
\be
E^{\rm B}_\sigma({\bf k}) \approx c_\sigma \hbar k, 
\ee 
where $c_\sigma$ (with $\sigma =$ASF$_{1,2,12}$, MSF$_{\rm p,fm}$,
AMSF$_{\rm p,fm}$) are the associated sound speeds with standard
Bogoliubov form $c_\sigma\approx\sqrt{g_\sigma n_\sigma/2m}$.  In each
of these SF states one Bogoliubov mode (and only one in the ASF$_i$
states) corresponds to the overall condensate phase fluctuations. In
addition, the MSF$_{\rm p}$ exhibits two degenerate ``transverse''
Bogoliubov orientational acoustic modes. The MSF$_{\rm fm}$ is also
additionally characterized by one ``ferromagnetic'' spin-wave mode,
$E_k^{\rm MSF_{fm}}\sim k^2$ and one gapped mode, consistent with the
characteristics of a conventional spinor-1
condensate~\cite{ho.spinor.98, ohmi.machida.98}.

Because MSF$_{\rm p,fm}$'s are paired MSFs, they also
exhibit gapped single atomlike quasiparticles (akin to Bogoliubov
excitations in a fermionic paired BCS state) that do not carry a
definite atom number.  These single-particle excitations are
``squeezed'' by the presence of the molecular condensate, offering a
mechanism to realize atomic squeezed states~\cite{kimble.90}, which can
be measured by interference experiments, similar to those reported in
Ref.~\onlinecite{kasevich.01}.  The low-energy nature of these
single-atom excitations is guaranteed by the vanishing of the gap at
the MSF-AMSF transition at $\nu_c^{\rm MSF_{p,fm}-AMSF_{p,fm}}$, with
$E^{\rm gap}_{\rm MSF}(\nu_c) = 0$.

We also note that inside the MSF$_{\rm p,fm}$, for $\nu > \nu^{\rm
  p,fm}_*
=-(g_{\rm p,fm}+C_{\rm p, fm} m\alpha^2/\hbar^2)n_m$,
where $C_{\rm p,fm}=2,1$ for polar and ferromagnetic phases, respectively,
the minimum of the
single-atom excitations (that for $\nu < \nu_*$ is at $k=0$) shifts to
a finite momentum, $k\approx Q$.  This is a precursor of the atomic
gap-closing MSF-AMSF transition at $\nu_c^{\rm MSF-AMSF}$, where atoms
also Bose condense at finite momentum $Q$.

We predict that in addition to the conventional Bogoliubov superfluid
mode associated with the phase common to the atomic and molecular
condensates, the AMSF also exhibits a Goldstone mode corresponding to
the fluctuation of a {\em relative} phase between the two atomic
condensate components. Furthermore, a spatially periodic, collinear
AMSF state, characterized by at least $\pm\ibQ$ momenta (but not just
single $\ibQ$) further exhibits the condensate phonon mode $u$
corresponding to the difference between phases of the $\pm\ibQ$
condensate components, akin to the Larkin-Ovchinnikov
state~\cite{lo.65,radzihovsky.vishwanath.08,FFLOlongLeo}.

For the single $\ibQ$ AMSF states, we predict the smecticlike
``phonon'' spectra in the polar and ferromagnetic cases,
\bse 
\bea
\omega_{\rm AMSF_p}(\ibk)&=& \sqrt{(B k_z^2+K k^4_\perp)/\chi_-},\\
\omega_{\rm AMSF_{fm}}(\ibk)&=& \sqrt{(B
  k_z^2+k^2(K_x k_x^2+K_y k_y^2))/\chi_-},\nonumber
\\
\eea 
\ese
as well as the conventional Bogoliubov modes associated
with superfluid order, and an orientational mode $\omega_{\rm
  fm}^\gamma$, associated with orientational symmetry breaking in
AMSF$_{\rm fm}$
\bse
\begin{align}
\omega_{+p}(\ibk) =&\sqrt{\frac{2\rho_s}{\chi_+m}}k, \\
\omega_{\rm fm}^\gamma(\ibk) =& \sqrt{\frac{Jk^2\left[Bk^2_z+k^2(K_xk_x^2+K_yk_y^2)\right]}{J\chi_-k^2+\kappa^2k_y^2}},
\end{align}
\ese
where $B=\frac{2\hbar^2 n_a}{m}$,
$K=K_x=2K_y=\frac{\hbar^6}{2m^3\alpha^2}$, $J=\frac{\hbar^2 n_m}{4m}$,
$\kappa = \frac{\hbar^2 \sqrt{n_m}}{\alpha m}$, and
$\chi_-^{-1}=\frac{1}{2}(\lambda-\lambda_{12})$.

Having summarized the results of our study, we next turn to the
definition of the two-component $p$-wave resonant Bose-gas model,
followed by its detailed analysis.

\section{Model}
\label{sec:2channelmodel}

We study a gas mixture of two distinguishable bosonic atoms (e.g.,
$^{85}$Rb, $^{87}$Rb)~\cite{papp.08}, created by field operators
$\psi_\sigma^\dagger(\br)=\left(\psi_1^\dagger(\br),\psi_2^\dagger(\br)\right)$
and interacting through a $p$-wave FR associated with
a tunable ``closed''-channel bound state. The corresponding $p$-wave
($\ell=1$) closed-channel hetero-molecule (e.g., $^{85}$Rb-$^{87}$Rb)
is created by a Cartesian vector field operator
${\bm\phi}^\dagger(\br)=(\phi_x^\dagger,\phi_y^\dagger,\phi^\dagger_z)$,
related to $\phi_\pm^\dagger =(\phi_x^\dagger \pm
i\phi_y^\dagger)/\sqrt{2}, \phi_z^\dagger=\phi_z^\dagger$ operators,
which create closed-channel molecules in the $\ell_z=\pm 1, 0$
eigenstates, respectively.  This system is governed by a
grand-canonical Hamiltonian density (with $\hbar=1$ throughout),
\bea
\curH &=&  
\sum_{\sigma=1,2}\hat\psi_{\sigma}^{\dag} \hat{\varepsilon}_{\sigma}
\hat\psi_\sigma 
+\hat{\bm\phi}^{\dag}\cdot\hat{\omega}\cdot\hat{\bm\phi} 
+ \curH_{bg} 
\label{Hmain} \\
&&+\frac{\alpha}{2}
\left(\hat{\bm\phi}^{\dag}\cdot
\left[\hat\psi_{1}(-i\grad)\hat\psi_{2} -
\hat\psi_{2}(-i\grad)\hat\psi_{1}\right]+h.c. \right),\nonumber
\eea
where single-particle atomic and molecular Hamiltonians are given by
\bse
\bea
\hat{\varepsilon}_{\sigma} &=&-\frac{1}{2m}\grad^2-\mu_\sigma, \\
\hat{\omega}&=&-\frac{1}{4m}\grad^2-\mu_m, 
\eea
\ese 
with the effective molecular chemical potential,
\be
\mu_m=\mu_1+\mu_2-\nu,
\label{mu}
\ee
adjustable by a magnetic-field-dependent
detuning $\nu$, the latter being the rest energy of the closed-channel
molecule relative to a pair of open-channel atoms. For simplicity we have
taken atomic masses to be identical (a good approximation for the
$^{85}$Rb-$^{87}$Rb mixture that we have in mind) and will focus on
the balanced case of $\mu_1=\mu_2=\mu$, with $\mu$ fixing the total
number of $^{85}$Rb and $^{87}$Rb atoms, whether in the (open-channel)
atomic or (closed-channel) molecular form.  The FR interaction encodes
a coherent interconversion between a pair of open-channel atoms $1,2$
(in a singlet combination of $1,2$ labels, as required by bosonic
statistics) and a closed-channel $p$-wave molecule, with amplitude
$\alpha$~\cite{commentFRequalmass}.

The FR coupling $\alpha$ and detuning $\nu$ are fixed experimentally
through measurements of the low-energy two-atom $p$-wave scattering
amplitude~\cite{regal.03,gaebler.07},
\be
f_p(k) = \frac{k^2}{-v^{-1}+\frac{1}{2}k_0k^2-ik^3},
\ee
where $v$ is the scattering volume (tunable via magnetic field
dependent detuning $\nu$) and $k_0$ (negative for the FR case) is the
characteristic wave vector~\cite{landau.QM,radzihovsky.07},
\bse
\bea
v^{-1} &=& -\frac{6\pi}{m\alpha^2}(\nu-c_1), \\
k_0 &=& -\frac{12\pi}{m^2\alpha^2}(1+c_2),
\eea
\ese
respectively analogous to the scattering length $a$ and the effective
range $r_0$ in $s$-wave scattering case. In the above equations, 
constants $c_{1,2}$
are determined by the details of the $p$-wave interaction at
short scales, which in the pseudopotential model above are given
by~\cite{radzihovsky.07}
\bse
\bea
c_1 &=& \frac{m \alpha^2}{9\pi^2}\Lambda^3, \\
c_2 &=& \frac{m^2 \alpha^2}{3\pi^2}\Lambda,
\eea
\ese
where $\Lambda=2\pi/d$ is the inverse size of the closed-channel
molecular bound state, on the order of the interatomic potential range.

The $p$-wave resonance and bound-state energy are determined by the
poles of $f_p(k)$.  At low energies (where $ik^3$ can be neglected)
the energy of the pole is given by
\be
E_p=\frac{k_p^2}{2m} \approx -\frac{1}{mv|k_0|},
\ee
which is real and negative and thus is a bound-state energy for $v>0$
(negative detuning) and a finite lifetime resonance for $v<0$
(positive detuning).

In the above, for simplicity we have focused on a rotationally invariant
FR interaction, with $\hat{\omega}$ and $\alpha$ independent of the
molecular component $i$. This is an approximation for our system of
interest, the $^{85}$Rb-$^{87}$Rb mixture, where, indeed, the $p$-wave FR
around $B = 257.8$ G~\cite{papp.thesis,papp.08} is split into a
doublet by approximately $\Delta B = 0.6$ G, similar to the
fermionic case of
$^{40}$K~\cite{regal.03,gaebler.07,radzihovsky.05,radzihovsky.07}. We
leave the more realistic, richer case for future studies.

The background (nonresonant) interaction density
\be
\curH_{bg}=\curH_{a} + \curH_{m} + \curH_{am}
\label{Hbg}
\ee
is given by
\bse
\bea
\curH_a &=&\sum_{\sigma=1,2}\frac{\lambda_\sigma}{2}\hat\psi^{\dag
    2}_{\sigma}\hat\psi^{2}_{\sigma}
+ \lambda_{12}\hat\psi^{\dag}_{1}\hat\psi^{\dag}_{2}\hat\psi_{2}\hat\psi_{1}, \\
\curH_m
&=&\frac{g_1}{2}(\hat{\bm\phi}^{\dag}\cdot\hat{\bm\phi})^2
+\frac{g_2}{2}\ve\hat{\bm\phi}\cdot\hat{\bm\phi}\ve^2 \label{Hm}, \\
\curH_{am} &=&\sum_{\sigma=1,2} g_{am}
\hat\psi^{\dag}_{\sigma}\hat{\bm\phi}^\dag\cdot\hat{\bm\phi}\hat\psi_{\sigma},
\eea
\label{HaHmHam}
\ese
where coupling constants $\lambda_\sigma$, $\lambda_{12}$, $g_{1,2}$,
$g_{am}$ are related to the corresponding {\em background} $s$-wave scattering lengths
($a_1$, $a_2$, etc.)  in a standard way and thus are fixed
experimentally through measurements on the gas in a dilute
limit~\cite{regal.04}.
Correspondingly, we take these background $s$-wave couplings to be independent of the $p$-wave detuning,
an approximation that we expect to be quantitatively valid in the narrow resonance
and/or dilute limits considered here.
A miscibility of a two-component atomic gas
requires~\cite{greene.97}
\be
a_1 a_2 > a_{12}^2
\ee
which may be problematic for the case of $^{85}$Rb-$^{87}$Rb due to the
negative background scattering length of $^{85}$Rb. 

The molecular interaction couplings $g_1$, $g_2$ (set by the $L=0$ and
$L=2$ channels of $p$-wave molecule-molecule scattering), and $g_{am}$
can be derived from a combination of $s$-wave atom-atom
($\lambda_\sigma$) and $p$-wave FR ($\alpha$) interactions. We present
lowest order of this analysis in Sec.\ref{MMscattering}, which shows
that these parameters can, in principle, be tuned via a magnetic field
through the $p$-wave FR detuning $\nu$.

The above two-channel model [\rfs{Hmain}] faithfully captures the
low-energy $p$-wave resonant and $s$-wave nonresonant scattering
phenomenology of the $^{85}$Rb-$^{87}$Rb $p$-wave Feshbach-resonant
mixture~\cite{papp.08}. Its analysis at nonzero balanced atomic
densities, which is our focus here, leads to the predictions summarized
in the previous section.

\subsubsection{Lattice model}
\label{sec:lattice}

As discussed in the Introduction, based on the experience for the
$s$-wave
case~\cite{radzihovsky.04.boson,radzihovsky.08.boson,romans.04,lee.lee.04},
it is likely that a stable realization of the above continuum $p$-wave
resonant two-species bosonic model will require an introduction of an
optical lattice~\cite{zoller1.10,zoller2.10}. This leads to
a two-component atomic Hubbard model, with standard tight-binding
atomic and molecular lattice-hopping kinetic energies, density-density
interactions, and a lattice projection of the $p$-wave Feshbach
resonant coupling that in a single-band Wannier basis is given by
\begin{align}
\curH_{\rm FR_p}^{\rm lattice} &=
\frac{\alpha}{2}\sum_{\br_i,\alpha} b_{\alpha\br_i}^\dag
(a_{1,\br_i}a_{2,\br_i-{{\bm \delta}_\alpha}} 
-a_{1,\br_i}a_{2,\br_i+{{\bm \delta}_\alpha}})  + h.c.,
\end{align}
where, for example, on a cubic lattice, ${\bm \delta}_\alpha$ are lattice vectors.
A related finite angular-momentum FR lattice model was proposed and studied
in an interesting paper by Kuklov~\cite{kuklov.06},
predicting a robust $p$-wave atomic condensate in an optical lattice.
As usual\cite{FisherWeichmanGrinsteinFisher}, at low lattice filling
this lattice model reproduces the phenomenology of the continuum
model. As an additional qualitative feature, at commensurate lattice
fillings we also expect it to admit a rich variety of zero-temperature
Mott insulating phases and quantum phase transitions from them to the
superfluid ground states exhibited by the continuum system studied
here. We leave the detailed analysis of the lattice model to future
studies.

\subsubsection{Coherence-state formulation of thermodynamics}
\label{sec:coherentstate}
With the model defined by $\hat\curH$ [Eqs.\ (\ref{Hmain}),
(\ref{Hbg}), and (\ref{HaHmHam})], the thermodynamics as a function of
the chemical potential $\mu$ (or equivalently total atom density,
$n$), detuning $\nu$, and temperature $T$ can be worked out in a
standard way by computing the partition function $Z =
\text{Tr}[e^{-\beta\hat{H}}]$ ($\beta \equiv 1/k_B T$) and the
corresponding free energy $F = -k_B T\ln Z$.  The trace over quantum
mechanical many-body states can be conveniently reformulated in terms
of an imaginary-time ($\tau=i t$) functional integral over
coherent-state atomic, $\psi_\sigma(\tau,\br)$ ($\sigma=1,2$), and
molecular, $\bm\phi(\tau,\br)$, fields:
\begin{equation}
Z = \int  D\psi^*_\sigma D\psi_\sigma D{\bm\phi^*} D\bm\phi\, e^{-S},
\label{Z}
\end{equation}
where the imaginary-time action is given by~\cite{Negele.Orland}
\bse
\begin{align}
S =& \int_0^{\beta} d\tau \int d{\br} \bigg[
\psi^*_\sigma\partial_\tau \psi_\sigma +
\bm\phi^*\cdot\partial_\tau\bm\phi\nonumber\\
&+
\curH(\psi_\sigma^*,\psi_\sigma,\bm\phi^*,\bm\phi) \bigg],
\label{S}
\\
=& \int_0^{\beta} d\tau \int d{\br} \curL .
\label{cL}
\end{align}
\label{ScurL}
\ese
The Lagrangian density is given by
\begin{widetext}
\begin{eqnarray}
\curL &=& 
\psi^*_\sigma(\partial_\tau-\frac{\nabla^2}{2m}-\mu_\sigma)\psi_\sigma +
\bm\phi^*\cdot(\partial_\tau-\frac{\nabla^2}{4m}-\mu_m)\cdot\bm\phi 
+\frac{\lambda_\sigma}{2}\ve\psi_\sigma\ve^4 \nonumber \\
&+&\lambda_{12}\ve\psi_{1}\ve^2\ve\psi_{2}\ve^2
+g_{am}\left(\ve\psi_{1}\ve^2+\ve\psi_{2}\ve^2\right)\ve{\bm\phi}\ve^2 
+\frac{g_1}{2}\ve{\bm\phi}^*\cdot{\bm\phi}\ve^2
+\frac{g_2}{2}\ve{\bm\phi}\cdot{\bm\phi}\ve^2 \nonumber \\
&+&\frac{\alpha}{2}
\left({\bm\phi}^*\cdot \left[\psi_1(-i\grad)\psi_2-\psi_2(-i\grad)\psi_1\right]+c.c.\right).
\label{curlL}
\end{eqnarray}
\end{widetext}
Above (and throughout), the summation over a repeated index, as for
$\sigma$ in the first term, is implied.  

We note that closely related models also arise in completely distinct
physical contexts. These include quantum magnets that exhibit
incommensurate spin-liquid states~\cite{sachdev} and bosonic atoms in
the presence of spin-orbit interactions~\cite{huizhai}.

The associated coherent-state action $S$ is the basis of all of
our analysis in subsequent sections for the computation of the phase
diagram, the nature of the phases and excitations in each of the
corresponding phases of a $p$-wave resonant Bose gas.
\section{Phases and their symmetries}
\label{sec:symmetries}
Before turning to a microscopic analysis, it is instructive to
consider the nature of the expected phases, corresponding Goldstone
modes and associated phase transitions based on the underlying
symmetries and their spontaneous breaking.

The fully disordered symmetric state of our two-component Bose gas
confined inside an isotropic and homogeneous~\cite{boxtrap} trap
exhibits the $U_N(1)\otimes U_{\Delta N}(1)\otimes O(3)\otimes
T_r\otimes{\cal T}$ symmetries. The first two $U(1)$ groups are
associated with the total (whether in atomic or molecular form) atom
number $N=N_1+N_2+2N_m$ and the atom species number difference $\Delta
N = N_1-N_2$ conservations. The $O(3)\times T_r$ symmetries correspond
to the Euclidean group of three-dimensional rotations and translations
(in a trap-free case), and ${\cal T}$ is a symmetry of time reversal.

Since our system is composed of {\em bosonic} atoms and molecules
confined to a large trap~\cite{commentNoMI}, at sufficiently low
temperature we expect it to be a superfluid that in three dimensions
exhibits BEC, characterized by complex scalar
atomic, $\Psi_\sigma$, and/or $3$-vector molecular, ${\bm\Phi}$, order
parameters. Thus, in addition to the high-temperature normal
(non-superfluid) state, where the above order parameters all vanish and
the full symmetry $U_N(1)\otimes U_{\Delta N}(1)\otimes O(3)\otimes{\cal
  T}\otimes T_r$ is manifest~\cite{commentSymmetries}, at low
temperature we expect the system to exhibit three classes of SF
phases,
\begin{enumerate}
\item Atomic Superfluid (ASF), $\Psi_\sigma \neq 0$ and ${\bm\Phi}=0$,
\item Molecular Superfluid (MSF), $\Psi_\sigma = 0$ and ${\bm\Phi}\neq 0$,
\item Atomic Molecular Superfluid (AMSF), $\Psi_\sigma \neq 0$ and
  ${\bm\Phi}\neq 0$,
\end{enumerate}
that spontaneously break one or more of the above symmetries.
Although these phase classes resemble the previously studied phases of
an $s$-wave Feshbach-resonant
system~\cite{radzihovsky.04.boson,radzihovsky.08.boson,romans.04}, as
is clear in the following discussion, there are important
qualitative differences.

\subsection{Atomic superfluid phases, ASF}
At large {\em positive} detuning $\nu$ it is clear that the molecules
are gapped and all atoms are in the unpaired open channel. In this
regime, the gapped molecules can be neglected (or integrated out) and
the Hamiltonian \rf{Hmain} reduces to that of two bosonic atom
species, that can exhibit BEC characterized by $\Psi_1$,
$\Psi_2$ condensates. Such a two-component system is characterized by
two types of phase-diagram topologies and has been extensively
studied in the statistical physics
community~\cite{LiuFisher,KNF,Vicari}.

\begin{figure}[thb]
\includegraphics[width=8cm]{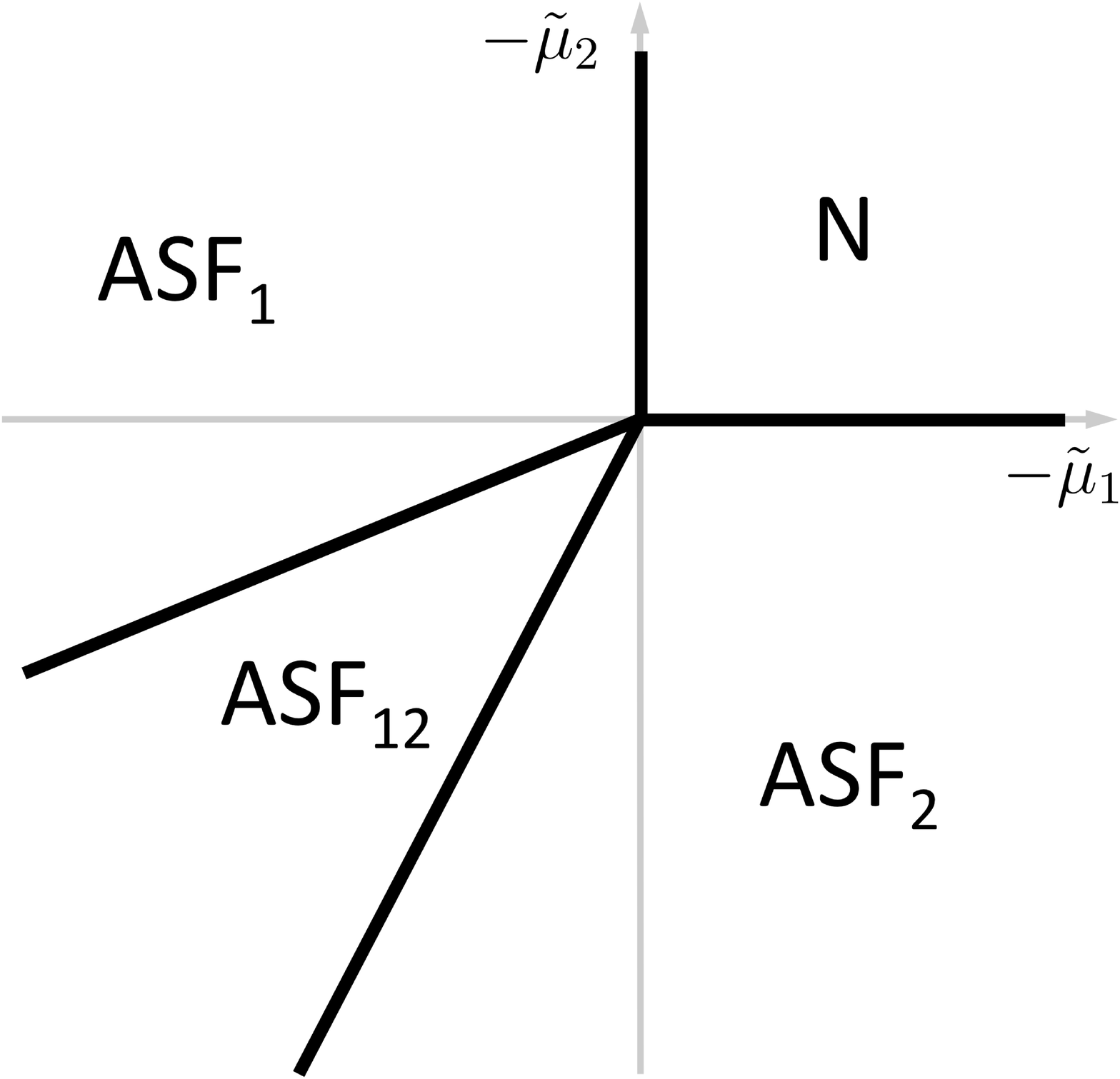}  \\
\caption{Mean-field phase diagram of a $p$-wave resonant two-component
  Bose gas for large positive detuning. Molecules are gapped, reducing
  the system to a conventional two-component Bose gas, for
  $\lambda_1\lambda_2 > \lambda_{12}^2$ displaying three types of ASF
  phases.}
\label{fig:asfphase}
\end{figure}

For $a_1 a_2 > a_{12}^2$ it admits three ASF phases,
\begin{enumerate}
\item ASF$_1$ ($\Psi_1\neq 0, \Psi_2=0$), 
\item ASF$_2$ ($\Psi_1 = 0, \Psi_2\neq 0$), 
\item ASF$_{12}$ ($\Psi_1\neq 0,\Psi_2\neq 0$),
\end{enumerate}
with ASF$_1$ and ASF$_2$ separated from ASF$_{12}$ and the normal
phases by continuous phase transitions driven by temperature and
density, or atomic polarization (or equivalently the chemical
potential imbalance) as illustrated in a mean-field phase diagram
(Fig.~\ref{fig:asfphase}).  These phases clearly break $U_1(1)$,
$U_2(1)$, or both of these symmetries, respectively, and are therefore
expected to exhibit conventional Bogoliubov modes corresponding to
these $U(1)$ symmetries.

Alternatively, for $a_1 a_2 < a_{12}^2$, the ASF$_{12}$ state is
unstable, with ASF$_{1}$ and ASF$_{2}$ separated by a first-order
transition and the associated phase separation visible in a trap.

We emphasize that, in contrast to the $s$-wave FR bosonic system (where
atomic condensation necessarily induces a molecular one, and therefore
the ASF phase is not qualitatively distinct from the $s$-wave AMSF phase,
being separated from it by a smooth
crossover)~\cite{radzihovsky.04.boson,radzihovsky.08.boson,romans.04},
for a $p$-wave FR, above $k=0$ atomic ASF condensates do {\em not}
automatically induce a $p$-wave molecular condensate since for $k=0$
the $p$-wave FR coupling vanishes. Thus, the ASF class of phases is
qualitatively distinct from the AMSF class that we discuss below.

\subsection{Molecular superfluid phases, MSF}

In the opposite limit of a large {\em negative} detuning, atoms are
gapped, tightly bound into heteromolecules that at low temperature
condense into a $p$-wave MSF. In this regime of
atomic vacuum, the gas reduces to that of interacting $p$-wave
molecules, a system quite clearly isomorphic to that of the
extensively studied $F=1$ spinor condensate~\cite{
  kurn.08,ho.spinor.98,ohmi.machida.98,
  bigelow.98,ho.yip.00,zhou.spinor.01,demler.zhou.02,barnett.demler.06,
  mukerjee.moore.06}, with the hyperfine spin $F$ here replaced by the
orbital $\ell=1$ angular momentum of two constituent atoms.

Like $F=1$ spinor condensates, the $p$-wave MSF
can exhibit two distinct phases depending on the sign of the
renormalized interaction coupling $g_2$ in \rfs{Hm}, or equivalently
the sign of the difference $a_0^{(m)}-a_2^{(m)}$ of the molecular
$L=0$ and $L=2$ channels $s$-wave scattering lengths~\cite{comment_g}.

\subsubsection{Ferromagnetic molecular superfluid, MSF$_{fm}$}
For $g_2>0$ the ground state is the so-called ``ferromagnetic''
molecular superfluid, MSF$_{\rm fm}$, characterized by an order
parameter, ${\bm\Phi}=\frac{\Phi_{\rm fm}}{\sqrt{2}}({\hat{\bm n}} \pm i{\hat{\bm m}})$, with
${\hat{\bm n}}$, ${\hat{\bm m}}$, $\hat{\bm\ell}\equiv{\hat{\bm
    n}}\times{\hat{\bm m}}$ a real orthonormal triad, $\Phi_{\rm fm}$ a real
amplitude, and the state corresponding to $\ell_z=\pm 1$ projection of the
internal molecular orbital angular momentum along the $\hat{\bm\ell}$
axis. MSF$_{\rm fm}$ spontaneously breaks the time reversal, the
$O(3)$ rotational and the global gauge symmetry $U_N(1)$, the latter
corresponding to a total atom number $N$ conservation. Inside
MSF$_{\rm fm}$ the low-energy order parameter manifold is that of the
$O(3)=SU(2)/\mathbb{Z}_2$ group, corresponding to orientations of the
orthonormal triad ${\hat{\bm n}},{\hat{\bm m}}, \hat{\bm\ell}$.

As its hyperfine spinor-condensate cousin, MSF$_{\rm fm}$, 
exhibits two gapless Goldstone modes, one linear ($\propto k$)
Bogoliubov mode associating with the broken global gauge symmetry and
another quadratic ($\propto k^2$) corresponding to the ferromagnetic
order, with associated spin waves~\cite{ho.spinor.98} reflecting the
precessional FM dynamics.

\subsubsection{Polar molecular superfluid, MSF$_{p}$}

Alternatively, for $g_2 < 0$ the ground state is the so-called
``polar''~\cite{commentTerminology} molecular superfluid, MSF$_{\rm
  p}$, characterized by a (collinear) order parameter,
${\bm\Phi}=\Phi_{\rm p}e^{i\varphi}{\hat{\bm n}}$, with ${\hat{\bm n}}$ a real
unit vector, $\varphi$ a (real) phase, and $\Phi_{\rm p}$ a (real)
order-parameter amplitude, with the state corresponding to $\ell_z=0$
projection of the internal molecular orbital angular momentum along
${\hat{\bm n}}$. MSF$_{\rm p}$ clearly spontaneously breaks rotational
symmetry by its choice of the $\ell_z=0$ quantization axis ${\hat{\bm n}}$
and the global gauge symmetry, corresponding to a total atom number
conservation. The low-energy order parameter manifold that
characterizes MSF$_{\rm p}$ is given by the coset space $(U(1)\otimes
S_2) /\mathbb{Z}_2$, admitting half-integer ``charge''
vortices~\cite{mukerjee.moore.06} akin to (but distinct from) the
$s$-wave
MSF~\cite{radzihovsky.04.boson,romans.04,radzihovsky.08.boson}.

As we demonstrate explicitly in Sec.~\ref{sec:fluctuations}, based on
symmetry we expect the polar MSF$_{\rm p}$ state to exhibit three
gapless Bogoliubov-like modes.  One corresponds to the breaking of the
global atom number conservation and two are associated with the breaking of
rotational $O(3)$ symmetry~\cite{ho.spinor.98}.

\subsection{Atomic-molecular superfluid phases, AMSF}

As detuning is increased from large negative values of the MSF$_{\rm
  p,fm}$ phases, for intermediate $\nu$ the gap to atomic excitations
decreases, closing at a critical value of $\nu_c^{\rm MSF-AMSF}$ at
which, in addition, an atomic BEC takes place. General
arguments show that this precedes the atomic condensation in the
absence of FR coupling; that is, $\nu_c^{\rm MSF-AMSF}(\alpha) < \nu_c^{\rm
  MSF-AMSF}(0)$. The features of this MSF-AMSF transition and
the AMSF phase are derived from the fact that at these intermediate
detuning, the atomic condensation necessarily takes place at a {\em
  finite} momentum $k = Q$, set by a balance of the $p$-wave FR
hybridization and the atomic kinetic energies.

We emphasize that in contrast to the $s$-wave Feshbach-resonant
bosons~\cite{radzihovsky.04.boson,radzihovsky.08.boson}, for which an
atomic condensate necessarily induces a molecular condensate, thereby
erasing a qualitative distinction between the AMSF and ASF states, for
the $p$-wave case, ASF and AMSF phases are qualitatively
distinct~\cite{radzihovsky.04.boson}. The latter is ensured by the
momentum-dependent nature of the $p$-wave coupling that breaks spatial
rotational invariance and vanishes for $\ibQ=0$.

As with other crystalline states of
matter~\cite{Chaikin.Lubensky,ff.64,lo.65}, the detailed nature of the
resulting AMSF states depends on the symmetry of the crystalline
order, set by the reciprocal lattice vectors, $\ibQ_n$, at which
condensation takes place. Determined by a detailed nature of
interactions and fluctuations, typically the nature of crystalline
order is challenging to determine generically.  Here we focus on
the {\em collinear} states, with a parallel set of $\ibQ_n=n\ibQ$, that
in the present system can be generically shown to be energetically
preferred in the AMSF$_{\rm p}$ state. There are two possible classes of
such collinear states, which are bosonic condensate analogs of the
Fulde-Ferrell (FF)~\cite{ff.64} and Larkin-Ovchinnikov
(LO)~\cite{lo.65} states, extensively studied in fermionic paired
superconductors and superfluids~\cite{SRprl.06, SRaop.07,
  RSreview.10}. The qualitative features of these classes of
finite-momentum superfluids are well captured by two simplest
representative states, one with a single $\ibQ$ and the other with a
pair of $\pm\ibQ$ condensate, which we respectively denote as
``vector'' (AMSF$^v$) and ``smectic'' (AMSF$^s$).
With a choice of $\ibQ$ the AMSF$^{v,s}$ states both
break spatial rotational symmetry. However, they are qualitatively
distinguished by the AMSF$^v$ also spontaneously breaking the
time-reversal symmetry, while remaining homogeneous, and the AMSF$^s$
instead also breaking the translational symmetry along $\ibQ$, while
remaining symmetric under the time reversal. Because within a
mean-field theory analysis it is the former, vector state that appears
to be favored, for simplicity we focus on the single $\ibQ$ AMSF
states.

The nature and symmetries of these AMSF states furthermore
qualitatively depends on the parent MSF, with
AMSF$_{\rm fm}$ and AMSF$_{\rm p}$ as two possibilities
depending on the sign of the renormalized interaction coupling
$g_2$. In addition to the symmetries already broken in its MSF parent,
by virtue of atomic condensation the AMSF state breaks the remaining
$U_{\Delta N}(1)$ global gauge symmetry associated with the
conservation of the difference in atom species number, $\Delta
N$. Other symmetries that it breaks depend on the detailed structure
of the AMSF$_{\rm fm,p}^{v,s}$ states.

\subsubsection{Polar atomic-molecular superfluid, AMSF$_{p}$}

\begin{figure}[thb]
\includegraphics[width=8.5cm]{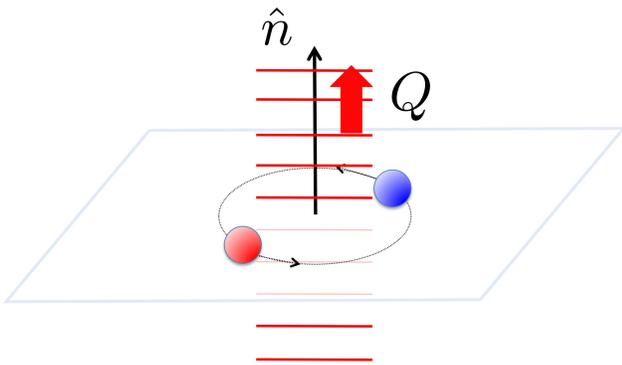}  \\
\caption{(Color online) Schematic of the AMSF$_{\rm p}$ state.  The thick arrow
  indicates the atomic condensate momentum $\ibQ$ and the $\hat n$
  arrow denotes the quantization axis along which the projection of
  molecular internal orbital angular momentum vanishes.}
\label{fig:AMSFp}
\end{figure}

The AMSF$_{\rm p}$ emerges from the MSF$_{\rm p}$. As we see in the next
section, in the AMSF$_{\rm p}$ the finite-momentum atomic condensate
orders with $\ibQ$ along the molecular condensate field $\bm\Phi$, and
therefore (as illustrated in Fig.\ref{fig:AMSFp}) for a single $\ibQ$
the vector superfluid does not break any additional spatial
symmetries.  With the molecular quantization axis, $\bm\Phi$ locked to
the atomic condensate momentum $\ibQ$, on general symmetry grounds
(simultaneous rotations of $\bm\Phi$ and $\ibQ$ is a zero-energy
Goldstone mode), we expect and indeed find that 
(see Sec.~\ref{sec:GM}) the superfluid phase
will be characterized by a smectic~\cite{Chaikin.Lubensky} Goldstone-mode 
Hamiltonian. The AMSF$_{\rm p}^s$ superfluid, in addition
breaks translational symmetry along $\bm\Phi$, with low-energy
fluctuations about this state described by a smectic phonon $u$ and a
superfluid phase $\varphi$ Goldstone modes.

\subsubsection{Ferromagnetic atomic-molecular superfluid, AMSF$_{fm}$}

\begin{figure}[thb]
\includegraphics[width=8.5cm]{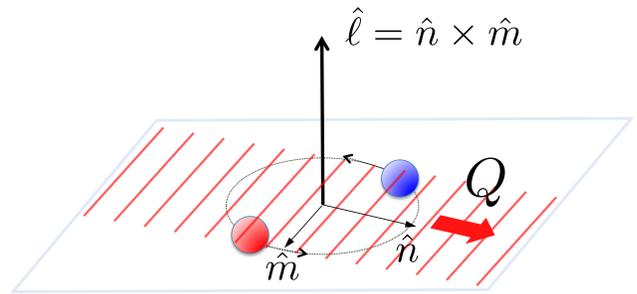}  \\\
\caption{(Color online) Schematic of the AMSF$_{\rm fm}$ state.  The
  thick arrow indicates the atomic condensate momentum $\ibQ$, lying
  in the plane transverse to the quantization axis $\hat{\bm\ell}$,
  along which the projection of the molecular internal orbital angular
  momentum is $\ell_z=+1$.}
\label{fig:AMSFfm}
\end{figure}

In contrast, a finite-momentum atomic condensation from the MSF$_{\rm
  fm}$ leads to the AMSF$_{\rm fm}$. 
In this state, a $p$-wave Feshbach resonant
interaction leads to the energetic preference for a transverse
orientation of the atomic condensate momentum $\ibQ$ to the molecular
quantization axis, $\hat{\bm \ell}=\hat{\bm n}\times\hat{\bm
  m}$. Consequently, as illustrated in Fig.~\ref{fig:AMSFfm}, the
AMSF$_{\rm fm}$ state breaks additional orientational symmetry of the
uniaxial molecular state in the plane transverse to the molecular
quantization axis $\hat{\bm\ell}$.  That is, the AMSF$_{\rm fm}$ state
is a biaxial nematic superfluid defined by $\ibQ$ and $\hat{\bm\ell}$
axes, with the superfluid phase described by a
smectic~\cite{Chaikin.Lubensky} Goldstone-mode Hamiltonian akin to that
of the FF state~\cite{FFLOlongLeo}. The latter form is enforced by the
symmetry associated with a simultaneous reorientation of atomic
momentum $\ibQ$ and molecular gauge transformation.  The biaxial
AMSF$_{\rm fm}^s$ superfluid, in addition breaks translational
symmetry along $\ibQ$, with low-energy fluctuations about this state
described by two Goldstone modes, which are a smectic phonon $u$ and a
superfluid phase $\varphi$.

\section{Mean Field Theory}
\label{sec:mft}

Our main goal in this paper is to establish the phase diagram and
nature of phase transitions exhibited by the $p$-wave
Feshbach-resonant two-component Bose gas. This requires a minimization
of the free energy which, in the presence of interactions and
fluctuations is a nontrivial function of a number of systems' physical
parameters.  However, outside the critical region, inside each phase
where fluctuations are small~\cite{commentNearTc}, we can approximate
the Landau free-energy functional $F[\Psi_\sigma,{\bm\Phi}]$ by
replacing the atomic and molecular coherent state fields with
the classical order parameters, $\Psi_\sigma(\br)$, ${\bm\Phi}(\br)$,
that minimize the action $S$ via the saddle-point method.  In the
simplest approximation, the Landau free-energy functional
$F[\Psi_\sigma,{\bm\Phi}]$ takes the form identical to
$H[\hat\psi_\sigma,\hat{\bm\phi}]$,
\begin{align}
F[\Psi_{\sigma},{\bm\Phi}] =
&\int d^3r \Bigg[\sum_{\sigma=1,2}\Big(\Psi_\sigma^*\tilde\varepsilon_{\sigma}\Psi_\sigma
+\frac{\tilde\lambda_\sigma}{2}\ve\Psi_\sigma\ve^4\Big) \nonumber \\
+&\tilde\lambda_{12}\ve\Psi_{1}\ve^2\ve\Psi_{2}\ve^2
+\tilde g_{am}\left(\ve\Psi_{1}\ve^2+\ve\Psi_{2}\ve^2\right)\ve{\bm\Phi}\ve^2 \nonumber \\
+&{\bm\Phi}^*\cdot\tilde\omega\cdot{\bm\Phi} +\frac{\tilde g_1}{2}\ve{\bm\Phi}^*\cdot{\bm\Phi}\ve^2
+\frac{\tilde g_2}{2}\ve{\bm\Phi}\cdot{\bm\Phi}\ve^2 \nonumber \\
+&\frac{\tilde\alpha}{2}
\left({\bm\Phi}^*\cdot
\left[\Psi_1(-i\grad)\Psi_2-\Psi_2(-i\grad)\Psi_1)\right]+c.c.\right)\Bigg],
\label{landauFE0}
\end{align}
with the effective couplings
$(\tilde\mu_\sigma,\tilde\mu_m,\tilde\lambda_\sigma,\dots)$, which are
functions of the microscopic parameters
$(\mu_\sigma,\mu_m,\lambda_\sigma,\dots)$ in Eq.~\eqref{Hmain},
encoding all the complexity of the fluctuations and interactions on short scales.
Though nontrivial, these parameters are, in principle, derivable from
the Hamiltonian. However, we are not concerned with this aspect of
the problem. Instead, our goal is to capture the qualitative form of
the phase diagram, taking fluctuations into account only when they
qualitatively modify the nature of the phases and phase transitions.
For simplicity of notation, we therefore neglect the distinction
between the microscopic and effective couplings, dropping tildes.

\subsection{Order parameters} 

We begin by introducing order parameters that in mean-field
approximation completely characterize the states of the system.  In
contrast to a conventional ($s$-wave interacting) Bose gas,
anticipating the energetics, we allow the atomic condensates
$\Psi_1(\br)$ and $\Psi_2(\br)$ to be complex {\em periodic} functions
characterized by momenta $\ibQ_n$, with the simplest {\em single}
$\ibQ_1=\ibQ$ form given by
\bse
\bea
\psi_{1}(\br) &\rightarrow& \Psi_1(\br)= \Psi_{1,\ibQ} e^{i\ibQ\cdot\br}, \label{atomOP1}\\
\psi_{2}(\br) &\rightarrow& \Psi_2(\br)= \Psi_{2,-\ibQ} e^{-i\ibQ\cdot\br}, \label{atomOP2}\\
{\bm\phi}(\br) &\rightarrow& {\bm\Phi} \label{molOP},
\eea
\ese
where $\bm\Phi$ is a complex 3-vector order parameter characteristic
of the $\ell=1$ molecular condensate and the choice of $\pm\ibQ$
momentum relation for the two atomic condensate fields is dictated by
momentum conservation. 

More generally, the atomic condensate order parameter is given by
\bse
\bea
\Psi_{\sigma}(\br)&=&
\begin{pmatrix}
\Psi_1(\br) \\ 
\Psi_2(\br)
\end{pmatrix},\\
&=&\sum_{\ibQ_n}
\begin{pmatrix}
\Psi_{1,\ibQ_n}e^{i\ibQ_n\cdot\br} \\
\Psi_{2,-\ibQ_n}e^{-i\ibQ_n\cdot\br}
\end{pmatrix}.
\eea
\ese
However, as alluded to in the previous section, based on the
energetics of the model, we expect that for most of the phase diagram
a single $\ibQ_n=\ibQ$ and double $\ibQ_n=\pm\ibQ$ collinear forms of the
atomic order parameters are sufficient to capture the ground-state
atomic condensates. The latter LO-like form can
equivalently, more simply be written as
\be
\Psi_\sigma(\br) = \Psi_{\sigma,\ibQ}e^{i\ibQ\cdot\br} + \Psi_{\sigma,-\ibQ}e^{-i\ibQ\cdot\br},
\ee
with $\Psi_{\sigma,\pm\ibQ}$, $\Phi$, and $\ibQ$ to be determined by the
minimization of the mean-field free energy.  As we demonstrate in
Appendix~\ref{OPstructure}, it is the single $\ibQ$ (FF-like)
condensate that is preferred energetically in a mean-field
approximation and is therefore the primary focus of the
analysis presented here.

The molecular condensate complex order parameter ${\bm\Phi}$ can, in
general, be decomposed in terms of orthonormal real 3-vectors $\bu$ and
$\bv$,~\cite{radzihovsky.07}
\be
{\bm\Phi} = \bu + i \bv.
\ee
As we demonstrate explicitly shortly, in this representation the 
two possible  $\ell=1$ MSFs, ferromagnetic and polar
condensates are described by
\bse
\begin{align}
\bu\perp\bv,\ &\mbox{``ferromagnetic'', $\ell_z=\pm1$ condensate},\\
\bu\parallel\bv,\ &\mbox{``polar'', $\ell_z=0$ condensate},
\end{align}
\ese
where for ferromagnetic state $u=v$ and the polar state can obviously
be equivalently characterized by a vanishing of one (but not both) of
$\bu$ and $\bv$.  These two molecular condensate states are the
bosonic analogs of the $p_x + ip_y$ and $p_x$ $p$-wave paired
superfluids~\cite{radzihovsky.05,radzihovsky.07}.  

We next consider the Landau free energy as a function of these atomic
and molecular order parameters and, by minimizing it for a range of
experimentally tunable parameters, compute the mean-field phase
diagram for this $p$-wave resonant two-component Bose gas.

\subsection{Atomic Superfluid (ASF)}
As is clear from Eqs.~\rf{mu} and \rf{landauFE0} for large {\it positive} detuning,
$\nu$, the molecular chemical potential $\mu_m < 0$ is negative, with
molecules gapped and therefore the ground state is a molecular vacuum.
We can thus safely integrate out the small Gaussian molecular
excitations, leading to an effective atomic free energy,
\bea
F_a[\Psi_\sigma] &\approx& F[\Psi_\sigma,0] \nonumber \\
&\approx&\int d^3r\Bigg[
\sum_{\sigma=1,2}\left(\Psi_\sigma^*\hat\varepsilon_\sigma\Psi_\sigma
+\frac{\lambda_\sigma}{2}\ve\Psi_\sigma\ve^4\right) \nonumber \\
&&+\lambda_{12}\ve\Psi_{1}\ve^2\ve\Psi_{2}\ve^2\Bigg],
\eea
with coefficients that are only slightly modified from their bare
values in Eq.~\rf{landauFE0}. This functional is a special $U(1)\otimes U(1)$
form of a $O(N)\otimes O(M)$ model, first studied many years ago by
Fisher {\it et al}. and more recently in magnetic and many other
contexts~\cite{LiuFisher,KNF,Vicari}. This free energy is clearly
minimized by a spatially uniform atomic order parameter,
$\Psi_\sigma$, giving
\bse
\begin{align}
f_{\rm asf} =& F[\ve\Psi_\sigma\ve,0]/V \\
=& \sum_{\sigma=1,2}\big[ -\mu_\sigma\ve\Psi_\sigma\ve^2
+\frac{\lambda_\sigma}{2}\ve\Psi_\sigma\ve^4\big]
+\lambda_{12}\ve\Psi_1\ve^2 \ve\Psi_2\ve^2
\end{align}
\ese
as the ASF free-energy density.

A minimization of $f_{\rm asf}$, leads to four states corresponding to
condensed and normal (nonsuperfluid) combinations of the two-component
Bose gas. For both negative chemical potentials, $\mu_1 < 0$,
$\mu_2<0$, both atoms are in the noncondensed, normal (N) phase
\be
\ve\Psi_1\ve=\ve\Psi_2\ve=0.
\ee
On a lattice (e.g., generated by a periodic optical
potential~\cite{Bloch.08}) at commensurate atom filling, this would
correspond to a Mott insulating phase extending down to zero
temperature. In a continuum (e.g., a trap), the normal state can only
be realized by heating the gas above its degeneracy temperature.

As physical parameters are varied (e.g., a weaker periodic potential,
lower temperature, and higher density for one of the atomic species)
for asymmetric mixture (different densities and/or masses), one of the
two atomic chemical potentials, $\mu_1,\mu_2$ can turn
positive, leading to a conventional normal-superfluid transition to
ASF$_1$ or ASF$_2$ states, respectively. The order parameters and
mean-field phase boundaries in each of these conventional
single-component atomic BECs are given by
\bse
\bea
\mbox{ASF$_1$:}\ \ \ \
\Psi_1 &=& \sqrt{\frac{\mu_1}{\lambda_1}}, 
\Psi_2 = 0, \text{for}\; \mu_1>0, 
\; \mu_2 <\frac{\lambda_{12}}{\lambda_1}\mu_1,
\nonumber\\
&&\\
\mbox{ASF$_2$:}\ \ \ \
\Psi_1 &=& 0, 
\Psi_2 = \sqrt{\frac{\mu_2}{\lambda_2}}, 
\text{for}\; \mu_2>0, 
\; \mu_1 <\frac{\lambda_{12}}{\lambda_2}\mu_2.
\nonumber\\
&&
\eea
\ese
We note that generically for a {\em symmetric} two-component Bose
mixture, these phases will be avoided by symmetry.

Further changes in the system's parameters, so as to drive both chemical
potentials positive, for $\lambda_1\lambda_2 > \lambda_{12}^2$ leads to
ASF$_1$ - ASF$_{12}$ or ASF$_2$ - ASF$_{12}$ transitions. The
resulting two-component condensate, ASF$_{12}$, is characterized by two
nonzero atomic condensates and mean-field phase boundaries given by
\bea
&&\mbox{ASF$_{12}$:} \nonumber\\
&&\Psi_1 =\left[ 
\frac{\lambda_2\mu_1-\lambda_{12}\mu_2}
{\lambda_1\lambda_2-\lambda_{12}^2}\right]^{\frac{1}{2}},
\Psi_2 =\left[ \frac{\lambda_1\mu_2-\lambda_{12}\mu_1}
{\lambda_1\lambda_2-\lambda_{12}^2}\right]^{\frac{1}{2}}, 
\nonumber \\
&&\mbox{for $\mu_1>0, \mu_2>0,
\frac{\lambda_2}{\lambda_{12}}>\frac{\mu_2}{\mu_1}
>\frac{\lambda_{12}}{\lambda_{1}}$}.
\eea
These classical phase transitions are generically continuous, in the
XY universality class, breaking the associated $U(1)$ symmetries.  The
N-ASF$_{12}$ transition only takes place in a fine-tuned balanced
mixture $\mu_1=\mu_2$ (which is our primarily focus here) going
directly through a tetracritical point, $\mu_1=\mu_2 = 0$.  Extensive
studies demonstrate it to be in the {\it decoupled} universality
class~\cite{LiuFisher,KNF,Vicari}.

For $\lambda_1\lambda_2 <\lambda_{12}^2$, the ASF$_1$ and ASF$_2$
energies cross before either becomes locally unstable. Consequently,
instead of continuous transitions to the ASF$_{12}$ state, the
two-component ASF$_{12}$ is absent and the ASF$_1$ and ASF$_2$ phases
are separated by a first-order transition, located at
\be
\mu_2 = \sqrt{\frac{\lambda_2}{\lambda_1}}\mu_1
\ee
which terminates at a bicritical point. On this critical line the
ASF$_1$ and ASF$_2$ states coexist and spatially phase separate.

\subsection{Molecular Superfluid (MSF)}

In the opposite limit of large {\it negative} detuning, that is, $-\nu \gg
\ve\mu\ve$, open-channel atoms are gapped and the ground state
is an atomic vacuum. Hence, for $\mu < 0 $ the free energy
$F[\Psi_\sigma, {\bm\Phi}]$ is minimized by $\Psi_\sigma=0$ and a
uniform molecular condensate $\bm\Phi$.  The free-energy density then
reduces to
\bse
\begin{align}
f_{\rm msf}[\bm\Phi] &= F[0,\bm\Phi]/V, \nonumber \\
&= -\mu_m\ve{\bm\Phi}\ve^2 +\frac{g_1}{2}\ve{\bm\Phi}^*\cdot{\bm\Phi}\ve^2
+\frac{g_2}{2}\ve{\bm\Phi}\cdot{\bm\Phi}\ve^2, \\
&= -\mu_m(u^2+v^2)+\frac{g_1}{2}(u^2+v^2)^2+\frac{g_2}{2}(u^2-v^2)^2,
\end{align}
\ese
identical to a spinor-1 bosonic condensate, corresponding to the
$\ell=1$ molecular Bose gas.  Thus, the thermodynamics and low-energy
excitations of the MSF are isomorphic to that of the well-studied
spin-1 Bose-Einstein condensate~\cite{ho.spinor.98, ohmi.machida.98}.

The minimization of $f_{\rm msf}[\bm\Phi]$ 
then leads to two superfluid phases, the MSF$_{\text{p}}$ for $g_2<0$ and 
the MSF$_{\text{fm}}$ for $g_2>0$ molecular
condensates.  For the polar MSF, the order parameter is given by
\be
\bm\Phi = 
\sqrt{\frac{\mu_m}{g_1+g_2}}{\hat{\bm n}} =\Phi_{\rm p}{\hat {\bm
    n}},\ \ \ \mbox{for $g_2 < 0$},
\ee
spanning the $[U(1)\times S_2]/{\mathbb Z}_2$ manifold of degenerate
ground states. For the ferromagnetic MSF, we instead find
\be 
\bm\Phi =\sqrt{\frac{\mu_m}{2g_1}}(\hat{\bm n}+i\hat{\bm m})
=\frac{\Phi_{\rm fm}}{\sqrt{2}}(\hat {\bm n}+i\hat{\bm m}),\ \ \ 
\mbox{for $g_2 > 0$},
\ee
spanning the $SO(3)$ manifold of states. In the above equation, $\hat{\bm n},
\hat{\bm m}, \hat{\bm l}\equiv\hat{\bm n}\times\hat{\bm m}$ is an
orthonormal triad and $\Phi_{\rm p,fm}$ are complex order-parameter
amplitudes, breaking the $SO(3)\times U_N(1)$ symmetry of the
disordered phase.  For finite $T$ the N-MSF transitions are in the
well-studied universality class of a complex $O(3)$
model~\cite{ho.spinor.98}. The MSF$_{\rm p}$ and MSF$_{\rm fm}$ are
separated by a first-order transition, at $g_2=0$ in mean-field
approximation.

\subsection{Atomic Molecular Superfluid (AMSF)}
For the intermediate detuning, we consider a condensation of both
atoms and molecules, for generality allowing atoms to condense at a
nonzero momentum. The latter is motivated by the discussion in the
Introduction of the $p$-wave atom-molecule Feshbach coupling, which
drives such finite momentum atom condensation~\cite{kuklov.06, RCprl.09}.

To analyze the phase boundaries and the behavior of the order
parameters in the AMSF phase, it is convenient to approach the AMSF
state from the MSF phase at negative detuning, where molecular
condensate is well formed, and study the atomic condensation upon the
increase of the detuning and of the atomic chemical potential.

We focus on the simpler case of a single momentum, $\ibQ$ atomic
condensate, that we also later find to be the preferred form of
the AMSF state.  We relegate to Appendix~\ref{OPstructure} the
conceptually straightforward, but technically slightly involved,
analysis of the more general $\pm\ibQ$ momenta state.

Using the order parameter form from Eqs.~\rf{atomOP1},~\rf{atomOP2}, and
\rf{molOP} inside the mean-field free-energy density $f_{\rm
  amsf}=F[\Psi_\sigma,\bm\Phi]/V=f_Q + f_{\rm msf}$, we obtain
\begin{widetext}
\bea
f_Q &=&
\varepsilon_Q
\left(\Psi_{1,\ibQ}^* \Psi_{1,\ibQ}+\Psi_{2,-\ibQ}^* \Psi_{2,-\ibQ}\right)
-\Delta_\ibQ\Psi_{1,\ibQ}^*\Psi_{2,-\ibQ}^*
-\Delta_\ibQ^*\Psi_{1,\ibQ}\Psi_{2,-\ibQ}\nonumber \\
&&+\frac{\lambda_1}{2}\ve\Psi_{1,\ibQ}\ve^4
+\frac{\lambda_2}{2}\ve\Psi_{2,-\ibQ}\ve^4
+\lambda_{12}\ve\Psi_{1,\ibQ}\ve^2\ve\Psi_{2,-\ibQ}\ve^2,
\label{ffmft} 
\eea
\end{widetext}
where $\varepsilon_Q=\frac{Q^2}{2m}-\mu+g_{am}\ve\bm\Phi\ve^2$,
$\Delta_\ibQ = \alpha\bm\Phi\cdot\ibQ\equiv |\Delta_\ibQ|e^{i\varphi_0}$,
and for simplicity we specialized to a balanced mixture set by
$\mu_1=\mu_2=\mu$.  
To determine the nature of the atomic condensate
in the AMSF state,
we diagonalize the quadratic part of the
free-energy density, $f_Q^0$ with a unitary transformation $U_0$,
\be
U_0 = \frac{1}{\sqrt{2}}
\begin{pmatrix} e^{i\varphi_0} & -e^{i\varphi_0} \\
                1           & 1              
\end{pmatrix},
\ee
obtaining
\bse
\begin{align}
f_{Q}^{0}&=
\begin{pmatrix} \Psi_{1,\ibQ}^* & \Psi_{2,-\ibQ}\end{pmatrix}
\begin{pmatrix}
\varepsilon_Q & -\Delta_\ibQ \\
-\Delta_\ibQ^* & \varepsilon_Q
\end{pmatrix} 
\begin{pmatrix} \Psi_{1,\ibQ} \\ \Psi_{2,-\ibQ}^*\end{pmatrix},\ \ \ \ \\
&=\begin{pmatrix} \Psi_{-}^* & \Psi_{+}\end{pmatrix}_\ibQ
U_0^\dag 
\begin{pmatrix}
\varepsilon_Q & -\Delta_\ibQ \\
-\Delta_\ibQ^* & \varepsilon_Q
\end{pmatrix} U_0
\begin{pmatrix} \Psi_{-} \\ \Psi_{+}^* \end{pmatrix}_\ibQ, \\
&=\epsilon^+_\ibQ\ve\Psi_{+}\ve^2+\epsilon^-_\ibQ\ve\Psi_{-}\ve^2,
\end{align}
\ese
where 
\begin{align}
\begin{pmatrix} \Psi_{-} \\ \Psi_{+}^* \end{pmatrix}_\ibQ &= 
U_0^\dag 
\begin{pmatrix} \Psi_{1,\ibQ} \\ \Psi_{2,-\ibQ}^* \end{pmatrix} =
\frac{1}{\sqrt{2}} \begin{pmatrix}
e^{-i\varphi_0}\Psi_{1,\ibQ}+\Psi_{2,-\ibQ}^* \\
-e^{-i\varphi_0}\Psi_{1,\ibQ}+\Psi_{2,-\ibQ}^* 
\end{pmatrix},
\label{pmOP}
\end{align}
and
\begin{align}
\epsilon^+_\ibQ &= \varepsilon_\ibQ + \ve\Delta_\ibQ\ve,\quad\quad 
\epsilon^-_\ibQ = \varepsilon_\ibQ - \ve\Delta_\ibQ\ve.
\end{align}
Expressing the quartic terms of the free-energy density in terms of
the diagonalized atomic condensate fields, $\Psi_{\pm}$, we find
\begin{widetext}
\bse
\bea
\ve\Psi_{1,\ibQ}\ve^4 &=&
\frac{1}{4}\Big(\ve\Psi_{+}\ve^4+\ve\Psi_{-}\ve^4
+4\ve\Psi_{+}\ve^2\ve\Psi_{-}\ve^2
+(\Psi_{+}\Psi_{-})^2+(\Psi_{+}^*\Psi_{-}^*)^2 \nonumber \\
&& -2\ve\Psi_{+}\ve^2(\Psi_{+}\Psi_{-}+\Psi_{+}^*\Psi_{-}^*)
-2\ve\Psi_{-}\ve^2(\Psi_{+}\Psi_{-}+\Psi_{+}^*\Psi_{-}^*)\Big),  \\
\ve\Psi_{2,-\ibQ}\ve^4 &=&
\frac{1}{4}\Big(\ve\Psi_{+}\ve^4+\ve\Psi_{-}\ve^4
+4\ve\Psi_{+}\ve^2\ve\Psi_{-}\ve^2
+(\Psi_{+}\Psi_{-})^2+(\Psi_{+}^*\Psi_{-}^*)^2 \nonumber \\
&&+2\ve\Psi_{+}\ve^2(\Psi_{+}\Psi_{-}+\Psi_{+}^*\Psi_{-}^*)
+2\ve\Psi_{-}\ve^2(\Psi_{+}\Psi_{-}+\Psi_{+}^*\Psi_{-}^*)\Big), \\
\ve\Psi_{1,\ibQ}\ve^2\ve\Psi_{2,-\ibQ}\ve^2 &=&
\frac{1}{4}\left(\ve\Psi_{+}\ve^4+\ve\Psi_{-}\ve^4-(\Psi_{+}\Psi_{-})^2
-(\Psi_{+}^*\Psi_{-}^*)^2\right).
\eea
\ese
\end{widetext}
Since $\epsilon_\ibQ^-<\epsilon_\ibQ^+$, the MSF-AMSF transition takes
place at $\epsilon_\ibQ^-=0$, tuned to this point by the FR detuning,
$\nu\rightarrow\nu_c^{\rm MSF-AMSF}$. At higher detuning, $\nu >
\nu_c^{\rm MSF-AMSF}$, a finite momentum $\ibQ$ atomic condensate
develops, characterized by a nonzero order parameter $\Psi_-\neq 0$,
and $\Psi_+ = 0$. From the latter condition, we deduce that
\be
\Psi_{2,-\ibQ}^*=e^{-i\varphi_0}\Psi_{1,\ibQ},
\label{Psi1Psi2amsf}
\ee
and
\be
\Psi_-=\sqrt{2}e^{-i\varphi_0}\Psi_{1,\ibQ},
\ee
leading to a considerable simplification of the AMSF Landau
free-energy density,
\bea
f_{\rm amsf} &=& \epsilon_\ibQ^-\ve\Psi_-\ve^2 
+\frac{1}{2}\lambda\ve\Psi_-\ve^4 \nonumber \\
&&-\mu_m\ve\bm\Phi\ve^2+\frac{g_1}{2}\ve\bm\Phi^*\cdot\bm\Phi\ve^2
+\frac{g_2}{2}\ve\bm\Phi\cdot\bm\Phi\ve^2,\ \ \ \ \quad
\eea
where $\lambda =\frac{1}{4}(\lambda_1+\lambda_2+2\lambda_{12})$.  The
minimization of $f_{\rm amsf}[\Psi_-,\bf\Phi]$ over the order
parameters and the atomic momentum $\ibQ$ is straightforward.
The optimum $\ve\ibQ_0\ve=Q_0$ is given by
\be
\frac{\partial f_{\rm amsf}}{\partial Q}=0,
\ee
and leads to 
\be
Q_0 = \alpha m \left[(u^2-v^2)\cos^2\theta_\ibQ + v^2\right]^{1/2},
\ee
with $\theta_\ibQ$ the angle between $\ibQ_0$ and $\bf u$.  Without loss
of generality, taking $u > v$ and putting $Q_0$ back into the free
energy shows that $f_{\rm amsf}$ is minimized by $\theta_\ibQ=0$, that is,
by $\ibQ_0$ aligned along the longest of the $\bf u$ and $\bf v$
components, giving
\be
Q_0 = \alpha m u\approx\alpha m \sqrt{n_m}.
\ee
Thus, as illustrated in Fig.~\ref{fig:Q}, the momentum $Q_0$ is at its
maximum value near the MSF-AMSF phase boundary and decreases
continuously to zero with the molecular condensate $n_m$ at the
AMSF-ASF transition, tunable with a magnetic field via detuning,
$\nu$.

As in the treatment of the MSF phases, it is convenient to express the
free energy in terms of the magnitudes of the real, $\bf u$, and
imaginary, $\bf v$, vector components of $\bf\Phi$. Minimizing it over
$\Psi_-$, we obtain
\begin{align}
f_{\rm amsf}
&=-\frac{1}{2\lambda}\left(\mu+\frac{m \alpha^2}{2} u^2 - 
g_{am}(u^2+v^2)\right)^2 \nonumber \\
&-\mu_m(u^2+v^2)+\frac{g_1}{2}(u^2+v^2)^2 +\frac{g_2}{2}(u^2-v^2)^2,
\nonumber \\
\label{famsf}
\end{align}
with the atomic condensate given by
\be
\ve\Psi_-\ve = \left[\big(\mu + \frac{m \alpha^2}{2}u^2 
 -g_{am}(u^2+v^2)\big)/\lambda\right]^{1/2}.
\ee
Minimization of $f_{\rm amsf}$ with respect to $u$ and $v$ gives a
number of solutions.  In addition to the normal ($\Psi_-=0$,
${\bm\Phi}=u=v=0$) and the ASF ($\Psi_-\neq 0$, ${\bm\Phi}=u=v=0$)
phases, we find the AMSF$_{\rm p}$ ($\Psi_-\neq 0$, $u\neq 0$,
$v=0$) and the AMSF$_{\rm fm}$ ($\Psi_-\neq 0$, $u >
v\neq 0$) phases that are the descendants of the MSF$_{\rm p}$ and
MSF$_{\rm fm}$ molecular condensates.

\begin{figure}[thb]
\includegraphics[width=7cm]{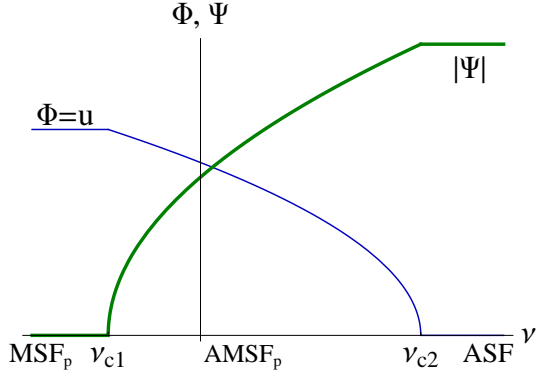} 
\caption{(Color online) Schematic atomic (thick) and molecular (thin) order
  parameters versus the FR detuning $\nu$ for the polar phase, with
  $\nu_{\rm c1}= \nu_{\rm c}^{\rm MSF_{\rm p}-AMSF_{\rm p}}$ and
  $\nu_{\rm c2}= \nu_{\rm c}^{\rm AMSF_{\rm p}-ASF}$.}
\label{polarOPplot}
\end{figure} 

\subsubsection{Polar AMSF: AMSF$_{p}$}
A straightforward minimization of $f_{\rm amsf}[u,v]$, Eq.\rf{famsf} for
$g_2 < 0$ leads to the AMSF$_{\rm p}$ phase, characterized by order
parameters,
\bse
\begin{align}
u_p &= \sqrt{\frac{\lambda\mu_m-\tilde g_{am}\mu}
{\lambda(g_1+ g_2)-\tilde g^2_{am}}},\ \ \ v_p=0, \\
\ve\Psi_{-,p}\ve &= \sqrt{\frac{(g_1+g_2)\mu-\tilde g_{am}\mu_m}
{\lambda(g_1+g_2)-\tilde g_{am}^2}},
\label{psi_minus}
\end{align}
\ese
where $\tilde g_{am} = g_{am}-m \alpha^2/2$. 
\begin{figure}[bht]
\begin{tabular}{c}
\includegraphics[width=8.5cm]{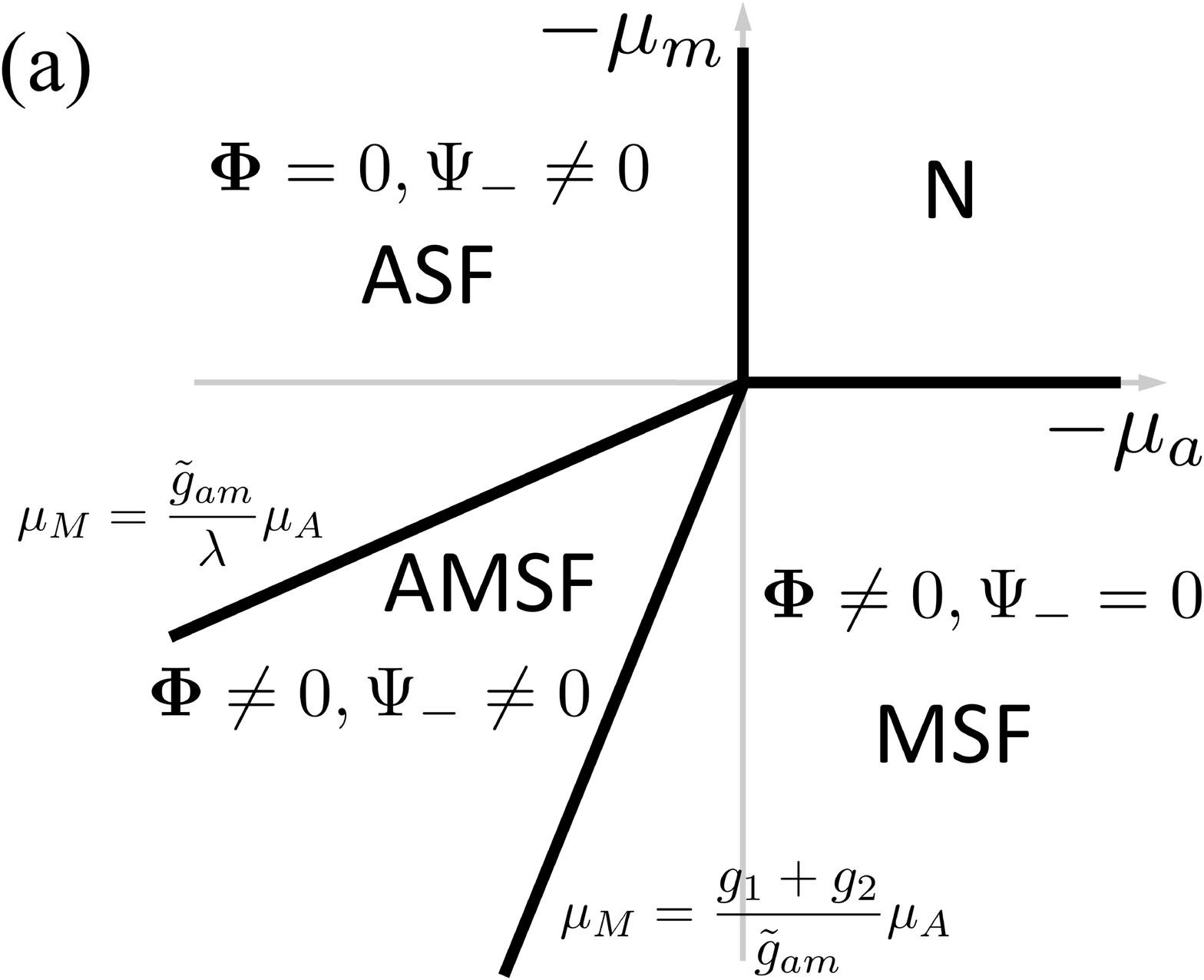} \\
\includegraphics[width=8.5cm]{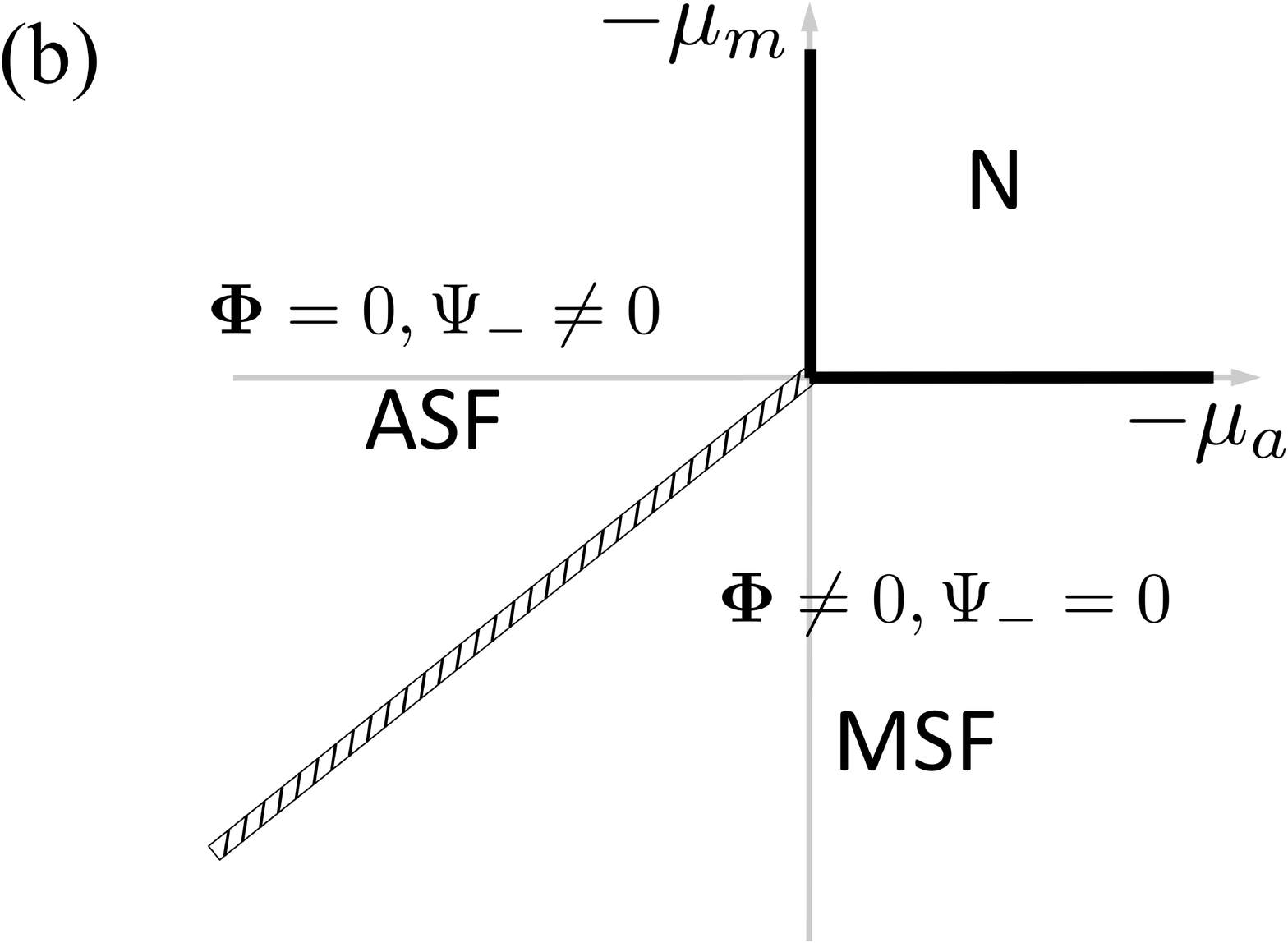}
\end{tabular}
\caption{Mean-field phase diagrams for polar phase as a function of atomic and
  molecular chemical potentials, $\mu_a$, $\mu_m$, respectively.  
  The ferromagnetic phase is similar but with different parameters. 
  (a) For $\lambda(g_1+g_2)-{\tilde g_{am}}^{2}>0$, all three superfluid
  phases---ASF, AMSF, and MSF---appear and are separated by continuous
  phase transitions (thick black lines). (b) For $\lambda(
  g_1+g_2)-{\tilde g_{am}}^{2}<0$, AMSF is unstable, and the ASF and
  MSF are separated by a first-order transition (hatched double
  line).}
\label{phase1}
\end{figure} 
The phase boundaries corresponding to the MSF$_{\rm p}$ - AMSF$_{\rm
  p}$ and AMSF$_{\rm p}$ - ASF transitions are also
easily worked out (set by the vanishing of the atomic and molecular
condensates, respectively) and are given by
\bse
\begin{align}
\nu_{\rm c}^{\rm MSF_p-AMSF_p} &= -\left(g_1+ g_2-2\tilde g_{am}\right)n_m, 
\label{pMSF-AMSF}\\ 
&\approx -\frac{1}{2}\left(g_1+ g_2-2\tilde g_{am}\right)n,\\ 
\nu_{\rm c}^{\rm AMSF_p-ASF} &= \left(2\lambda-\tilde g_{am}\right)n_a, 
\label{pASF-AMSF}\\
&\approx\left(2\lambda-\tilde g_{am}\right)n, 
\end{align}
\ese
where we used $\mu_m=2\mu-\nu = (g_1+ g_2)n_m$ and $\mu = \lambda n_a$
to eliminate the molecular and atomic chemical potentials in favor of
the molecular condensate $n_m$, the atomic condensate $n_a$, and the
detuning $\nu$. We also used the fact that at low temperature and for
weak interactions, $n_m\approx n/2$ and $n_a \approx n$ in the MSF and
ASF, respectively.

It is clear from Fig.~\ref{phase1} (a) and Eq.\rf{psi_minus} for $\Psi_{-,p}$
that the condition
\be
\lambda(g_1+g_2)-\tilde g_{am}^2 > 0
\ee
is necessary for the stability of AMSF$_{\rm p}$.  We observe that in
addition to setting the value of the finite momentum, $Q_0$ of the
atomic condensate, the $p$-wave FR coupling, $\alpha$,
expands the stability of the AMSF phase. Within the mean-field
approximation, the MSF$_{\rm p}$-AMSF$_{\rm p}$ and AMSF$_{\rm p}$-ASF
transitions are of second order. This will be qualitatively modified,
as we see when we discuss fluctuation effects in
Sec.\ref{sec:PhaseTransition}.

For $\lambda(g_1+ g_2)-\tilde g_{am}^2<0$ [Fig.~\ref{phase1} (b)],
the AMSF$_{\rm p}$ state is unstable, replaced by a direct first-order
ASF-MSF$_{\rm p}$ transition. The corresponding phase boundary is
given by the degeneracy condition of the ASF and MSF$_{\rm p}$
free energies
\be
f_{\rm asf} = -\frac{\mu^2}{2\lambda} = -\frac{\mu_m^2}{2(g_1+g_2)} 
= f_{\rm msf_{\rm p}}.
\ee

\subsubsection{Ferromagnetic AMSF: AMSF$_{fm}$}
A minimization of the free energy, $f_{\rm amsf}[u,v]$ for a range of
couplings shows that for intermediate detuning, the low-temperature
state is the ferromagnetic AMSF$_{\rm fm}$, characterized by
\begin{widetext}
\bse
\begin{align}
u_{fm} &= \sqrt{\frac{2\lambda g_2 \mu_m- g_{am}^2\mu_m
-(g_1+g_2)\tilde g_{am}\mu+(g_1-g_2)g_{am}\mu
+g_{am}\tilde g_{am} \mu_m}
{4\lambda g_1 g_2-4g_2 g_{am}\tilde g_{am}
-(g_1+g_2)(m\alpha^2/2)^2}},\quad\quad\quad\quad \label{u.ferro} \\
v_{fm} &= \sqrt{\frac{2\lambda g_2\mu_m-{\tilde g_{am}}^{2}\mu_m
-(g_1+g_2)g_{am}\mu
+(g_1-g_2)\tilde g_{am}\mu+g_{am}\tilde g_{am}\mu_m}
{4\lambda g_1 g_2-4g_2 g_{am}\tilde g_{am}
-(g_1+g_2)(m\alpha^2/2)^2}},\quad\quad\quad\quad \label{v.ferro} \\
\ve \Psi_{fm} \ve &=\sqrt{\frac{g_2(4g_1\mu -4g_{am}\mu_m+m\alpha^2\mu_m)}
{4\lambda g_1 g_2-4g_2 g_{am}\tilde g_{am}
-(g_1+g_2)(m\alpha^2/2)^2}}.
\end{align}
\ese
\end{widetext}

\begin{figure}[thb]
\includegraphics[width=8.5cm]{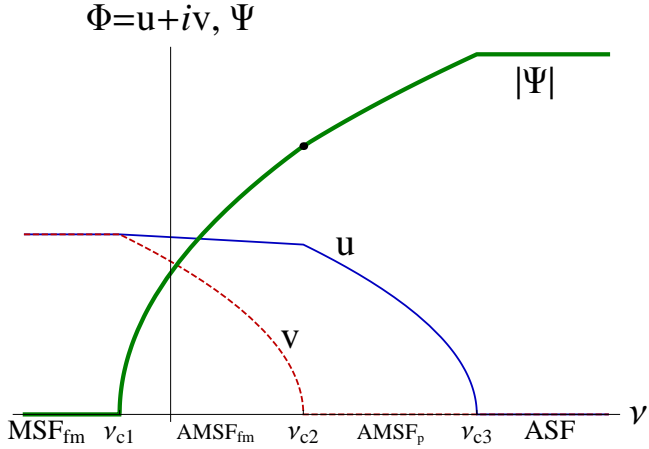} 
\caption{(Color online) Schematic atomic (thick) and molecular (thin and
  dashed) order parameters versus the FR detuning $\nu$ for
  ferromagnetic phases.  The AMSF$_{\rm fm}$-AMSF$_{\rm p}$ phase
  transition at $\nu_{c2}$ leads to kinks (change in slope) in the
  molecular ($u$) and atomic ($\Psi$) order parameter, later indicated
  by a black dot. Without loss of generality we choose the $\hat n$ axis
  (component of $u$) to lie along $\ibQ_0$. The critical detunings are
  denoted by $\nu_{\rm c1}= \nu_{\rm c}^{\rm MSF_{\rm fm}-AMSF_{\rm
      fm}}$, $\nu_{\rm c2}= \nu_{\rm c}^{\rm AMSF_{\rm fm}-AMSF_{\rm
      p}}$, and $\nu_{\rm c3}= \nu_{\rm c}^{\rm AMSF_{\rm p}-ASF}$.  }
\label{ferroOP}
\end{figure} 

The behavior of these order parameters as a function of detuning,
$\nu$, is illustrated in Fig.~\ref{ferroOP}.  With increasing detuning,
the component $v$ (being smaller than $u$) vanishes
first, signaling a transition of the ferromagnetic AMSF$_{\rm fm}$ to the
polar AMSF$_{\rm p}$ state. Depending on the value of other
parameters, upon further increase of $\nu$ the system either
continuously transitions at $\nu_{\rm c}^{\rm AMSF_p - ASF}$ to one
of the three ASF states or undergoes a first-order AMSF$_{\rm fm}$-ASF
transition with $u$ discontinuously jumping to zero when $v$
vanishes. As we discuss in Sec.\ref{sec:fluctuations}, on general
grounds, beyond the mean-field approximation, we expect the
transitions from such smectic like superfluid phases (AMSF$_{\rm
  p,fm}$) to homogeneous and isotropic ASF states to be driven
first-order by fluctuations.

The detuning phase boundaries corresponding to the MSF$_{\rm fm}$ -
AMSF$_{\rm fm}$ and the AMSF$_{\rm fm}$ - AMSF$_{\rm p}$ transitions,
determined by a vanishing of the atomic and the $v$ (transverse to
$\ibQ_0$) component of the molecular condensates, respectively, are
given by
\begin{widetext}
\bse
\begin{align}
\nu_{\rm c}^{\rm MSF_{\rm fm}-AMSF_{\rm fm}} &= -\left(g_1-2g_{am}+m\alpha^2/2\right)n_m, 
\label{fmMSF-AMSF}\\ 
&\approx -\frac{1}{2}\left(g_1-2g_{am}+m\alpha^2/2\right)n,\\
\nu_{\rm c}^{\rm AMSF_{\rm fm}-AMSF_{\rm p}}&=
\frac{8\lambda g_2+ g_{am}\left(2m\alpha^2-4g_2\right)
-m\alpha^2\left(g_1- g_2+m\alpha^2\right)}
{4g_2+2m\alpha^2}n_a.
\label{fmAMSF-pAMSF}
\end{align}
\label{fmCriticalnu}
\ese
\end{widetext}

\begin{figure}[thb]
\includegraphics[width=7cm]{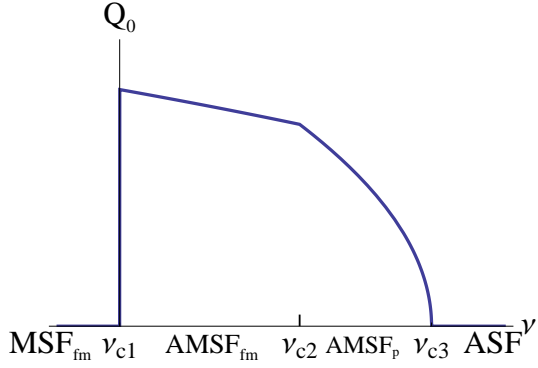}
\caption{(Color online) Schematic detuning dependence of the momentum $\ibQ_0$ of the atomic
  condensate starting with the MSF$_{\rm fm}$, with
  $\nu_{\rm c1}= \nu_{\rm c}^{\rm MSF_{\rm fm}-AMSF_{\rm fm}}$,
  $\nu_{\rm c2}= \nu_{\rm c}^{\rm AMSF_{\rm fm}-AMSF_{\rm p}}$, and
  $\nu_{\rm c3}= \nu_{\rm c}^{\rm AMSF_{\rm p}-ASF}$.}
\label{Qoferro}
\end{figure} 

As with the polar state, the stability of the AMSF$_{\rm fm}$ is
dictated by a condition on the interaction couplings, given by
\be
4\lambda g_1 g_2-4g_2 g_{am}\tilde g_{am}
-(g_1+g_2)(m\alpha^2/2)^2 > 0.
\ee
In the opposite regime of 
$4\lambda g_1 g_2-4g_2 g_{am}\tilde g_{am} -(g_1+g_2)(m\alpha^2/2)^2 <
0$ [Fig.~\ref{phase1} (b)], the AMSF$_{\rm fm}$ state is unstable,
replaced by a direct first-order MSF$_{\rm fm}$-ASF transition. The
corresponding phase boundary is given by the degeneracy condition of
the ASF and MSF$_{\rm fm}$ free energies,
\be
f_{\rm asf} = -\frac{\mu^2}{2\lambda} = -\frac{\mu_m^2}{g_1} 
= f_{\rm msf_{\rm fm}}.
\ee
\subsubsection{Renormalized molecular interactions couplings} 
\label{MMscattering}

\begin{figure}[h]
\includegraphics[width=4.5cm]{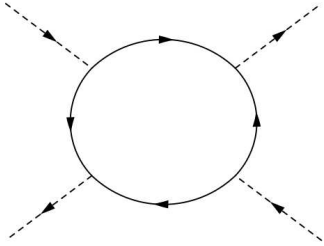}
\caption{A lowest-order diagrammatic correction to molecular
  interaction coupling $g_i$.}
\label{MOLcorrection}
\end{figure}

\begin{figure}[h]
\includegraphics[width=4cm]{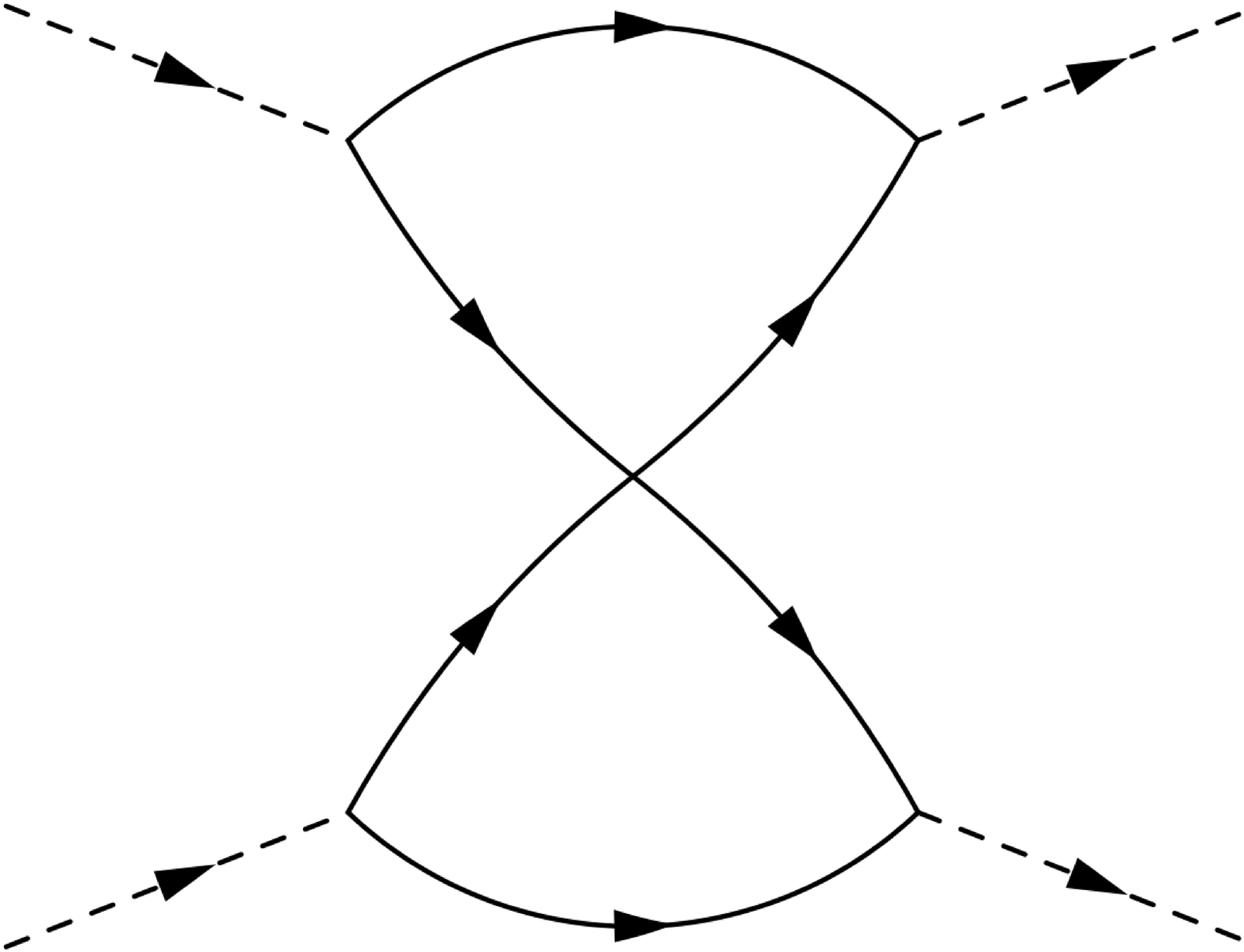}
\includegraphics[width=4cm]{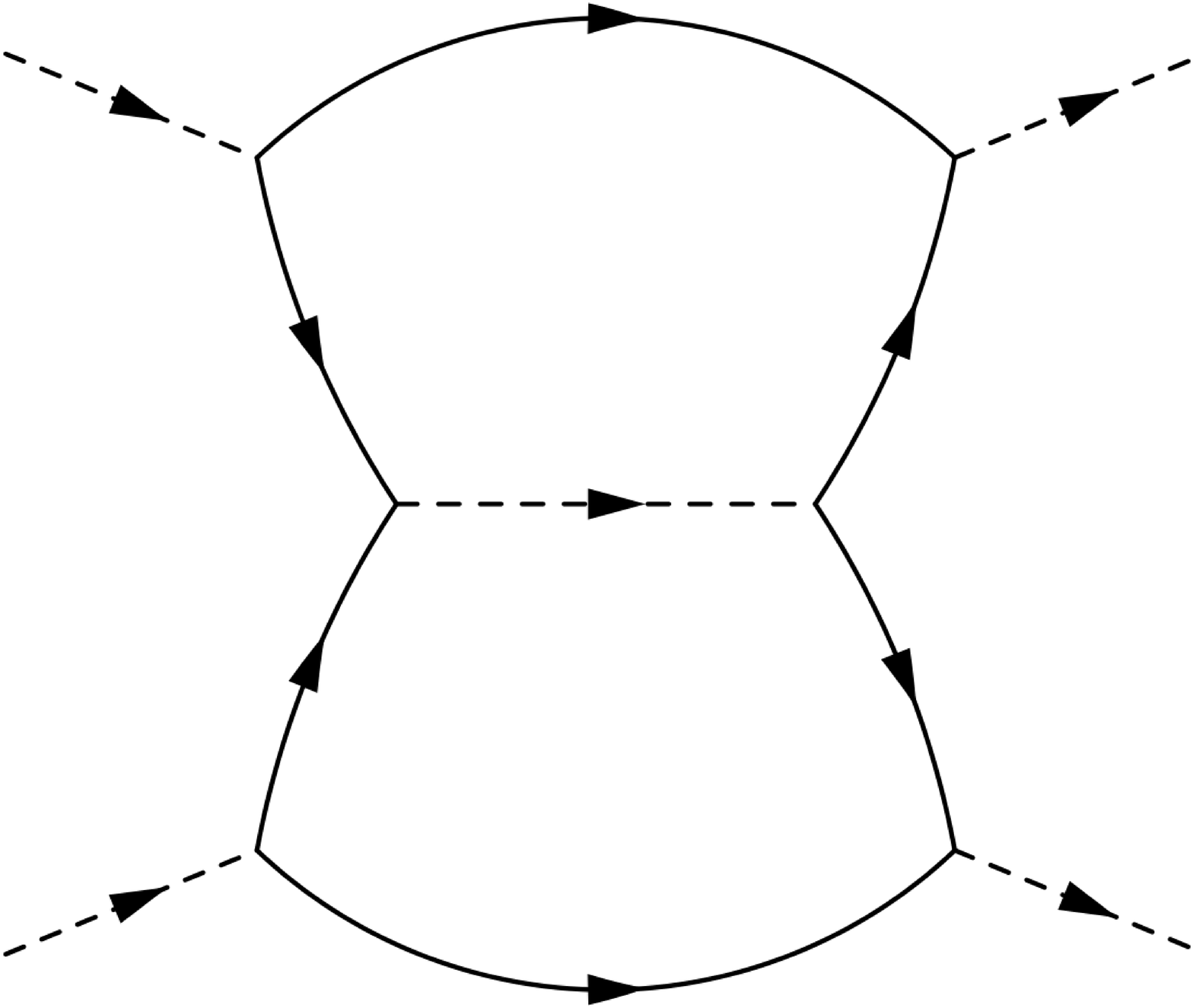}
\caption{Next-lowest-order diagrammatic corrections
  to molecular interaction couplings $g_i$.}
\label{MOLcorrection2}
\end{figure}

We conclude this section by noting that near a FR the
microscopic pseudopotentials $g_i,\lambda_i$ are modified by quantum
fluctuations, replaced by corresponding experimentally determined
scattering lengths.  To lowest order (Born approximation, valid at low
densities) in the FR coupling $\alpha$, the diagrammatic corrections
illustrated in Figs.~\ref{MOLcorrection} and Fig.~\ref{MOLcorrection2}
are given by
\bse
\begin{align}
\delta g^R_1 &= \frac{m^4\alpha^4\Lambda^2}{\pi^4}
\left(-\frac{2\pi^2}{15m\Lambda}+\frac{a_{bg}}{9\pi}-\frac{m\alpha^2}{16}(0.468) \right), \\
\delta g^R_2 &= \frac{m^4\alpha^4\Lambda^2}{\pi^4}
\left(-\frac{\pi^2}{15m\Lambda}-\frac{m\alpha^2}{16}(0.0489) \right),
\end{align}
\ese
where $a_{bg} = a_{1}+a_{2}+2a_{12}$,
the scattering lengths are defined by a standard relation,
$\lambda_{\sigma\sigma'} = \frac{4\pi\hbar^2 a_{\sigma\sigma'}}{m}$,
and $\Lambda\approx 2\pi/d$ is the ultraviolet cutoff set by the
interatomic potential range. In the large $\Lambda$ limit, $\delta g^R_i$
reduce to
\bse
\begin{align}
\delta g^R_1 &\simeq \frac{m^4\alpha^4\Lambda^2}{\pi^4}
\left(\frac{a_{bg}}{9\pi}-\frac{m\alpha^2}{16}(0.468) \right), \\
\delta g^R_2 &\simeq -\frac{m^5\alpha^6\Lambda^2}{(2\pi)^4}(0.0489).
\end{align}
\ese
This two-loop approximation (though valid only in the narrow FR
limit), which finds $ \delta g^R_2<0$, suggests that in the
broad-resonance limit it is the MSF$_{\rm p}$ that prevails.

More generally, the importance of these fluctuation corrections to
molecular interactions is that they provide a mechanism to tune and,
in principle, even change the sign of the effective $g_2$, thereby
allowing a detuning-driven MSF$_{\rm p}$-MSF$_{\rm fm}$ transition.

\section{Elementary excitations}
\label{sec:fluctuations}

Having established the existence of a variety of superfluid ground
states, we now turn our attention to the nature of low-energy
excitations in each of these phases.  As long as fluctuations remain
finite for a range of a system's parameters, the phases detailed in the
previous section are self-consistently guaranteed to be stable in
these regimes and to retain their qualitative form.

We study quantum fluctuations within each of the ASF, MSF and AMSF
classes of phases established above. To this end we expand the atomic
and molecular bosonic operators around their mean-field condensate
values $\Psi_\sigma$, $\bm\Phi$,
\bse
\bea
\psi_\sigma  &=& \Psi_\sigma + \delta\psi_\sigma, \\
\phi_{i} &=& \Phi_{i} + \delta\phi_{i},
\eea
\ese
where $\delta\psi_\sigma$ ($\sigma=1,2$) are fluctuation fields for
atoms of flavors $1$ and $2$, respectively, and $\delta\phi_i$ ($i =
x,y,z$) are triplet of the $\ell = 1$ molecular fluctuation fields.

For some of the analysis it is convenient to work in momentum space,
\bse
\bea
\delta\psi_\sigma &=& \frac{1}{\sqrt{V}}\sum_{\ibk} a_{\sigma,\ibk}\  
e^{i\ibk\cdot\ibr}, \\
\delta\phi_i &=& \frac{1}{\sqrt{V}}\sum_\ibk b_{i,\ibk}\ e^{i\ibk\cdot\ibr}.
\eea
\ese

Using the above momentum representation inside the Hamiltonian
[Eq.~\eqref{Hmain}] and expanding to second order in the fluctuations
operators $a_{\sigma,\ibk}$, $b_{i,\ibk}$, we obtain $H=H_{\text
  {mft}}[\Psi_\sigma,\bm\Phi] + H_f$, with
\begin{widetext}
\bse
\begin{align}
H_f =&
\sum_{\ibk}\Bigg[\sum_{\sigma=1,2}
\left(
\frac{1}{2}\tilde\eps_{\sigma,\ibk+\ibQ_\sigma}
a^\dag_{\sigma,\ibk+\ibQ_\sigma}a_{\sigma,\ibk+\ibQ_\sigma}
+\tilde\lambda_\sigma a_{\sigma,-\ibk+\ibQ_\sigma} a_{\sigma,\ibk+\ibQ_\sigma} \right)
+ t_1 a^{\dag}_{1,\ibk+\ibQ} a_{2,\ibk-\ibQ}
+ t_{2,\ibk+\ibQ} a_{1,\ibk+\ibQ} a_{2,-\ibk-\ibQ} \nonumber \\
&+\sum_{i=x,y,z}
\left(
\frac{1}{2}\tilde\omega_{i,\ibk}b^\dag_{i,\ibk}b_{i,\ibk}+\delta_i b_{i,-\ibk}b_{i,\ibk}\right)
+\oh\sum_{\sumfrac{i,j=x,y,z}{i\neq j}}
\left(g_{ij}b^{\dag}_{j,\ibk}b_{i,\ibk} 
+\gamma_{ij}b_{i,-\ibk}b_{j,\ibk}\right) \nonumber \\
&-\sum_\sigma
\bm\alpha_{\overline\sigma,\ibk}\cdot{\bf b}^\dag_{\ibk}a_{\sigma,\ibk+\ibQ_\sigma}
+h.c. \Bigg],
\label{main_fluct}\\
&\equiv \sum_{\ibk,\alpha,\beta}c^\dag_{\alpha,\ibk}\tilde h_\ibk^{\alpha\beta}c_{\beta,\ibk}
\end{align}
\ese
\end{widetext}
where $\tilde h_\ibk^{\alpha\beta}$ is a Bogoliubov Hamiltonian matrix
defined by matrix elements
\bse
\begin{align}
\tilde\eps_{\sigma,\bk} &= 
\epsilon_\bk-\mu_\sigma + 2 \lambda_\sigma \ve\Psi_\sigma\ve^2 + 
\lambda_{12}\ve\Psi_{\overline{\sigma}}\ve^2 + g_{am}\ve\bm\Phi\ve^2, \\
\tilde\omega_{i,\bk}&=\oh\epsilon_\bk-\mu_m +g_1\ve\bm\Phi\ve^2
+(g_1+2g_2)\ve\Phi_i\ve^2, \nonumber \\
&+g_{am}(\ve\Psi_1\ve^2+\ve\Psi_2\ve^2),\\
\tilde\lambda_{\sigma} &= \frac{1}{2} \lambda_{\sigma}\Psi^{*2}_\sigma,\\
t_1 &= \lambda_{12}\Psi_1\Psi^*_2,\\
t_{2,\ibk} &= \lambda_{12}\Psi_1^*\Psi_2^*-\alpha\bm\Phi^*\cdot\ibk,\\
\delta_{i} &= \frac{1}{2}g_1\Phi^*_{i}\Phi^*_{i}
+\frac{1}{2}g_2\bm\Phi^*\cdot\bm\Phi^*,\\
g_{ij} &= g_1\Phi^*_{i}\Phi_{j} 
+2g_2\Phi^*_i\Phi_j,\\
\gamma_{ij} &=\oh g_1\Phi^*_{i}\Phi^*_{j},\\
\bm\alpha_{\sigma=(1,2),\ibk} 
&=\pm\alpha\Psi_{\sigma,\ibQ_\sigma}(\ibQ_\sigma-\ibk/2),
\end{align}
\ese
$\epsilon_\bk=\frac{k^2}{2m}$, $\overline 1 = 2, \overline 2 =
1$, and we suppressed the $\ibQ$ subscript on the atomic condensate
order parameter, $\Psi_{\sigma,\ibQ}$. The ten-dimensional bosonic
Nambu spinor $c_{\alpha,\ibk}$ is given by
\begin{eqnarray}
c_{\alpha,\ibk}\equiv\left(a_{\sigma,\ibk + \ibQ_\sigma},b_{i,\ibk},
a^\dag_{\sigma,-\ibk + \ibQ_\sigma},b^\dag_{i,-\ibk}\right).
\end{eqnarray}

A diagonalization of this ten-dimensional Bogoliubov Hamiltonian,
preserving bosonic commutation relations of the $c_{\alpha,\ibk}$
components gives the spectrum of the five modes throughout the phase
diagram. This can be done numerically, but is not very
enlightening. Instead, we study the problem one phase at a time,
which allows a significantly more revealing solution of the problem.

\subsection{ASF phases}
In the simplest limit of a large positive detuning, $\nu >
\nu^{\rm AMSF-ASF}$, the molecules are gapped, one or both species of the
atoms are condensed at zero momentum, $\ibQ=0$, and the system is in
the ASF phases. As discussed in Sec.~\ref{sec:symmetries}, these are conventional
well-studied superfluids, characterized by one Bogoliubov mode for
each of the atomic $U(1)$ symmetry that is broken. In the ASF phases
$\bm\Phi=0$, the three molecular modes are gapped and can therefore be
integrated out (adiabatically eliminated). Away from the transition,
this leads to only a small renormalization (that we will neglect) of
effective parameters in the resulting $H_f$. From Eq.~\rf{main_fluct} the atomic
sector of the Bogoliubov Hamiltonian is then given by
\begin{widetext}
\begin{align}
H^{\rm ASF_\sigma}_f =&
\sum_{\ibk}\left[\sum_{\sigma=1,2}
\left(\frac{1}{2}\tilde\eps_{\sigma,\ibk} a^\dag_{\sigma,\ibk}a_{\sigma,\ibk}
+\tilde\lambda_\sigma a_{\sigma,-\ibk} a_{\sigma,\ibk}\right)
+ t_1 a^{\dag}_{\sigma,\ibk} a_{\overline\sigma,\ibk}
+ t_{2,\ibk} a_{\sigma,-\ibk} a_{\overline\sigma,\ibk}\right] + h.c..
\end{align}
\end{widetext}

\subsubsection{ASF$_{\sigma}$: single atomic species BEC}
In the regime where only a single atomic species of $\psi_{1,2}$
condenses (i.e., $\Psi_\sigma\neq 0, \Psi_{\overline\sigma} = 0$), the
system is in an ASF$_{\sigma}$ phase. Standard analysis then leads to
a conventional, gapless atomic Bogoliubov sound mode for species
$\sigma$
\bse
\bea
E^{(a)}_{k\sigma} &=& \sqrt{\frac{k^2}{2m}\Big(\frac{k^2}{2m}+\lambda_{\sigma}n\Big)},\\
&\approx& c_a k,
\eea
\ese
with $c_a \approx \sqrt{\frac{\lambda_\sigma n}{2m}}$, and a gapped
atomic mode for the complementary atomic species $\overline\sigma$:
\bea
E^{(a)}_{k\overline\sigma} &\simeq& 
\frac{k^2}{2m^*_-}-\mu_{\overline\sigma}+\lambda_{12}\frac{n}{2}
\eea
where $\frac{n}{2}\simeq n_1=n_2$ for a balanced case.
Above, the coupling parameters are those from Eq.~\rf{landauFE0}, with
$\bm\Phi=\Psi_{\overline\sigma}=0$, and $m^*_\pm$ are effective atomic
masses renormalized by interaction
\bse
\bea
\frac{1}{m^*_+}&=&\frac{1}{m}+
\frac{3n\alpha^2}
{2(\nu-\lambda n+\frac{g_{am}}{2}n)}, \\
\frac{1}{m^*_-}&=&\frac{1}{m}-
\frac{3n\alpha^2}
{4(\nu-\lambda n+\frac{g_{am}}{2}n)}.\hspace{1cm} 
\eea
\ese
The remaining three molecular-like modes (corrected by coupling to
atoms) are gapped and in a $k\rightarrow0$ limit are given by
\bse
\bea
E^{(m)}_{k1} &=& E^{(m)}_{k2}=
\frac{k^2}{4m}+\nu-\frac{\lambda_\sigma}{2} n+\frac{g_{am}}{2}n-\mu_{\overline\sigma}, \quad\quad\quad \\
E^{(m)}_{k3}&\simeq& \frac{k^2}{4m^*_+}+\nu-\frac{\lambda_\sigma}{2} n+\frac{g_{am}}{2}n-\mu_{\overline\sigma}. 
\eea
\ese
%
%

\subsubsection{ASF$_{12}$: double atomic species BEC}
In the regime where both atomic species of $\psi_{1,2}$ condense,
that is, $\Psi_{1,2}\neq 0$, the system is in a two-species ASF$_{12}$
phase. Standard analysis, consistent with two $U(1)$ symmetries
spontaneously broken, then leads to two gapless atomic Bogoliubov
sound modes for species $1$ and $2$. Together with the gapped molecular
excitations this leads to spectra of the five modes:
\bse
\bea
E^{(a_{12})}_{k1} &=& \sqrt{\frac{k^2}{2m}\Big(\frac{k^2}{2m}+2\lambda n\Big)}, \label{asf.double.sound1} \\
E^{(a_{12})}_{k2} &\simeq& c^{(a_{12})} k,
\label{asf.double.sound2}\\
E^{(m_{12})}_{k1} &=& E^{(m_{12})}_{k2}
=\frac{k^2}{4m}+\nu-2\lambda n+g_{am}n, \quad\quad \,\\
E^{(m_{12})}_{k3} &\simeq& \frac{k^2}{4m^*} + \nu-2\lambda n + g_{am}n,
\eea
\ese
where for $E^{(a_{12})}_{k2}$ and $E^{(m_{12})}_{k3}$ we took $k
\rightarrow 0$ and $\alpha \rightarrow 0$ limit and defined the sound
velocity and effective atomic mass:
\bse
\bea
c^{(a_{12})}&=&\sqrt{\frac{(\lambda - \lambda_{12} )n}{m}}
-\frac{3n\alpha^2\sqrt{(\lambda-\lambda_{12})mn}}{4(\nu-2\lambda n+g_{am}n)}, \quad\quad\quad \\
\frac{1}{m^*}&=&
\frac{1}{m}+\frac{3(\nu-(\lambda+\lambda_{12})n+g_{am}n)n\alpha^2}
{(\nu-2\lambda n+g_{am}n)^2}. \quad\quad
\eea
\ese
$E^{(a_{12})}_{k1}$ and $E^{(a_{12})}_{k2}$ are atomlike, gapless,
in-phase and out-of-phase modes, respectively.  $E^{(a_{12})}_{k2}$
and $E^{(m_{12})}_{k3}$ are modified by the FR interaction between
atoms and molecules.  
The ASF-AMSF phase boundary is determined by the point where the
molecular gap
%
\be
E^{\rm ASF}_{\rm gap} = \nu - 2\lambda n + g_{am} n \label{gapASF}
\ee
closes, and is consistent with the critical detuning determined by the development of
the molecular order parameter that we found in Sec.~\ref{sec:mft}.

\subsection{MSF phases}
In the opposite limit of a large negative detuning, $\nu < \nu^{\rm
  MSF-AMSF}$ both atomic species are gapped, $\Psi_\sigma=0$, and
$p$-wave molecules are condensed into one of the two ($\ell = 1$)
$\ell_z=0$ MSF$_{\rm p}$ and $\ell_z=\pm1$ MSF$_{\rm fm}$,
isomorphic to spinor-1 condensates with
well-studied
properties~\cite{ho.spinor.98,ho.yip.00,demler.zhou.02}. To see this,
we note that the atomic Bogoliubov excitations are gapped and can
therefore be integrated out. Away from the transition, they lead to
only a small renormalization of effective parameters. Neglecting these
small effects, the vanishing of $\bm\alpha_{\sigma,\ibk} =
\pm\alpha\Psi_{\sigma,\ibQ_\sigma}(\ibQ_\sigma-\ibk/2)=0$ decouples
the Hamiltonian, $H_f=H_a + H_m$ into atomic and molecular parts, that
then are straightforwardly diagonalized.

The atomic sector, $H_a$ is of standard Bogoliubov form, simplified to
a $2\times2$ form by $t_1=\tilde\lambda_\sigma=0$ inside the MSF
phases, leading to the atomic excitation spectrum, that for the
symmetric case of $\mu_1=\mu_2\equiv\mu$ is given by
\be
E^{\rm MSF}_{a,\ibk}
=\sqrt{(\tilde\eps_{\ibk} + \ve\alpha\bm\Phi\cdot\ibk\ve)
(\tilde\eps_{\ibk} - \ve\alpha\bm\Phi\cdot\ibk\ve)},
\ee
where $\tilde\eps_{\ibk} = k^2/2m -\mu + g_{am}\ve\bm\Phi\ve^2$.

One key observation is that already inside the MSF phases the atomic
spectrum, $E^{\rm MSF}_{a,\ibk}$ (degenerate for $\sigma=1,2$
species) develops a minimum at a nonzero momentum
$\ibk_{\rm min}=\ibQ_{\rm p,fm}$, with the corresponding atomic gap minimum,
$E^{\rm MSF_{p,fm}}_{a,{\rm gap}}$, given by a value dependent on the nature of
the MSF$_{\rm p,fm}$ phase.

\subsubsection{MSF$_p$ state : $g_2<0$}
As analyzed in Sec.~\ref{sec:mft}, the MSF$_{\rm p}$ phase is defined by a
molecular condensate order parameter, which can be taken to be a
three-dimensional real vector, $\bm\Phi=\bm u = \Phi_{\rm p}\nh$, with
$n_m = |\Phi_{\rm p}|^2$.  In terms of the molecular condensate density
$n_m\approx n/2$ the atomic chemical potential for the symmetric case,
$\mu_1=\mu_2=\mu$ is given by
\be
\mu = \frac{1}{2}(\mu_m+\nu) = \frac{1}{2}\left((g_1+g_2) n_m+\nu\right),
\ee
controlled by the FR detuning, $\nu$.

For this symmetric case $\mu_1=\mu_2=\mu$ (easily generalizable for
the asymmetric, imbalanced case), the atomic spectrum minimum is
characterized by
\bse
\bea
k_{\rm min}&=&Q_{\rm p},\nonumber\\
&=& \alpha m \sqrt{n_m},\label{kminPolar}\\
E^{\rm (MSF_p,a)}_{\rm gap} &=& 
-\mu +g_{am}n_m-\frac{m\alpha^2 n_m}{2}, \quad\quad \label{gapMSFpolar}
\eea
\ese
where in an isotropic trap the orientation of $\ibk_{\rm min}$ is
spontaneously chosen.  The MSF$_{\rm p}$-AMSF$_{\rm p}$ phase transition boundary
is set by the closing of this atomic gap and is given by
\be
\nu^{\rm MSF_p-AMSF_p}_{c} = -\left(g_1+g_2-2g_{am}+m\alpha^2\right)n_m.
\ee
Reassuringly, this is identical to the critical detuning for this
phase boundary, which we obtained in Sec.~\ref{sec:mft} from the value
of detuning at which the finite-momentum atomic order-parameter became
nonzero.

The diagonalization of molecular part $H_m$ is also straightforward,
and is identical to the case of the spinor-1
condensates~\cite{ho.spinor.98,ho.yip.00,demler.zhou.02}, with effective parameters of our
physically distinct, $p$-wave resonant scalar Bose gas.  Substituting
characteristics of the polar phase MSF$_{\rm p}$ (order parameters, $\mu$,
$\mu_m \approx (g_1+g_2)n_m$, $g_2<0$, etc. from above) into $H_m$, we
obtain
\begin{align}
H^{\rm MSF_p}_m =&\sum_\ibk\Bigg[\left(\oh\epsilon_{\ibk}
+(g_1+g_2) n_m \right)b^{\dag}_{\parallel,\ibk}b_{\parallel,\ibk}\nonumber\\
&+\bigg(\oh\epsilon_{\ibk}+|g_2| n_m\bigg)
{\bf b}^{\dag}_{\perp,\ibk}\cdot{\bf b}_{\perp,\ibk}\nonumber\\
&+\Bigg(\frac{1}{2}(g_1+g_2)n_m b_{\parallel,-\ibk}b_{\parallel,\ibk}
\nonumber \\
& +\frac{1}{2}g_2 n_m {\bf b}_{\perp,-\ibk}\cdot{\bf b}_{\perp,\ibk} + h.c.\Bigg)\Bigg],
\end{align}
where ${\bf b}_{\perp,\ibk}$ are two degenerate transverse (to $\ibQ_{\rm p}$)
molecular modes. This leads to three Bogoliubov-type dispersions,
\bse
\bea
E^{\rm MSF_p}_{\parallel,k} 
&=& \oh\sqrt{\epsilon_{\ibk}^2+4(g_1+g_2)n_m\epsilon_{\ibk}},\\
&\simeq& \sqrt{\frac{(g_1+g_2)n_m}{2m}} k, \label{msf.polar.sound}\\
E^{\rm MSF_p}_{\perp,k} &=& \oh\sqrt{\epsilon_{\ibk}^2+4\ve g_2\ve n_m\epsilon_{\ibk}},\\
&\simeq& \sqrt{\frac{\ve g_2\ve n_m}{2m}} k,
\eea
\label{msfp_atom}
\ese
where the longitudinal mode, $E^{\rm MSF_p}_{\parallel,k}$ describes the
conventional MSF phase fluctuations and the
doubly degenerate transverse mode, $E^{\rm MSF_p}_{\perp,k}$ is the
dispersion for the $\ell=1$ molecular orientational spin-waves. From
the second set of $k\rightarrow 0$ expressions we read off the
corresponding phase and spin-wave velocities, given by
\bse
\bea
c^{\rm MSF_p}_\parallel &=& \sqrt{\frac{(g_1+g_2)n_m}{2m}},\\
c^{\rm MSF_p}_\perp &=& \sqrt{\frac{|g_2|n_m}{2m}}.
\label{polarspeed}
\eea
\ese

\subsubsection{MSF$_{fm}$ state : $g_2>0$}
Inside the MSF$_{\rm fm}$ state, the molecular condensate
order parameter is given by $\bm\Phi = \frac{\Phi_{\rm fm}}{\sqrt{2}}(\hat{\bm
  n}+i\hat{\bm m})$, expressed in terms of an orthonormal triad,
$\hat{\bm n}\times\hat{\bm m}=\hat{\bm \ell}$.  From the earlier
mean-field analysis, the molecular
condensate density is given by $n_m = |\bm\Phi|^2 = \mu_m/g_1$,
leading for the symmetric case, $\mu_1=\mu_2=\mu$
\be
\mu = \frac{1}{2}\left(g_1 n_m+\nu\right).
\ee
To lowest order, the atomic spectrum inside MSF$_{\rm fm}$ has identical structure as that
of the MSF$_{\rm p}$ state [Eq.~\rf{msfp_atom}], but with the replacement
$g_1+g_2\rightarrow g_1$ and $\alpha^2 \rightarrow \alpha^2/2$,
\bse
\bea
k_{\rm min}&=&Q_{\rm fm},\nonumber\\
&=& \frac{1}{\sqrt{2}}\alpha m \sqrt{n_m},\label{kminPolar}\\
E^{\rm (MSF_{fm},a)}_{\rm gap} &=& 
-\mu +g_{am}n_m-\frac{m\alpha^2 n_m}{4}. \quad\quad \label{gapMSFpolar}
\eea
\ese
The MSF$_{\rm fm}$-AMSF$_{\rm fm}$ phase transition boundary
is determined by the vanishing of the atomic gap, and is given by
\be
\nu^{\rm MSF_{fm}-AMSF_{fm}}_{c} = -\left(g_1-2g_{am}+\oh m\alpha^2\right)n_m,
\ee
identical to the critical detuning obtained from mean-field theory for
the order parameter in Sec.~\ref{sec:mft}.

Using the above parameters characteristic of the
MSF$_{\rm fm}$ phase inside $H_m$, the molecular sector of the Hamiltonian
reduces to
\begin{widetext}
\bea
H^{\rm MSF_{fm}}_m &=& 
\sum_\ibk\Big[(\oh\epsilon_{\ibk} + 2g_2 n_m)b^{\dag}_{+,\ibk} b_{+,\ibk}
+(\oh\epsilon_{\ibk}+g_1 n_m) b^{\dag}_{-,\ibk}b_{-,\ibk}
+\oh\epsilon_{\ibk} b^{\dag}_{z,\ibk} b_{z,\ibk} 
+\oh g_1 n_m b_{-,\ibk}b_{-,\ibk}
+\oh g_1n_m b^{\dag}_{-,\ibk}b^{\dag}_{-,\ibk}\Big],\nonumber\\
&&
\eea
\end{widetext}
where
\bse
\bea
b_+ &=& \frac{1}{\sqrt{2}}(b_n+i b_m), \\
b_- &=& \frac{1}{\sqrt{2}}(b_n-ib_m),
\eea
\ese
are expressed in terms of operators $b_n$, $b_m$, that are components
of $\bf b$ along $\hat{\bm n}$, $\hat{\bm m}$, respectively.
Diagonalization of the above Hamiltonian then gives the following spectrum
\bse
\bea
E^{\rm MSF_{fm}}_{z,k} &=& \oh\epsilon_{\ibk}=\frac{k^2}{4m},\\
E^{\rm MSF_{fm}}_{+,k} &=& \oh\epsilon_{\ibk}+2g_2 n_m, \\
E^{\rm MSF_{fm}}_{-,k} &=&\oh\sqrt{\epsilon_{\ibk}^2+4g_1n_m\epsilon_{\ibk}},\\
&\simeq& \sqrt{\frac{g_1 n_m}{2m}} k,
\eea
\label{msf.ferro.sound}
\ese
where the Bogoliubov sound speed is given by $c_{\rm MSF_{fm}} =
\sqrt{g_1 n_m/2m}$.

We note that despite a three-dimensional coset space, $SO(3)$
characterizing MSF$_{\rm fm}$, only {\it two} modes (linear and quadratic in
$k$) exhibit a spectrum that vanishes in $k\rightarrow 0$ limit. The
spectrum $E^{\rm MSF_{fm}}_{-,k}$ is that of a conventional
Bogoliubov superfluid phase, here associated with the $U(1)$ broken
gauge symmetry of the molecular condensate. The quadratic in $k$
gapless spectrum is that of the ferromagnetic spin waves, where the
two components of the spinor are canonically conjugate and, as a result,
combine into a single low-frequency mode.

\subsection{AMSF phases}
\label{sec:GM}
To obtain the spectrum inside the AMSF phases requires a solution of
the fully general Hamiltonian, $H_f$ [Eq.~\rf{main_fluct}]. Because in
this superfluid state all atomic and molecular modes are coupled, a
direct BdG analysis generically involves a diagonalization of a
$10\times 10$ Bogoliubov matrix. This can be done
numerically. However, instead, below we take a complementary
coherent-state path-integral approach that allows us to obtain the
modes and dispersions analytically, leading to more insight into their
structure. Using the formulation of the problem introduced in
Sec.~\ref{sec:coherentstate}, we analyze the low-energy fluctuations
in the AMSF states using the coherent-state Lagrangian density,
$\curL[\psi_\sigma,\bm\phi]= \curL_{\rm MFT}[\Psi_\sigma,\bm\Phi]+
\delta\curL$ [Eq.~\rf{ScurL}], where $\curL_{\rm
  MFT}[\Psi_\sigma,\bm\Phi]$ is the mean-field Lagrangian defining the
AMSF phase and $\delta\curL$ is the Lagrangian density of the
quadratic fluctuations. To obtain $\delta\curL$ we expand the atomic
and molecular bosonic fields $\psi_\sigma,\bm\phi$ about their
mean-field values 
(for clarity of notation in this section we choose to use $\rho$ instead of $n$ 
of the previous sections, where $\rho_\sigma = n_a/2$,
$\rho_m = n_m$, and $\rho_s = n$),
\bse
\bea
\psi_\sigma &=&\sqrt{\rho_\sigma}e^{i\theta_\sigma + i\ibQ_\sigma\cdot\ibr},\\
\bm\phi &=&\sqrt{\rho_m}\bm{\hat\phi} e^{i\varphi},
\eea
\ese
where $\ibQ_\sigma=\pm\ibQ$ for $\sigma=1,2$, respectively,
$\rho_m=\rho_{m0}+\delta\rho_m$ and $\rho_\sigma = \rho_0 +
\delta\rho_\sigma$ are the molecular and atomic densities, with the
mean-field values $\rho_{m0}=|\bm\Phi|^2$ and $\rho_0=|\Psi_\sigma|^2$,
and, based on Eq.~\rf{Psi1Psi2amsf}, with the latter $\sigma$ independent in
the AMSF phase. In addition to the density fluctuations, $\delta\rho_m,
\delta\rho_\sigma$, and two atomic and one molecular superfluid
phases, $\theta_\sigma$, $\varphi$, the molecular Goldstone modes are
characterized by a unit vector, $\bm{\hat\phi}$, whose form depends on
the polar or ferromagnetic nature of the AMSF state:
\bse
\bea
\bm{\hat\phi} &=& {\hat{\bm n}},\ \ \ \mbox{for AMSF$_{\rm p}$},
\label{pAMSFphi}\\
 &=& \frac{1}{\sqrt{2}}({\hat{\bm n}}+i{\hat{\bm m}}),
\ \ \ \mbox{for AMSF$_{\rm fm}$}.\label{fmAMSFphi}\nonumber\\
&&
\eea
\ese
Substituting these parametrizations of the atomic and molecular fields
into the Lagrangian, Eq.~\rf{curlL}, we obtain $\delta\curL$ that
controls fluctuations in the AMSF phases. 

\subsubsection{AMSF$_p$}
Focusing first on the polar state, with
$\bm\phi=\sqrt{\rho_m}{\nh} e^{i\varphi}$, we find
\begin{widetext}
\bse
\begin{eqnarray}
\delta\curL_{\rm p} &=& 
\rho_\sigma(i\partial_\tau\theta_\sigma - \mu_\sigma) +
\frac{\rho_\sigma}{2m}(\grad\theta_\sigma + \ibQ_\sigma)^2 +
\rho_m(i\partial_\tau\varphi  - \mu_m) + 
\frac{\rho_m}{4m}(\grad\varphi)^2 +
\frac{\rho_{m}}{4m}(\grad\nh)^2\nonumber\\
&-&\alpha\sqrt{\rho_m\rho_1\rho_2}\;
\nh\cdot(\grad\theta_1-\grad\theta_2 + 2\ibQ)\cos(\varphi-\theta_1-\theta_2)
+\frac{1}{8m\rho_\sigma}(\grad\rho_\sigma)^2
+\frac{1}{16m\rho_m}(\grad\rho_m)^2
+\frac{\lambda_\sigma}{2}\rho_\sigma^2\nonumber \\
&+&\lambda_{12}\rho_{1}\rho_{2}
+g_{am}(\rho_{1}+\rho_{2})\rho_m
+\frac{g}{2}\rho_m^2 - \curL_{\rm MFT}[\rho_0,\rho_{m0},\nh_0,\ibQ],
\label{deltaLpolarfull}\\
 &=& 
i\delta\rho_+\partial_\tau\theta_+ +
\frac{\rho_0}{m}(\grad\theta_+)^2 +
i\delta\rho_-\partial_\tau\theta_- +
\frac{\rho_0}{m}(\grad\theta_-+\ibQ)^2 +
i\delta\rho_{m}\partial_\tau\varphi + 
\frac{\rho_{m0}}{4m}(\grad\varphi)^2 +
\frac{\rho_{m0}}{4m}(\grad\nh)^2\nonumber\\
&-&2\alpha\rho_0\sqrt{\rho_{m0}}\;
\nh\cdot(\grad\theta_- + \ibQ)\cos(\varphi-2\theta_+)\nonumber\\
&+&\frac{1}{16m\rho_0}(\grad\rho_+)^2+\frac{1}{16m\rho_0}(\grad\rho_-)^2
+\frac{1}{16m\rho_{m0}}(\grad\rho_m)^2
+\frac{\lambda}{4}\delta\rho_+^2
+\frac{\lambda}{4}\delta\rho_-^2
+\frac{\lambda_{12}}{4}(\delta\rho_+^2-\delta\rho_-^2)
+g_{am}\delta\rho_+\delta\rho_m
+\frac{g}{2}\delta\rho_m^2,\label{deltaLpolar0}\nonumber\\
\end{eqnarray}
\label{deltaLpolar}
\ese
\end{widetext}
where $g\equiv g_1 + g_2$, $\lambda=\lambda_1=\lambda_2$ for simplicity, and 
\bse
\begin{eqnarray}
\theta_\pm &=&\oh(\theta_1 \pm \theta_2),\\
\delta\rho_\pm &=&\delta\rho_1 \pm \delta\rho_2,\\
\mu &=&\oh(\mu_1 + \mu_2),\\
h &=&\oh(\mu_1 - \mu_2),\\
\ibQ &=& \alpha m\sqrt{\rho_{m0}}{\hat{\bm n}_0}\label{ibQn}.
\end{eqnarray}
\ese
In the second form [Eq.~\rf{deltaLpolar0}], we expanded the Lagrangian
about its mean-field value $\curL_{\rm MFT}$ to quadratic order in
fluctuations, $\theta_\sigma,\varphi,\delta\rho_\sigma,\delta\rho_m$,
and neglected the constant and subdominant contributions, that are
negligible at long scales and low energies. We note that, as usual, the
linear terms in $\delta\curL_{\rm p}$ [Eq.~\rf{deltaLpolar0}] vanish identically,
enforced by the saddle-point equations for the condensates, $\rho_{-0}$, $\rho_{m0}$, and $\ibQ$.

Examining the last form of $\delta\curL_{\rm p}$, it is clear that important
simplifications take place at long scales. In particular, the Feshbach
resonant (Josephson-like) coupling, $-\alpha\cos(\varphi-2\theta_+)$,
between the closed-channel molecules and atoms (which is always
relevant in three dimensions and therefore acts like a ``mass'') locks
their phases together at low energies giving
\begin{eqnarray}
\varphi=2\theta_+.
\label{deltaLp}
\end{eqnarray}
Integrating $\varphi$ out and completing the square for the
$\grad\theta_-+\ibQ$ and $\nh$, to lowest order then gives
\begin{widetext}
\bse
\begin{eqnarray}
\delta\curL_{\rm p} &=& 
i(\delta\rho_++2\delta\rho_m)\partial_\tau\theta_+ +
\frac{\rho_{s0}}{m}(\grad\theta_+)^2 +
i\delta\rho_-\partial_\tau\theta_- +
\frac{\rho_0}{m}(\grad\theta_-+\ibQ - \alpha m\sqrt{\rho_{m0}}\nh)^2 +
\frac{\rho_{m0}}{4m}(\grad\nh)^2\nonumber\\
&+&\frac{1}{16m\rho_0}(\grad\rho_+)^2+\frac{1}{16m\rho_0}(\grad\rho_-)^2
+\frac{1}{16m\rho_{m0}}(\grad\rho_m)^2
+\frac{\lambda}{4}\delta\rho_+^2
+\frac{\lambda}{4}\delta\rho_-^2
+\frac{\lambda_{12}}{4}(\delta\rho_+^2-\delta\rho_-^2)
+g_{am}\delta\rho_+\delta\rho_m
+\frac{g}{2}\delta\rho_m^2,\label{deltaLpolar2}\nonumber\\
&&\\
&=&i(\delta\rho_++2\delta\rho_m)\partial_\tau\theta_+ +
\frac{\rho_{s0}}{m}(\grad\theta_+)^2 +
i\delta\rho_-\partial_\tau\theta_- +
\frac{\rho_0}{m}(\grad\theta_- - \alpha m\sqrt{\rho_{m0}}\delta\nh)^2 +
\frac{\rho_{m0}}{4m}(\grad\nh)^2\nonumber\\
&+&\frac{1}{16m\rho_0}(\grad\rho_+)^2+\frac{1}{16m\rho_0}(\grad\rho_-)^2
+\frac{1}{16m\rho_{m0}}(\grad\rho_m)^2
+\frac{\lambda}{4}\delta\rho_+^2
+\frac{\lambda}{4}\delta\rho_-^2
+\frac{\lambda_{12}}{4}(\delta\rho_+^2-\delta\rho_-^2)
+g_{am}\delta\rho_+\delta\rho_m
+\frac{g}{2}\delta\rho_m^2,\label{deltaLpolar3}\nonumber\\
&&\\
&=&i(\delta\rho_++2\delta\rho_m)\partial_\tau\theta_+ +
\frac{\rho_{s0}}{m}(\grad\theta_+)^2 +
i\delta\rho_-\partial_\tau\theta_- +
\frac{\rho_0}{m}(\partial_z\theta_-)^2+
\frac{1}{4m^3\alpha^2}(\grad\grad_\perp\theta_-)^2\nonumber\\
&+&\frac{1}{16m\rho_0}(\grad\rho_+)^2+\frac{1}{16m\rho_0}(\grad\rho_-)^2
+\frac{1}{16m\rho_{m0}}(\grad\rho_m)^2
+\frac{\lambda}{4}\delta\rho_+^2
+\frac{\lambda}{4}\delta\rho_-^2
+\frac{\lambda_{12}}{4}(\delta\rho_+^2-\delta\rho_-^2)
+g_{am}\delta\rho_+\delta\rho_m
+\frac{g}{2}\delta\rho_m^2,\label{deltaLpolar4}\nonumber\\
&&
\end{eqnarray}
\label{deltaLpolar5}
\ese
\end{widetext}
where
\bse
\begin{eqnarray}
\rho_{s0}&=&\rho_0+\rho_{m0},\\
\zh&=&\hat\ibQ,
\end{eqnarray}
\ese
and in the second form [Eq.~\rf{deltaLpolar3}] we used the minimum value of
$\ibQ$ [Eq.~\rf{ibQn}] characterizing the AMSF$_{\rm p}$ phase, which leads to
a minimal-like coupling between $\grad\theta_-$ and $\delta\nh$,
the latter transverse ($\perp$) to $\nh_0$ and $\ibQ$.  Subsequently, to
obtain our final expression, we integrated out $\delta\nh$ that to
lowest order via a Higgs-like mechanism introduced a low-energy
constraint
\bse
\begin{eqnarray}
\delta\nh&=&\frac{1}{\alpha m\sqrt{\rho_{m0}}}\grad_\perp\theta_-, \\
&=&\frac{1}{Q}\grad_\perp\theta_-.
\label{Higgslike}
\end{eqnarray}
\ese
Using it inside the $(\nabla\nh)^2$ term then leads to a quantum
smecticlike ``elasticity'' for the $\theta_-$ Goldstone mode, with
$\hat{\bm z}$ chosen to lie along $\ibQ$, that is, $\hat{\bm z} = \hat{\ibQ}$.
This smectic dispersion is expected based on the underlying rotational
symmetry, which is spontaneously broken by the periodic AMSF$_{\rm p}$
state. It is closely related to other periodic superfluids, such as,
for example, the Fulde-Ferrell-Larkin-Ovchinnikov pair-density wave
states~\cite{ff.64,lo.65,radzihovsky.vishwanath.08,FFLOlongLeo}.

As a final step we now integrate out the densities $\delta\rho_\pm$
fluctuations, obtaining at long scales (where $\grad\rho_\pm$ can be
neglected) our final form for the Goldstone mode Lagrangian in the
AMSF$_{\rm p}$ state:
\begin{widetext}
\begin{eqnarray}
\delta\curL_{\rm p} &=& 
\oh\chi_+(\partial_\tau\theta_+)^2 +
\frac{\rho_{s0}}{m}(\grad\theta_+)^2 +
\oh\chi_-(\partial_\tau\theta_-)^2 +
\frac{\rho_0}{m}(\partial_z\theta_-)^2+
\frac{1}{4m^3\alpha^2}(\grad^2_\perp\theta_-)^2,\label{deltaLpolarFinal}
\end{eqnarray}
\end{widetext}
where the compressibilities are given by
\bse
\begin{eqnarray}
\chi_-&=&\frac{2}{\lambda -\lambda_{12}},\\
\chi_+&=&\frac{g + 4 g_+ - 4 g_{am}}{g_+ g - g_{am}^2},
\end{eqnarray}
\label{chis}
\ese
with $g_+ = \oh(\lambda + \lambda_{12})$. 

Thus, the in-phase and out-of-phase Goldstone modes are characterized
by dispersions:
\bse
\begin{eqnarray}
\omega_{+p}(\ibk)&=&c_+ k,\label{omega+AMSFp}\\
\omega_{-p}(\ibk)&=&\sqrt{(B k_z^2 + K k_\perp^4)/\chi_-},\label{omega-AMSFp}
\end{eqnarray}
\label{omegasAMSFp}
\ese
with defined parameters
\bse
\begin{eqnarray}
c_+ &=& \sqrt{\frac{2\rho_{s0}}{\chi_+ m}},\label{v+AMSFp}\\
B&=&\frac{2\rho_0}{m},\label{B_AMSFp}\\
K&=&\frac{1}{2m^3\alpha^2}.\label{K_AMSFp}
\end{eqnarray}
\label{parametersAMSFp}
\ese
The linear $\omega_+(k)$ dispersion of the superfluid phase $\theta_+$
is the expected Bogoliubov mode corresponding to the superfluid
order. The anisotropic smecticlike dispersion of the ``phonon''
$\theta_-$ is a reflection of the uniaxial finite-momentum order in
the AMSF$_{\rm p}$ state, akin to the FF
superconductor~\cite{ff.64,FFLOlongLeo}.

\begin{figure}[thb]
\includegraphics[width=7cm]{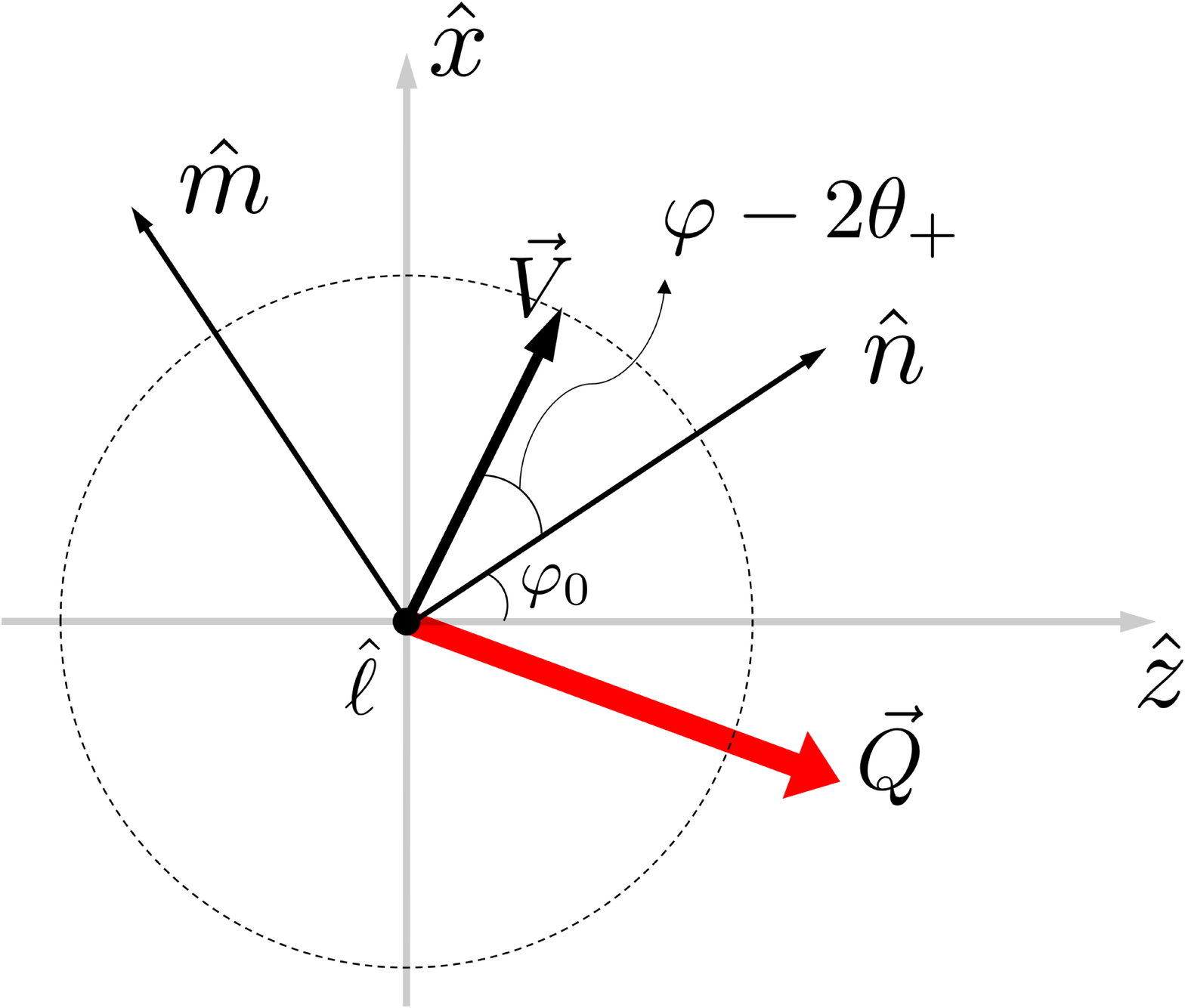} 
\caption{(Color online) The diagram defining various vectors appearing in Eq.~\rf{deltaLfm0}.
$\vec V = \nh \cos(\varphi-2\theta_+)-\mh\sin(\varphi-2\theta_+)$,
while $\varphi-2\theta_+$ is measured relative to the $\nh$ axis and $\varphi_0$ is measured relative to $\zh$.
}
\label{ferroangle}
\end{figure} 

\subsubsection{AMSF$_{fm}$}
The analysis for the AMSF$_{\rm fm}$ phase is very similar,
with only a single modification of the MSF$_{\rm fm}$ order parameter,
given instead by $\bm{\hat\phi}$ in Eq.~\rf{fmAMSFphi}. The
corresponding fluctuations Lagrangian density is given by
\begin{widetext}
\bse
\begin{eqnarray}
\delta\curL_{\rm fm} &\approx& 
i\delta\rho_+\partial_\tau\theta_+ +
\frac{\rho_0}{m}(\grad\theta_+)^2 +
i\delta\rho_-\partial_\tau\theta_- +
\frac{\rho_0}{m}(\grad\theta_-+\ibQ)^2 +
i\delta\rho_{m}\partial_\tau(\varphi-\varphi_0) + 
i\rho_{m0}\nh\cdot\partial_\tau\mh +
\frac{\rho_{m0}}{4m}(\grad\varphi)^2\nonumber\\
&+&\frac{\rho_{m0}}{8m}(\grad\nh)^2 +
\frac{\rho_{m0}}{8m}(\grad\mh)^2+
\frac{1}{16m\rho_{m0}}(\grad\rho_m)^2
-\sqrt{2}\alpha\rho_0\sqrt{\rho_{m0}}\;
(\grad\theta_- + \ibQ)\cdot\left[\nh\cos(\varphi-2\theta_+)
-\mh\sin(\varphi-2\theta_+) \right] \nonumber\\
&+&\frac{1}{16m\rho_0}(\grad\rho_+)^2+\frac{1}{16m\rho_0}(\grad\rho_-)^2
+\frac{\lambda}{4}\delta\rho_+^2
+\frac{\lambda}{4}\delta\rho_-^2
+\frac{\lambda_{12}}{4}(\delta\rho_+^2-\delta\rho_-^2)
+g_{am}\delta\rho_+\delta\rho_m
+\frac{g_1}{2}\delta\rho_m^2,\label{deltaLfm0}\\
&\approx&i(\delta\rho_++2\delta\rho_m)\partial_\tau\theta_+ +
\frac{\rho_{s0}}{m}(\grad\theta_+)^2 +
i\delta\rho_-\partial_\tau\theta_- +
\frac{\rho_0}{m}\left(\grad\theta_- - \frac{1}{\sqrt{2}}\alpha m\sqrt{\rho_{m0}}\delta\nh\right)^2 +
i\rho_{m0}\delta\nh\cdot\partial_\tau\mh\nonumber\\
&+&\frac{\rho_{m0}}{8m}(\grad\nh)^2+\frac{\rho_{m0}}{8m}(\grad\mh)^2
+\frac{1}{16m\rho_0}(\grad\rho_+)^2+\frac{1}{16m\rho_0}(\grad\rho_-)^2
+\frac{1}{16m\rho_{m0}}(\grad\rho_m)^2\nonumber\\
&+&\frac{\lambda}{4}\delta\rho_+^2
+\frac{\lambda}{4}\delta\rho_-^2
+\frac{\lambda_{12}}{4}(\delta\rho_+^2-\delta\rho_-^2)
+g_{am}\delta\rho_+\delta\rho_m
+\frac{g_1}{2}\delta\rho_m^2,\hspace{-0.7cm}\label{deltaLfm1}
\end{eqnarray}
\label{deltaLfm}
\ese
\end{widetext}
where to get the second form we performed a gauge transformation to
absorb the $\nh-\mh$ planar rotations angle
$\mh\cdot\partial_\tau\nh\equiv\partial_\tau\varphi_0$ into
$\partial_\tau\varphi$ and to simplify the FR term, as
well as subsequently integrated out $\varphi$, completed the square
into a minimal-like coupling for $\grad\theta_-$, and chose
$\ibQ=\alpha m \sqrt{\frac{\rho_{m0}}{2}}\nh_0$, similar to the
polar state analysis of the previous section.

Integrating out $\delta\nh$, with the effective minimal-coupling
constraint [Eq.~\rf{Higgslike}] and the constraint on the in-plane ($\nh-\mh$)
component of $\delta\mh$,
\begin{eqnarray}
\nh\cdot\delta\mh&=&-\mh\cdot\delta\nh,
\end{eqnarray}
at long scales we find
\begin{widetext}
\bse
\begin{eqnarray}
\delta\curL_{\rm fm} &\approx& 
i(\delta\rho_++2\delta\rho_m)\partial_\tau\theta_+ +
\frac{\rho_{s0}}{m}(\grad\theta_+)^2 +
i\delta\rho_-\partial_\tau\theta_- +
\frac{\rho_0}{m}(\partial_z\theta_-)^2+
\frac{1}{4m^3\alpha^2}(\grad\grad_\perp\theta_-)^2+
\frac{1}{4m^3\alpha^2}(\grad\partial_x\theta_-)^2\nonumber\\
&+&
i\frac{\sqrt{2\rho_{m0}}}{\alpha m}\partial_y\theta_-\partial_\tau\gamma
+\frac{\rho_{m0}}{8m}(\grad\gamma)^2
+\frac{1}{16m\rho_0}(\grad\rho_+)^2+\frac{1}{16m\rho_0}(\grad\rho_-)^2
+\frac{1}{16m\rho_{m0}}(\grad\rho_m)^2\nonumber\\
&+&\frac{\lambda}{4}\delta\rho_+^2
+\frac{\lambda}{4}\delta\rho_-^2
+\frac{\lambda_{12}}{4}(\delta\rho_+^2-\delta\rho_-^2)
+g_{am}\delta\rho_+\delta\rho_m
+\frac{g_1}{2}\delta\rho_m^2,\label{deltaLfm}\\
&=& 
\oh\chi_+(\partial_\tau\theta_+)^2 +
\frac{\rho_{s0}}{m}(\grad\theta_+)^2 +
\oh\chi_-(\partial_\tau\theta_-)^2 +
\oh B(\partial_z\theta_-)^2+
\oh K_x (\grad\partial_x\theta_-)^2+
\oh K_y (\grad\partial_y\theta_-)^2\nonumber\\
&+&
i\kappa\partial_y\theta_-\partial_\tau\gamma 
+ \oh J (\grad\gamma)^2,
\label{deltaLpolarFinal}
\end{eqnarray}
\label{deltaLferro}
\ese
\end{widetext}
where we used $[\grad(\delta\mh)]^2 = [\grad(\nh\cdot\delta\mh)]^2 +
[\grad(\lh\cdot\delta\mh)]^2$, introduced couplings 
\bse
\begin{eqnarray}
\kappa&=&\frac{\sqrt{2\rho_{m0}}}{\alpha m},\label{kappa_AMSFfm}\\
K_x&=&\frac{1}{m^3\alpha^2}=K,\label{Kx_AMSFfm}\\
K_y&=&\frac{1}{2m^3\alpha^2},\label{KyJ_AMSFfm} \\
J &=&\frac{\rho_{m0}}{4m} = K_y Q^2,
\end{eqnarray}
\label{kappaKxyAMSFfm}
\ese
defined a real scalar field
\begin{eqnarray}
\gamma\equiv\lh\cdot\delta\mh,
\end{eqnarray}
for fluctuations of $\mh$ outside of the $\nh-\mh$ plane, and chose
axes
\bse
\begin{eqnarray}
\xh &=&\mh,\\
\yh &=&\lh.
\end{eqnarray}
\ese

We note that the Goldstone-modes action [Eq.~\rf{deltaLferro}] exhibits a {\em
  biaxial} smectic energetics in the smectic phonon, $\theta_-$, in
addition to the $xy$-model energetics of the superfluid phase,
$\theta_+$. The biaxiality is expected and arises due to a smectic in-plane
polar ($p$-wave) order, characterized by a spinor, $\bm\phi_{fm}$, with
the quantization axis, $\lh$. The finite angular momentum, $\ell_z=\pm
1$ along $\lh$ distinguishes AMSF$_{\rm fm}$ from AMSF$_{\rm p}$ and leads to
an additional Goldstone mode $\gamma$. 

A straightforward diagonalization of the above Lagrangian leads to
dispersions for three Goldstone modes inside the AMSF$_{\rm fm}$ state:
\bse
\begin{eqnarray}
  \omega^+_{fm}(\ibk)&=&c_+ k,\label{omega+AMSFfm}\\
  \omega^-_{fm}(\ibk)&=&\sqrt{[B k_z^2 + k^2(K_x k_x^2 +
K_yk_y^2)]/\chi_-},\hspace{1cm}\label{omega-AMSFfm}\\
  \omega^\gamma_{fm}(\ibk)&=&
\sqrt{\frac{J k^2[B k_z^2 + k^2(K_x k_x^2 + K_yk_y^2)]}
{J\chi_-k^2 +\kappa^2 k_y^2}}.\,\label{omega_gammaAMSFfm}
\end{eqnarray}
\label{omegasAMSFfm}
\ese
The anisotropic $\omega^\gamma_{fm}(\ibk)$ dispersion corresponds to
the ferromagnetic spin waves in the plane of atomic condensate
phase fronts (``smectic layers'') of the $p$-wave atomic-molecular
condensate, AMSF$_{\rm fm}$, reducing to the dispersion of MSF$_{\rm fm}$ in
Eq.~\rf{msf.ferro.sound} for a vanishing smectic order, with $B=0$.

\section{Phase Transitions}
\label{sec:PhaseTransition}

In this section, we study the quantum MSF - AMSF phase transitions
beyond earlier mean-field approximation, demonstrating that they are
described by a d+1-dimensional quantum de Gennes (Abelian Higg's)
model\cite{degennes.prost.95} akin to that for a
normal-to-superconductor and nematic-to-smectic-A transitions. Based
on the extensive work for these systems
\cite{HLM74, Coleman73}, in three (spatial)
dimensions ($d=3$) we predict that the effective gauge-field
fluctuations drive this transition first order. The derivation is most
transparent via a coherent-state Lagrangian, Eq.~\rf{ScurL},
\begin{widetext}
\begin{eqnarray}
\curL &=& 
\psi^*_\sigma(\partial_\tau-\frac{\nabla^2}{2m}-\mu_\sigma)\psi_\sigma +
\bm\phi^*\cdot(\partial_\tau-\frac{\nabla^2}{4m}-\mu_m)\cdot\bm\phi 
+\frac{\lambda_\sigma}{2}\ve\psi_\sigma\ve^4 \nonumber \\
&+&\lambda_{12}\ve\psi_{1}\ve^2\ve\psi_{2}\ve^2
+g_{am}\left(\ve\psi_{1}\ve^2+\ve\psi_{2}\ve^2\right)\ve{\bm\phi}\ve^2 
+\frac{g_1}{2}\ve{\bm\phi}^*\cdot{\bm\phi}\ve^2
+\frac{g_2}{2}\ve{\bm\phi}\cdot{\bm\phi}\ve^2 \nonumber \\
&+&\frac{\alpha}{2}
\left({\bm\phi}^*\cdot \left[\psi_1(-i\grad)\psi_2-\psi_2(-i\grad)\psi_1\right]+c.c.\right),
\label{curLpt}
\end{eqnarray}
\end{widetext}
working in polar representation similar to that of the previous
section.

\subsection{MSF$_{\rm p}$-AMSF$_{\rm p}$ polar transition}
It is convenient to analyze the transition from the MSF side,
where the atomic and molecular order parameters are given by
\bse
\begin{align}
\psi_\sigma =& \psi_{\sigma,\ibQ_\sigma}e^{i\ibQ_\sigma\cdot\br}, \ \ \  \ibQ_\sigma = \pm \ibQ,
\ \ \ \text{for} \ \  \sigma = 1,2 \\
\bm\phi =& \sqrt{\rho_{m0}}e^{i\varphi(\br)}\nh.
\end{align}
\ese
Using these forms inside $\curL$ [Eq.~\rf{curLpt}] and for simplicity
focusing on the balanced case with
$\tilde\mu=\tilde\mu_\sigma=\mu_\sigma-g_{am}\rho_{m0}$, we obtain
\begin{widetext}
\begin{align}
\curL_{\rm p} =
&\psi_{1,\ibQ}^*\partial_\tau\psi_{1,\ibQ}
+\psi_{2,-\ibQ}^*\partial_\tau\psi_{2,-\ibQ}
+\left(\frac{Q^2}{2m}-\tilde\mu\right)(\ve\psi_{1,\ibQ}\ve^2+\ve\psi_{2,-\ibQ}\ve^2)
+\frac{1}{2m}\ve\grad\psi_{\sigma,\ibQ}\ve^2  \nonumber \\
+&\left(\frac{1}{2m}\ibQ_\sigma\cdot\psi_{\sigma,\ibQ}^*(-i\grad)\psi_{\sigma,\ibQ}
-\alpha\sqrt{\rho_{m0}}e^{-i\varphi}\nh\cdot
\Big(\ibQ\psi_{1,\ibQ}\psi_{2,-\ibQ}
+\frac{1}{2}\left[\psi_{1,\ibQ}(-i\grad)\psi_{2,-\ibQ}-\psi_{2,-\ibQ}(-i\grad)\psi_{1,\ibQ}\right]\Big)
+c.c.\right) \nonumber \\
+&i\delta\rho_m\partial_\tau\varphi
+i\rho_{m0}\delta{\bm m}\cdot\partial_\tau\nh
+\frac{\rho_{m0}}{4m}(\grad\varphi)^2
+\frac{\rho_{m0}}{4m}(\grad\nh)^2
+\frac{g}{2}\delta\rho_m^2
+\frac{g_2}{2}\rho_{m0}^2\ve \delta {\bm m}\ve^2
+\curL_{\rm int}+\curL_{\rm msf},
\end{align}
\end{widetext}
where terms linear in fields vanish by virtue of the saddle-point
equations. The contribution $\curL_{\rm msf}$ is the mean-field part
analyzed in Sec.~\ref{sec:mft} and $\curL_{\rm int}$ is the higher order term.
Defining
\bse
\begin{align}
\varepsilon_Q &= \frac{Q^2}{2m}-\tilde\mu, \\
\Delta_\ibQ &=\alpha \sqrt{\rho_{m0}}\nh\cdot\ibQ, 
\end{align}
\ese
and introducing atomic eigenfields $\psi_\pm$,
\bse
\begin{align}
\psi_+&=\frac{1}{\sqrt{2}}(-\psi_{1,\ibQ}+\psi_{2,-\ibQ}^*), \\
\psi_-&=\frac{1}{\sqrt{2}}(\psi_{1,\ibQ}+\psi_{2,-\ibQ}^*),
\end{align}
\ese
a mean-field version of which was obtained in Sec.~\ref{sec:mft}, the
Lagrangian simplifies considerably to,
\begin{widetext}
\begin{align}
\curL_{\rm p}=
&-\psi_+^*\partial_\tau\psi_- +\psi_+\partial_\tau\psi_-^*
+\frac{1}{2m}\ve\left(-i\grad+\ibQ + \alpha m \sqrt{\rho_{m0}}\nh\cos\varphi\right) \psi_+\ve^2
+\frac{1}{2m}\ve\left(-i\grad+\ibQ - \alpha m \sqrt{\rho_{m0}}\nh\cos\varphi\right) \psi_-\ve^2 \nonumber \\
&+\left[\epsilon_+-\frac{1}{2m}\left(\ibQ+\alpha m \sqrt{\rho_{m0}}\nh\cos\varphi\right)^2\right]\ve\psi_+\ve^2
+\left[\epsilon_--\frac{1}{2m}\left(\ibQ-\alpha m \sqrt{\rho_{m0}}\nh\cos\varphi\right)^2\right]\ve\psi_-\ve^2
\nonumber \\
&+i\alpha\sqrt{\rho_{m0}}\nh\cdot(\psi_+(-i\grad)\psi_-^*-\psi_+^*(-i\grad)\psi_-)\sin\varphi
+\frac{1}{2g}(\partial_\tau\varphi)^2
+\frac{\rho_{m0}}{4m}(\grad\varphi)^2
+\frac{1}{2g_2}(\partial_\tau\nh)^2
+\frac{\rho_{m0}}{4m}(\grad\nh)^2 \nonumber \\
&+\curL_{\rm int}
+\curL_{\rm msf},
\end{align}
\end{widetext}
where
\begin{align}
\epsilon_{\pm} &= \varepsilon_Q\pm\ve\Delta_\ibQ\ve
\end{align}
and we completed the square in $\curL_{\rm p}$.  It can be shown that near a
critical point the $\sin\varphi$ contribution leads to an irrelevant
quartic correction to $\ve\psi_-\ve^4$ and renormalization of
$(\partial_\parallel\psi_-)^2$ stiffness.  Furthermore, it is clear
that the canonically conjugate field $\psi_+$ (it appears as a
canonical momentum for the critical field $\psi_-$) remains massive at
the MSF-AMSF transition, defined by the vanishing of the coefficient
of $\ve\psi_-\ve^2$ term, consistent with Sec.~\ref{sec:mft}.
Therefore, safely integrating out $\psi_+$ and making a choice $\ibQ = \alpha
m\sqrt{\rho_{m0}}\nh_0$ that minimizes the energy, leads to
\begin{align}
\curL_{\rm p}=&
\eps_+^{-1}\ve\partial_\tau\psi_-\ve^2
+\frac{1}{2m}\ve\left(-i\grad - \alpha m \sqrt{\rho_{m0}}\delta\nh\right) \psi_-\ve^2 \nonumber \\
&+\epsilon_-\ve\psi_-\ve^2
+\frac{\lambda}{2}\ve\psi_-\ve^4
+\frac{1}{2g_2}(\partial_\tau\nh)^2
+\frac{\rho_{m0}}{4m}(\grad\nh)^2 \nonumber \\
&+\frac{1}{2g}(\partial_\tau\varphi)^2
+\frac{\rho_{m0}}{4m}(\grad\varphi)^2,
\end{align}
with $\lambda=\frac{1}{4}(\lambda_1+\lambda_2+2\lambda_{12})$, and we
dropped the mean-field part and irrelevant interactions.

Thus, as anticipated on symmetry grounds, the zero-temperature
MSF$_{\rm p}$-AMSF$_{\rm p}$ transition is indeed described by a quantum
($(d+1)$-dimensional) de Gennes model (or equivalently the
Ginzburg-Landau) Lagrangian~\cite{degennes.prost.95}, where the role
of the nematic director (gauge-field) is played by the $\ell_z=0$
quantization axis of the $p$-wave molecular condensate.  

\subsection{MSF$_{\rm fm}$-AMSF$_{\rm fm}$ ferromagnetic transition}
Using the field forms appropriate for the ferromagnetic case
\bse
\begin{align}
\psi_\sigma =& \psi_{\sigma \ibQ_\sigma}e^{i\ibQ_\sigma\cdot\br}, \ \ \  \ibQ_\sigma = \pm \ibQ,
\ \ \ \text{for} \ \  \sigma = 1,2 \\
\bm\phi =& \sqrt{\frac{\rho_{m0}}{2}}(\nh +i\mh),
\end{align}
\ese
a very similar analysis leads to 
\begin{widetext}
\bse
\begin{align}
\curL_{\rm fm} =& 
\epsilon_+\ve\psi_+\ve^2 +\epsilon_-\ve\psi_-\ve^2
+\frac{1}{2m}\ve\grad\psi_+\ve^2+\frac{1}{2m}\ve\grad\psi_-\ve^2 
+i\rho_{m}\nh\cdot\partial_\tau\mh
+\frac{\rho_{m0}}{8m}(\grad\nh)^2 +\frac{\rho_{m0}}{8m}(\grad\mh)^2 \nonumber \\
&+\alpha\frac{\sqrt{\rho_{m0}}}{\sqrt{2}}(\nh-i\mh)\cdot
\left(\psi_+^*(-i\grad)\psi_+ -\psi_-^*(-i\grad)\psi_-\right) \nonumber \\
&+\frac{\ibQ}{m}\cdot
\left(\psi_+^*(-i\grad)\psi_+ +\psi_-^*(-i\grad)\psi_-\right)
-\psi_+^*\partial_\tau\psi_- +\psi_+\partial_\tau\psi_-^*
+\frac{g}{2}\delta\rho_m^2
+\curL_{\rm int}+\curL_{\rm msf}, \\
=&-\psi_+^*\partial_\tau\psi_- +\psi_+\partial_\tau\psi_-^*
+\frac{1}{2m}\Bigg\ve\left(-i\grad+\ibQ + \frac{1}{\sqrt{2}}\alpha m \sqrt{\rho_{m0}}\nh\right) \psi_+\Bigg\ve^2
+\frac{1}{2m}\Bigg\ve\left(-i\grad+\ibQ - \frac{1}{\sqrt{2}}\alpha m \sqrt{\rho_{m0}}\nh\right) \psi_-\Bigg\ve^2 
\nonumber \\
&+\left[\epsilon_+-\frac{1}{2m}\left(\ibQ+\frac{1}{\sqrt{2}}\alpha m \sqrt{\rho_{m0}}\nh\right)^2\right]\ve\psi_+\ve^2
+\left[\epsilon_--\frac{1}{2m}\left(\ibQ-\frac{1}{\sqrt{2}}\alpha m \sqrt{\rho_{m0}}\nh\right)^2\right]\ve\psi_-\ve^2
+i\rho_{m}\nh\cdot\partial_\tau\mh \nonumber \\
&+\frac{\rho_{m0}}{8m}(\grad\nh)^2 +\frac{\rho_{m0}}{8m}(\grad\mh)^2
+\alpha\frac{\sqrt{\rho_{m0}}}{\sqrt{2}}(-i\mh)\cdot
\left(\psi_+^*(-i\grad)\psi_- -\psi_+(+i\grad)\psi_-^*\right) 
+\frac{g}{2}\delta\rho_m^2
+\curL_{\rm int}+\curL_{\rm msf},
\end{align}
\ese
\end{widetext}
where to obtain the final form we rotated $\nh$ and $\mh$ by
$-\varphi$ and completed the square.  Similarly to the treatment of
the polar case in the previous section, here it can be shown that
the linear $(-i\mh)$ term only leads to irrelevant quartic coupling
and can therefore be neglected.  Integrating out the noncritical
conjugate field $\psi_+$ gives the final Lagrangian
form
\begin{align}
\curL_{\rm fm}=&
\eps_+^{-1}\ve\partial_\tau\psi_-\ve^2
+\frac{1}{2m}\Bigg\ve\left(-i\grad - \frac{\alpha m \sqrt{\rho_{m0}}}{\sqrt{2}}\delta\nh\right)\psi_-\Bigg\ve^2 
\nonumber \\
&+\epsilon_-\ve\psi_-\ve^2
+\frac{\lambda}{2}\ve\psi_-\ve^4
+\frac{\rho_{m0}}{8m}(\grad\nh)^2
+\frac{\rho_{m0}}{8m}(\grad\mh)^2 \nonumber \\
&+i\rho_{m}\nh\cdot\partial_\tau\mh
\end{align}
of the quantum de Gennes-Ginzburg-Landau form that controls the
MSF$_{\rm fm}$-AMSF$_{\rm fm}$ transition.  In the above we dropped the
mean-field part and irrelevant interactions. As anticipated by
symmetry, it is distinguished from the polar case by the additional
biaxial order whose fluctuations are characterized by $\mh$.

\section{Topological Defects}
\label{sec:topol_defects}

Having established the nature of the ordered states, characterized by
Landau order parameters, and the associated Goldstone modes, we now
turn to a brief discussion of the corresponding topological
defects. As usual, these singular excitations are crucial to a
complete characterization of the states and their disordering,
particularly in the case of non-mean-field (e.g., partially disordered)
states that are not uniquely characterized by a Landau order
parameter.

\subsection{Defects in ASF}

As discussed in Sec.\ref{sec:mft}, the ASF$_i$ states (with
$i=1,2,12$) are characterized by two atomic condensate order
parameters, $\psi_\sigma =\sqrt{\rho_\sigma}
e^{i\theta_\sigma}$. Correspondingly, as in an ordinary superfluid,
because $\theta_\sigma$ are {\em compact} phase fields
($\theta_\sigma$ and $\theta_\sigma + 2\pi$ are physically
identified), in addition to their smooth Goldstone mode
configurations, there are vortex topological excitations,
corresponding to nonsingle-valued configurations of
$\theta_\sigma(\br)$. These are defined by two corresponding
integer-valued closed line integrals enclosing a
vortex line
\begin{eqnarray}
  \oint d\vec{\ell}\cdot\vec{\nabla}\theta_\sigma = 2\pi p_\sigma.
\end{eqnarray}

In a differential form, the line defects are equivalently encoded as
\begin{equation}
  \nabla\times \nabla\theta_\sigma={\bf m}_\sigma\;,\label{vortex_sigma}
\end{equation}
with vortex line topological ``charge'' density given by
\begin{equation}
{\bf m}_\sigma(\br)=2\pi\sum_i\int p^i_\sigma \hat{\bf t}_i(s_i)
\delta^3({\bf r}-{\bf r}_i(s_i)) d s_i\;,
\label{m_sigma}
\end{equation}
where $s_i$ parametrizes the $i$'th vortex line (or loop), ${\bf
  r}_i(s_i)$ gives its positional conformation, $\hat{\bf t}_i(s_i)$
is the local unit tangent, and vortex ``charges'' $p^i_\sigma$ are
independent of $s_i$, since the charge of a given line is constant
along the defect. Furthermore,
\begin{equation}
\nabla\cdot{\bf m}({\bf r})=0
\label{continuity}
\end{equation}
enforces the condition that vortex lines cannot end in the bulk of the
sample; they must either form closed loops or extend entirely through
the system.

Thus, vortices in the single-component ASF$_\sigma$ states are
characterized by a $n_\sigma$ integer, and in the two-component
ASF$_{12}$ the defects are specified by a pair of integers
$(p_1,p_2)$.  These are associated with the fundamental group $\pi_1$
of the torus $U(1)\otimes U(1)$, that characterizes
the low-energy manifold of Goldstone modes of the ASF$_{12}$ state. It
is therefore closely related to other $U(1)\otimes U(1)$ systems, such
as easy-plane spinor-1 condensates~\cite{PodolskyS1prb} and two-gap
superconductors, for example, MgB$_2$~\cite{Babaev02prl}. 

As in conventional superfluids vortices appear in response to imposed
rotation and proliferate with enhanced quantum and thermal
fluctuations, providing a complementary description of phase
transitions out of the ASF$_i$ states.

\subsection{Defects in MSF}

Because of its finite angular momentum, $\ell=1$, structure the defects
in the MSF states are somewhat more complicated. However, relying on
the aforementioned relation of the MSF to the well-explored spinor-$1$
condensates
~\cite{stenger.ketterle.98,mukerjee.moore.06,ho.spinor.98,ohmi.machida.98,zhou.spinor.01,demler.zhou.02},
we inherit a clear characterization of defects in the two MSF
phases. As discussed in Sec.~\ref{sec:mft} the polar MSF$_{\rm p}$ and
the ferromagnetic MSF$_{\rm fm}$ states are respectively characterized
by $[S_2 \times U_N(1)]/{\mathbb Z}_2$ (the mod out by ${\mathbb Z}_2$
corresponds to the identification of $\nh\rightarrow -\nh$ 
with $\varphi\rightarrow\varphi + \pi$) and $SO(3)$
order-parameter (Goldstone mode) manifolds. The defects are
characterized by the homotopy group of the corresponding manifolds. In
the MSF$_{\rm fm}$ case the SO(3)=S$_3/{\mathbb Z}_2$
manifold also appears in the dipole-locked A phase of helium-3 with
topological defects well understood~\cite{Mermin}.

The nature of defects in the MSF$_{\rm p}$ state was a subject of some
debate, until it was definitively resolved by Mukerjee, {\it et
al}.~\cite{mukerjee.moore.06}. These are characterized by elements of the
homotopy groups $\pi_n(S_2 \times U_N(1)/{\mathbb Z}_2)={\mathbb Z}$. The key new
feature is the appearance of a composite defect that is a $\pi$ vortex
and $\nh$ texture where $\nh\rightarrow -\nh$, keeping the molecular
order parameter single-valued at long scales.  The consequences of
this were discussed and explored through Monte Carlo simulations by
Mukerjee, {\it et al}.~\cite{mukerjee.moore.06}, and is quite closely related to
other realizations of composite half-integer
defects~\cite{radzihovsky.04.boson,romans.04,radzihovsky.08.boson,radzihovsky.vishwanath.08,FFLOlongLeo}. We expect
the MSF$_{\rm p}$ to exhibit similar phenomenology, which we do not explore
further here.

\subsection{Defects in AMSF}

As discussed in Sec.~\ref{sec:mft}, in addition to the
molecular condensate $\bm\phi$, the two AMSF states are characterized
by a finite momentum two-component atomic condensate order parameter,
with a nonzero amplitude
\begin{eqnarray}
\Psi_-&=&e^{-i\varphi}\Psi_{1,\ibQ}+\Psi_{2,-\ibQ}^*,
\label{psim}
\end{eqnarray}
and a vanishing amplitude $\Psi_+=0$ [Eq.~\rf{pmOP}]. The latter is
consistent with the locking of the atomic condensate phase
$\theta_{+}=\oh(\theta_1+\theta_2)$ to a MSF phase
$\varphi/2$, imposed by the FR coupling [Eq.~\rf{deltaLpolar}]. It also
locks the atomic condensate magnitudes to be equal,
$|\Psi_{1,\ibQ}|=|\Psi_{2,-\ibQ}|$.

Using the phase representation, the atomic condensate order parameter
reduces to
\begin{eqnarray}
\Psi_-&\sim&e^{i\theta_-}e^{-i\varphi/2}\cos(\theta_+ - \varphi/2).
\end{eqnarray}
From this form it is clear that, as a conventional superfluid, the
AMSF admits $2\pi$ vortices in $\theta_-=\oh(\theta_1-\theta_2)$, and
$\varphi=0$, corresponding to a $2\pi$ ``spin'' vortex,
\bse
\begin{eqnarray}
\theta_-(\br) &=& \theta_1(\br)=-\theta_2(\br),\\
&=&\theta,\ \ \ \mbox{$2\pi$ ``spin'' vortex},
\end{eqnarray}
\ese
with equal counterpropagating (atomic species $1$ and $2$) currents
and a vanishing ``charge'' (atomic number) current. Above, $\theta$ is
a polar coordinate angle.

Another type of a defect is topologically equivalent to a
$2\pi$ vortex in $\theta_+(\br)$,
\bse
\begin{eqnarray}
\theta_+(\br) &=& \theta_1(\br)=\theta_2(\br),\\
&=&\theta,\ \ \ \mbox{$2\pi$ ``charge''-vortex},
\end{eqnarray}
\ese
with equal copropagating (atomic species $1$ and $2$) currents and a
vanishing ``spin'' current. However, as is clear from the Feshbach
interaction form in Eq.~\rf{deltaLpolar0},
\begin{eqnarray}
\delta\curL_{FBR} &\sim & \cos(\varphi-2\theta_+),
\end{eqnarray}
for vortex-free molecular order parameter (e.g. $\varphi=0$),
inside the AMSF phase the ``charge'' vortex $2\pi$ winding and
currents are confined to a domain wall whose thickness is set by the
ratio of the superfluid stiffness and FR coupling $\alpha$, on the
order of $1/Q$. As a result of this current confinement the energy of
such domain-wall scales linearly in two dimensions and as a surface in
three dimensions. Consequently, such $\pm2\pi$ ``charge'' vortices are confined into
neutral pairs inside the AMSF phase.
However, in the presence of a molecular $4\pi$ vortex,
with $\varphi(\br)=2\theta_+(\br)=2\theta$,
no domain wall appears and
conventional $\pm2\pi$ ``charge'' vortices can deconfine.

Finally, as with other analogous physical systems~\cite{radzihovsky.vishwanath.08,FFLOlongLeo},
the product form of the atomic condensate order parameter, $\Psi_-$
[Eq.~\rf{psim}], admits composite defects with half-integer topological
charge. These are characterized by a bound state of a $\pi$-``spin''
and $\pm\pi$-``charge'' vortices, with latter (as above) confined by
FR interaction into a $\pm\pi$ domain wall. The simplest (topologically
faithful) realization of this is a vortex only in one (but not both)
atomic species,
\bse
\begin{align}
\theta_+(\br) &= \theta_-(\br)=\oh\theta_1(\br)=\oh\theta,\ \ \theta_2(\br)=0,\nonumber\\
&\ \ \ \mbox{$(+\pi) -(+\pi)$ vortex domain wall},\\
\theta_+(\br) &= -\theta_-(\br)=\oh\theta_2(\br)=\oh\theta,\ \ \theta_1(\br)=0,\nonumber\\
&\ \ \ \mbox{$(-\pi)-(+\pi)$ vortex domain wall}.
\end{align}
\ese
Again, in the presence of a $\pm2\pi$ molecular vortex, $\varphi(\br)=\pm\theta$,
the $\pi$-``spin'', $\pi$-``charge'' composite vortex, 
$\theta_-(\br)=\pm\theta_+(\br)=\frac{\theta}{2}$
no longer exhibits a domain wall,
since $\varphi-2\theta_+=2\pi p$. 
It is therefore not confined inside the AMSF state.

Clearly, out of the above six types of defects, the $2\pi$-``spin''
vortex is least energetically costly, because it does not involve a
``charge'' domain wall in $\theta_+$ nor require an additional molecular vortex.
On the other hand, it is the
two half-integer vortex domain-wall defects that are the
elementary ones. This therefore opens up a possibility of
unconventional nonsuperfluid states in the two-species $p$-wave
resonant Bose systems, driven by unbinding of composite topological
defects, like the $2\pi$-``spin'' vortex. We leave the discussion of
the resulting states to future work.

\begin{figure}[bth]
\includegraphics[width=85mm]{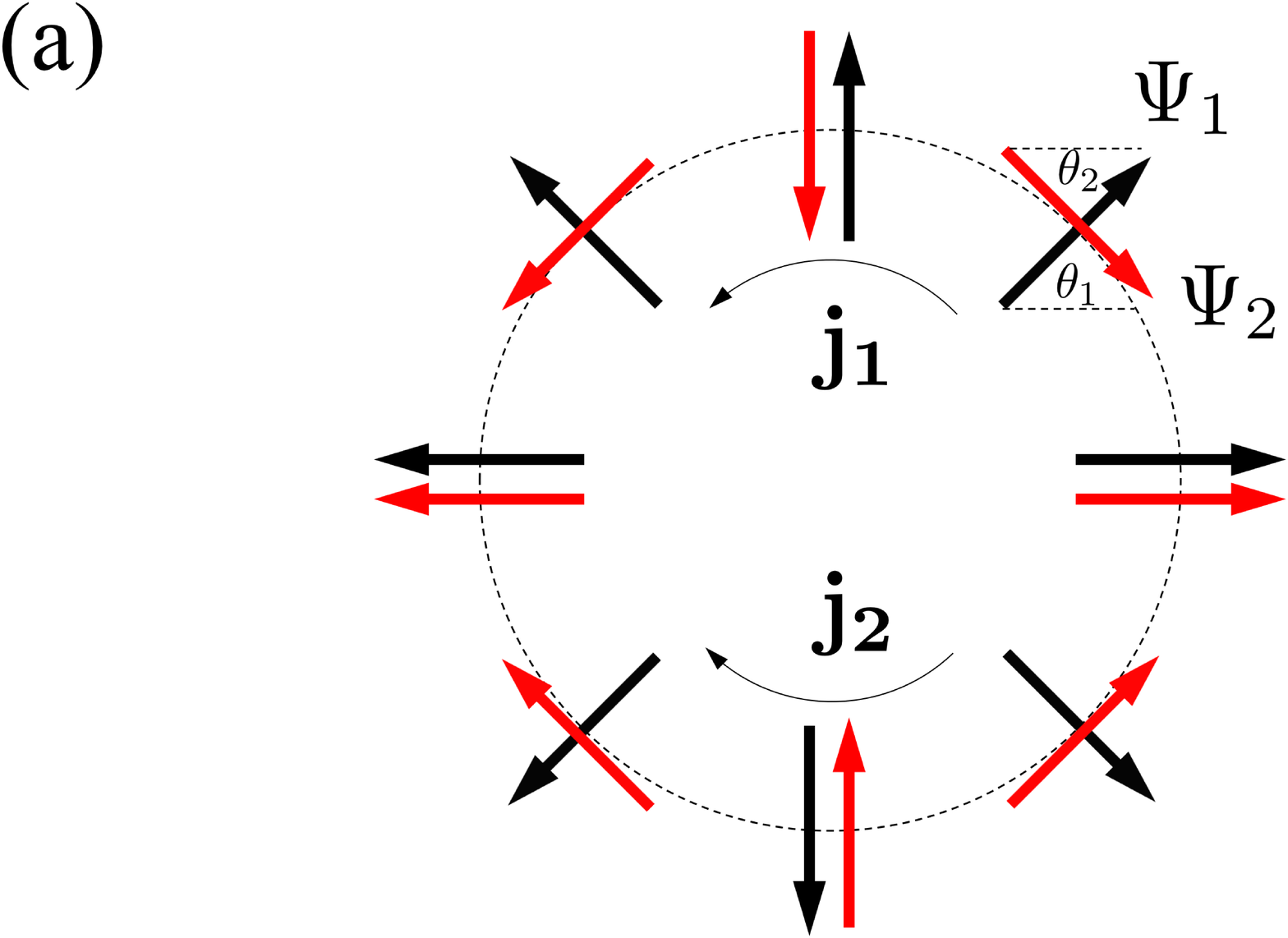}
\includegraphics[width=85mm]{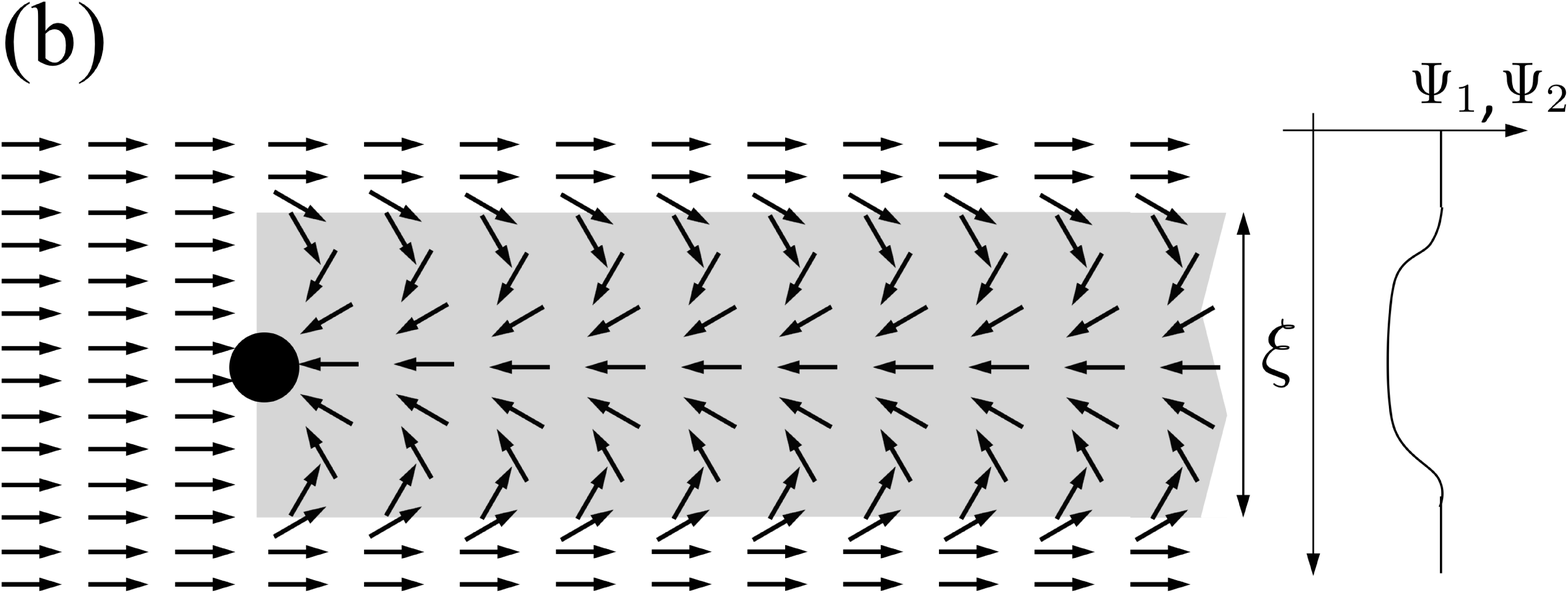}
\caption{(Color online) (a) $2\pi$-``spin'' vortex with equal counterpropagating
  atomic currents, $\bf j_1$, $\bf j_2$.  (b) $2\pi$-``charge'' vortex
  with equal copropagating currents, confined to a
  domain wall (gray area) of width $\xi \sim 1/Q$, with atomic order
  parameter suppressed.  
  In the presence of a molecular $4\pi$ vortex a domain wall is no longer required,
  and the ``charge'' vortex is deconfined.}
\label{fig:vortex}
\end{figure}
\begin{figure}[bth]
\includegraphics[width=85mm]{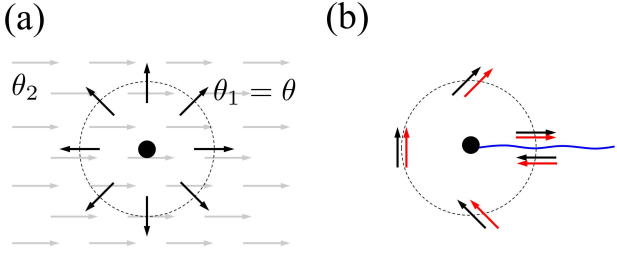}
\caption{(Color online) $\pi-\pi$ vortex in $\alpha=0$ limit.  (a) In
  $\theta_1-\theta_2$ representation; black (gray) arrows indicate
  $\theta_1$ ($\theta_2$).  (b) In $\theta_+-\theta_-$
  representation, the pair of arrows indicate $\theta_+$ and $\theta_-$,
  while the wavy line indicates a domain wall.  For $\alpha\neq 0$, the
  FR coupling ``squeezes'' the $\theta_+$ $\pi$-vortex
  textures into a domain wall of width $\xi\sim 1/Q$.  
  In the presence of an additional molecular $2\pi$ vortex, 
  the domain wall is absent and the composite defect is deconfined.}
\label{fig:vortex2}
\end{figure}

\section{Local Density Approximation}

\label{sec:LDA}

Because the primary experimental application of our predictions is to
degenerate atomic gases it is important to extend our analysis to
include the trapping potential $V_t(\br)$, which in a typical
experiment is well-approximated by a harmonic potential.  A full
analysis of the effect of the trap is beyond the scope of this
paper, and here we limit our treatment to a LDA.

Closely related to the WKB approximation~\cite{landau.QM}, LDA amounts
to the bulk system predictions, but with the chemical potential
replaced by an effective local chemical potential
$\mu(r)=\mu-V_t(r)$. The validity of the LDA relies on the smoothness
of the trap potential, with the criterion that $V_t(r)$ varies slowly
on the scale of the {\it longest\/} physical length $\lambda$ in the
problem, i.e., $(\lambda/V_t(r)) dV_t(r)/dr\ll 1$. Its accuracy can be
equivalently controlled by a ratio of the single-particle trap level
spacing $\delta E$ to the smallest characteristic energy $E_c$ of the
studied phenomenon (e.g, the chemical potential, condensation energy,
etc.), by requiring $\delta E/E_c \ll 1$.  For our system the longest
natural length scale is the period $2\pi/Q$, Eq.~\rf{Q} of the
finite-momentum atomic condensate inside the AMSF state. Thus, away
from the AMSF-ASF phase boundary, where $Q$ vanishes (see
Fig.~\ref{fig:Q}), we expect an LDA treatment of the effects of the
trap to be trustworthy.

A generalization of a resonant Bose-gas model [Eq.~\rf{Hmain}] to
include a trap is straightforward, accounted for by the additional
Hamiltonian density
\begin{eqnarray}
\curH_{\rm trap} &=& 
\sum_{\sigma=1,2} V_t(\br) \hat\psi_{\sigma}^{\dag}\hat\psi_\sigma 
+  2V_t(\br)\hat{\bm\phi}^{\dag}\cdot\hat{\bm\phi}, 
\label{eq:Htrap}
\end{eqnarray}
with $\curH\rightarrow \curH + \curH_{\rm trap}$. In the above, for
simplicity we specialized to an atomic species-independent trapping
potential and approximated the closed-channel molecular trapping
potential by twice the atomic one, valid for the interaction range
$r_0$ (typically less than 50 \AA) much smaller than the cloud size $R$
(typically larger than a micron).

Henceforth, to be concrete, we shall focus on an isotropic harmonic
trap (although this simplification can easily be relaxed) with
\bse
\begin{eqnarray}
V_t(\br) &=& \oh m \omega_t^2 r^2,\\
&\equiv&\mu\frac{r^2}{R^2},
\end{eqnarray}
\ese
the latter expression defining the cloud size $R$.  Within LDA, locally
the system is taken to be well-approximated as {\it uniform\/}, but
with a local chemical potential given by
\bse
\begin{eqnarray}
\label{eq:mulda}
\mu(r) &\equiv& \mu - \frac{1}{2} m \omega_t^2 r^2,\\
&=& \mu\left(1-\frac{r^2}{R^2}\right),
\end{eqnarray}
\ese
where $\mu$ is the true chemical potential (a Lagrange multiplier)
enforcing the total atom number $N$.  The spatially varying species
$1$ and $2$ chemical potentials are then given by:
\bse
\begin{eqnarray}
\mu_1(r) &=& \mu(r) +h ,
\\
\mu_2(r) &=& \mu(r) - h,
\end{eqnarray}
\ese
with a {\it uniform\/} chemical potential difference $h$ set by the
atomic species imbalance~\cite{SRprl.06,SRaop.07,FFLOlongLeo}.

\begin{figure}[bth]
\includegraphics[width=85mm]{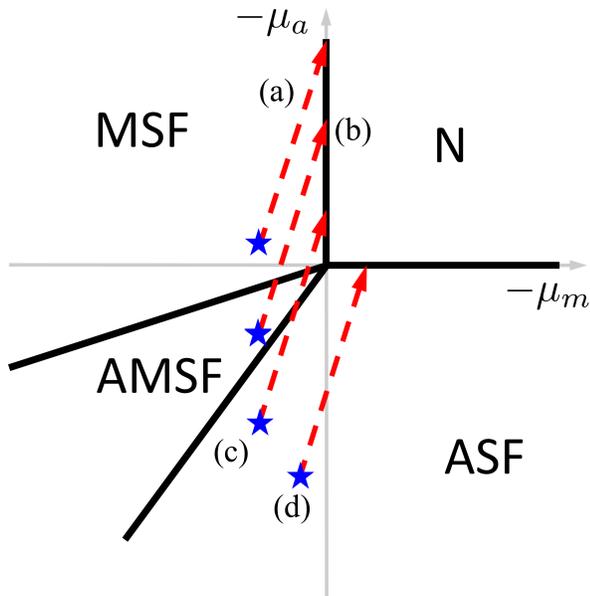}
\caption{(Color online) $N,\nu,T$-dependent cuts through the bulk phase diagram with
  increasing radial position $r$ through the atomic cloud.  Stars
  indicate system's parameters (local chemical potentials $\mu_a$,
  $\mu_m$) at the trap center.}
\label{fig:AMSFshells}
\end{figure}

Consequently, within LDA the system's energy density is approximated
by that of a uniform bulk system [Eq.~\rf{landauFE0}], with the spatial
dependence entering only through $\mu(r)$.  The ground-state energy is
then simply a volume integral of this energy density.  Thus, the phase
behavior of a uniform system as a function of the chemical potential,
$\mu$, translates into a spatial cloud profile through $\mu(r)$, with
the critical phase boundaries $\mu_c$ corresponding to critical radii
defined by $\mu_c = \mu(r_c,h)$~\cite{SRprl.06,SRaop.07}. As
predicted~\cite{weddingCakeMI,SRprl.06} and
observed~\cite{GreinerNature,Zwierlein06Science,Partridge06Science,
  Shin2006prl, Navon2009prl} in other systems, this leads to a
shell-like cloud structure ``imaging'' of the bulk phase diagram as  illustrated in Fig.~\ref{fig:AMSFshells}.

\begin{figure}[bth]
\includegraphics[width=85mm]{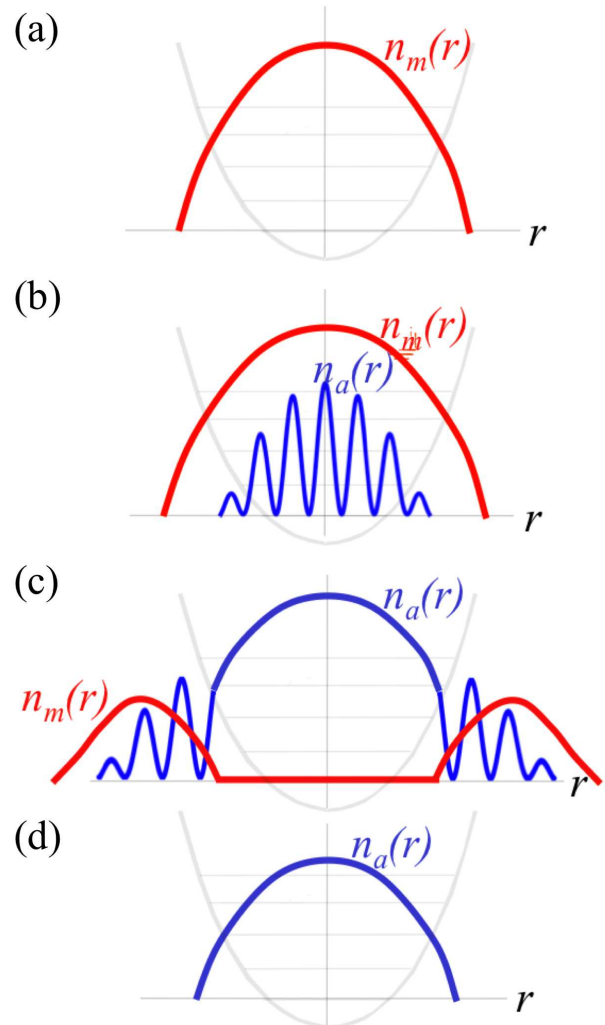}
\caption{(Color online) Schematic illustration of the shell structure expected in
  the $p$-wave resonant atomic cloud, corresponding to the phase
  diagram cuts in Fig.~\ref{fig:AMSFshells}.}
\label{fig:AMSFshellscut}
\end{figure}

Applying this LDA analysis to our system leads to a prediction of
rich, magnetic-field, atom-number, and temperature-tunable shell
structures in a $p$-wave resonant Bose gas, schematically illustrated
in Fig.~\ref{fig:AMSFshellscut}. For a range of atom number, detuning,
and temperature admitting the AMSF phase, we expect a cloud shell with
an $r$-dependent atomic condensate wavevector $q(r)$, given by
\bse
\begin{eqnarray}
  \label{eq:qr}
\hspace{-1cm}  q(r) &=& \alpha m \sqrt{n_m(r)},\\
  &\approx& q_0\sqrt{1-\frac{r^2}{R^2}},\ \ \mbox{for $r_{\rm MSF}< r < r_{\rm ASF}$}, \quad
\end{eqnarray}
\ese
where $r_{\rm MSF}(T,N,\nu)$ and $r_{\rm ASF}(T,N,\nu)$ are the boundaries of
the AMSF shell. 


\section{Summary and Conclusions}
\label{sec:summary}
To summarize, we studied a degenerate gas of two-species bosonic atoms
interacting through a $p$-wave FR, as realized, for
example, in a $^{85}$Rb-$^{87}$Rb mixture. We mapped out the
corresponding phase diagram and thermodynamic properties of the phases
as a function of temperature, atom number, and FR
detuning, and analyzed the nature of corresponding phase
transitions. We showed that at intermediate detuning such atomic
quantum gas generically exhibits an AMSF state with
atoms condensed at a finite tunable momentum $\ibQ(\nu)$ along a
direction set by the angular momentum axis of the molecular
condensate. This AMSF state undergoes quantum phase transitions
described by a quantum de Gennes model into a $p$-wave
(orbital spinor-1) MSF and into an $s$-wave
ASF at large negative and positive detunings,
respectively. A magnetic field can be used to tune the modulation
wavevector of the AMSF between zero and a value set by interactions as
well as to drive quantum phase transitions in this rich system.

\section{Acknowledgments}
\label{sec:acknowledge}
We thank V. Gurarie for discussions, and acknowledge financial support
by the National Science Foundation through grants No.\ DMR-1001240 and 
MRSEC DMR-0820579. L.R. also acknowledges support by the Miller Foundation and the University of Colorado,
and  thanks Berkeley Physics
Department for its hospitality during the initial part of this work.


\appendix
\section{Order parameter structure}
\label{OPstructure}
As discussed in the main text, the detailed nature of the AMSF states
depends on the structure (the set of reciprocal lattice vectors,
$\ibQ_n$) of the finite-momentum atomic order parameter. However,
because $\Psi_{\ibQ_n}$ depends on the details of the inter atomic
interactions and fluctuations, to determine its form in general is a
nontrivial problem, as exemplified by the FFLO problem and the
conventional crystallization. However, as seen in Sec.\ref{sec:mft},
for the case of the AMSF$_{\rm p}$ state, the problem simplifies
considerably as the energy is clearly minimized by a collinear state,
with $\ibQ_n$ parallel to $\bm\Phi$. Such collinear states fall into
two universality classes~\cite{FFLOlongLeo}, represented by the
FF-like~\cite{ff.64} and the
LO-like~\cite{lo.65} single harmonic forms,
\begin{align}
\Psi_\sigma^{\rm FF} &= \Psi_{\sigma, \ibQ_\sigma}e^{i\ibQ_\sigma\cdot\ibr}, \\
\Psi_\sigma^{\rm LO} &=
\Psi_{\sigma,\ibQ}e^{i\ibQ\cdot\ibr}+\Psi_{\sigma,-\ibQ}e^{-i\ibQ\cdot\ibr}.
\end{align}
In the FF-like (LO-like) state each species is characterized by a
single $\ibQ$ (double $\pm\ibQ$) momentum, exhibiting a uniform
(periodic) atomic density.

Focusing on these two collinear (FF and LO) states, in this appendix
we demonstrate that, generically (at least within the mean-field
theory), it is the FF state that is energetically selected by the
interactions.

To this end, we reexpress the mean-field energy densities for FF and
LO in terms of the corresponding eigenmodes, $\Psi_\pm^{\pm Q}$,
the latter involving two ($\pm Q$) critical modes,
\begin{align}
{\cal E}_{\rm FF} &= (\varepsilon_\ibQ-\ve\Delta_\ibQ\ve)\ve\Psi_-^Q\ve^2 
+ \frac{1}{2}\lambda\ve\Psi_-^Q\ve^4, \\
{\cal E}_{\rm LO} 
&=(\varepsilon_\ibQ-\ve\Delta_\ibQ\ve)(\ve\Psi_-^Q\ve^2+\ve\Psi_{-}^{-Q}\ve^2) 
\nonumber \\
&+\frac{1}{2}\lambda(\ve\Psi_{-}^Q\ve^2+\ve\Psi_-^{-Q}\ve^2)^2 
+\lambda'\ve\Psi_-^Q\ve^2\ve\Psi_-^{-Q}\ve^2,
\label{lolike2}
\end{align} 
where $\lambda = \frac{1}{4}(\lambda_1+\lambda_2+2\lambda_{12})$ and 
$\lambda' = \frac{1}{4}(\lambda_1+\lambda_2-2\lambda_{12})$.

These free energies thus show that the energetically preferred form of the
AMSF state is determined by the coefficient $\lambda'$ of last term in
Eq.~\eqref{lolike2}.  For $\lambda' > 0$, that is,
$\lambda_{1}+\lambda_{2} > 2\lambda_{12}$, the single $Q$ FF-like
state is selected.  On the other hand, for $\lambda'<0$, that is,
$\lambda_{1}+\lambda_{2} < 2\lambda_{12}$, it is the LO-like state
that has the lowest energy.

Combining the above requirement on $\lambda'$ for the stability of the
LO-like state with the condition for two-species miscibility,
$\lambda_1\lambda_2 > \lambda_{12}^2$, we find an inequality,
\be
\frac{\lambda_1+\lambda_2}{2} < \lambda_{12} <  \sqrt{\lambda_1\lambda_2}
\ee
which for positive couplings $\lambda_i$ can be shown to have a
zero range of stability.  Thus, as advertised, within mean-field
approximation it is the single $Q$ FF-like AMSF state that is always
energetically selected. Perhaps the LO-like AMSF form can be realized
for a metastable atomic gas with $\lambda_i < 0$, as, for example, realized by
a $^{87}\text{Rb}$-$^{85}\text{Rb}$ mixture.


\begin{thebibliography}{47}
\bibitem{commentFRequalmass} 
For unequal masses FR interaction is given by
$H_{FR}=\alpha\hat{\bm\phi}^{\dag}\cdot
\left[\frac{m_1}{m_1+m_2}\hat\psi_{1}(-i\grad)\hat\psi_{2} 
-\frac{m_2}{m_1+m_2}\hat\psi_{2}(-i\grad)\hat\psi_{1}
\right]+h.c.$
as required to preserve Galilean invariance.
\bibitem{boxtrap} For simplicity it is convenient to first consider a
  homogeneous bulk system or a ``box'' trap, and take into account the
  trap-induced local inhomogeneity using LDA, valid for large
  condensates and smooth trapping potential.
\bibitem{commentNoMI} In the absence of an optical potential or
  quenched disorder that can ``localize'' atom number and induce Mott-
  or Bose-glass insulating phases, only Bose-condensed, superfluid
  ground states are expected at zero temperature.
\bibitem{commentSymmetries} $U_N(1)$ and $U_{\Delta N}(1)$ symmetries
  are associated with the total atom number $N=N_1+N_2+2N_m$ and
  difference $\Delta N = N_1 - N_2$ conservations, respectively.
\bibitem{comment_g} 
More accurately, it is not $g_{2}$ but T-matrix, i.e., scattering length $a$'s that matter.
\bibitem{commentTerminology} In the context of conventional spinor
  condensate, the polar state is also termed ``nematic'' or
 ``antiferromagnetic'' superfluid.
\bibitem{commentNearTc} To keep the form of the Landau free energy
  simple, one needs to be sufficiently close to the transition
  boundaries (but staying outside the critical region, where local
  Landau mean-field expansion fails) to allow a local expansion in the
  order parameters.
\end{thebibliography}

\end{document}